\documentclass{article}

\usepackage{microtype}
\usepackage{graphicx}
\usepackage{subfigure}
\usepackage{booktabs} 

\usepackage{hyperref}

\usepackage{color, colortbl}
\definecolor{LightCyan}{rgb}{0.88,1,1}

\usepackage[table]{xcolor}
\definecolor{LightCyan}{rgb}{0.88,1,1}

\usepackage[accepted]{icml2025}

\usepackage[ruled, lined, longend, linesnumbered, algo2e]{algorithm2e}

\usepackage{amsmath}
\usepackage{amssymb}
\usepackage{mathtools}
\usepackage{amsthm}

\usepackage{enumitem}
\usepackage{wrapfig}

\usepackage{multirow}
\usepackage{wrapfig,lipsum,booktabs}

\let\norm\undefined 
\DeclarePairedDelimiter\norm{\lVert}{\rVert}

\let\abs\undefined 
\DeclarePairedDelimiter\abs{\lvert}{\rvert}

\usepackage{booktabs,nicematrix}

\usepackage[capitalize,noabbrev]{cleveref}

\theoremstyle{plain}
\newtheorem{theorem}{Theorem}[section]
\newtheorem{proposition}[theorem]{Proposition}

\theoremstyle{definition}

\theoremstyle{remark}

\usepackage[textsize=tiny]{todonotes}

\icmltitlerunning{RestoreGrad: Signal Restoration Using Conditional Denoising Diffusion Models with Jointly Learned Prior}

\begin{document}

\twocolumn[
\icmltitle{RestoreGrad: Signal Restoration Using Conditional Denoising \\ Diffusion Models with Jointly Learned Prior}




\begin{icmlauthorlist}
\icmlauthor{Ching-Hua Lee}{comp}
\icmlauthor{Chouchang Yang}{comp}
\icmlauthor{Jaejin Cho}{comp}
\icmlauthor{Yashas Malur Saidutta}{comp}
\icmlauthor{Rakshith Sharma Srinivasa}{comp}
\icmlauthor{Yilin Shen}{comp}
\icmlauthor{Hongxia Jin}{comp}
\end{icmlauthorlist}

\icmlaffiliation{comp}{Artificial Intelligence Center$-$Mountain View, Samsung Electronics, Mountain View, CA, USA}

\icmlcorrespondingauthor{Ching-Hua Lee}{chinghua.l@samsung.com}

\icmlkeywords{Machine Learning, ICML}

\vskip 0.3in
]



\printAffiliationsAndNotice{}  

\begin{abstract}
    Denoising diffusion probabilistic models (DDPMs) can be utilized to recover a clean signal from its degraded observation(s) by conditioning the model on the degraded signal. The degraded signals are themselves contaminated versions of the clean signals; due to this correlation, they may encompass certain useful information about the target clean data distribution. However, existing adoption of the standard Gaussian as the prior distribution in turn discards such information when shaping the prior, resulting in sub-optimal performance. In this paper, we propose to improve conditional DDPMs for signal restoration by leveraging a more informative prior that is jointly learned with the diffusion model. The proposed framework, called RestoreGrad, seamlessly integrates DDPMs into the variational autoencoder (VAE) framework, taking advantage of the correlation between the degraded and clean signals to encode a better diffusion prior. On speech and image restoration tasks, we show that RestoreGrad demonstrates faster convergence (5-10 times fewer training steps) to achieve better quality of restored signals over existing DDPM baselines and improved robustness to using fewer sampling steps in inference time (2-2.5 times fewer), advocating the advantages of leveraging jointly learned prior for efficiency improvements in the diffusion process.
\end{abstract}

\section{Introduction}
\label{sec: intro}

Denoising diffusion probabilistic models (DDPMs) \citep{ho2020denoising,sohl2015deep} are latent variable generative models consisting of i) the \textit{forward process}, where the original data samples are gradually corrupted by adding Gaussian noise to eventually become a standard normal prior; ii) the \textit{reverse process}, in which a neural network is responsible for recovering the original data from the corrupted samples by learning to sequentially reverse the diffusion process. With their exceptional capabilities of generating high-quality data, DDPMs can be applied to various signal restoration tasks -- recovering the missing components in a signal due to contamination (e.g., audio recorded with environmental noises \citep{lu2021study,lu2022conditional,tai2023revisiting}, images obstructed by various measurement noises \citep{ozdenizci2023restoring,croitoru2023diffusion}), by conditioning the DDPM on the degraded observations.

However, for the diffusion model to adequately learn the reverse process, a large number of training iterations may be required, leading to potentially slow convergence. Such inefficiency was related to the discrepancy between the real data distribution and the accustomed choice of the standard Gaussian prior by \citet{lee2021priorgrad}. They therefor proposed a simple yet effective approach called \textit{PriorGrad}, aiming to construct a better prior by using rule-based approaches to extract useful information from the conditioner data. However, despite improving performance on some generative speech tasks, handcrafting a ``better'' prior requires certain knowledge about the data characteristics, and such guidance may not always exist for a given task or application. 

\begin{figure}[!t]
    \centering
     \includegraphics[width=\linewidth]{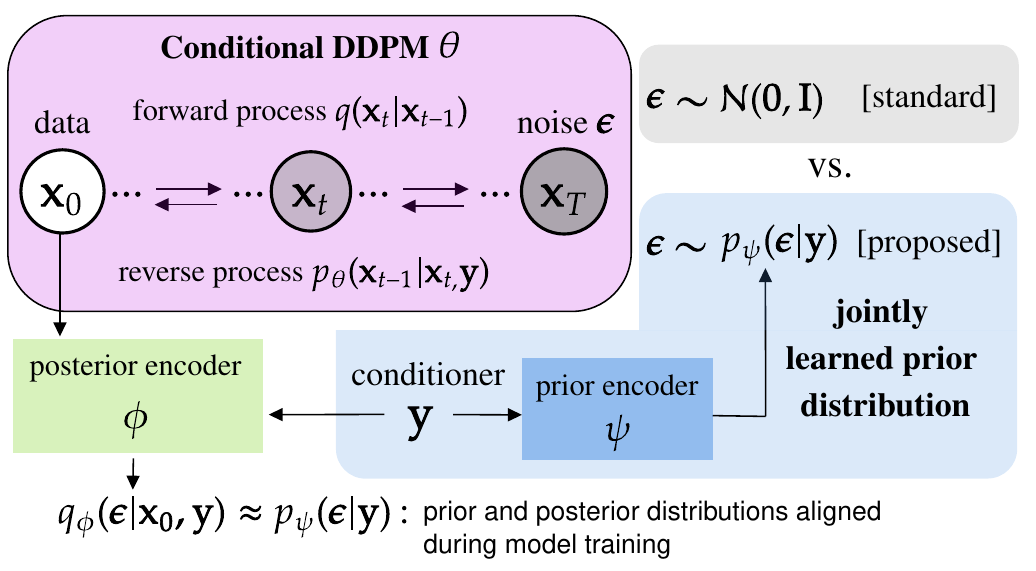}
     \vspace{-0.6cm}
    \caption{High-level view of the proposed method.} 
\label{fig: overview}
\vspace{-0.1cm}
\end{figure}

In this paper, our main focus is to investigate the question: \textit{Can we systematically obtain a better prior distribution that improves the efficiency of the diffusion generative process?} In other words, we aim to develop a framework of \textit{learning-based} diffusion priors for improved DDPM efficiency. A high-level view of our approach is depicted in Figure \ref{fig: overview}, where the conditional DDPM (parameterized by $\theta$) samples the latent noise $\boldsymbol{\epsilon}$ from a learned prior distribution estimated by a \textit{prior encoder} $\psi$, which takes the conditioner $\mathbf{y}$ as input. 
The prior encoder is jointly trained with the DDPM $\theta$ to synthesize the data $\mathbf{x}_0$, and a \textit{posterior encoder} $\phi$ that exploits information from both $\mathbf{x}_0$ and $\mathbf{y}$, to align the prior and posterior distributions. The main idea here is that, if there is a certain correlation between the conditioner $\mathbf{y}$ and the target data $\mathbf{x}_0$, e.g., in signal restoration problems where $\mathbf{y}$ is typically a degraded version of $\mathbf{x}_0$, our framework can exploit such correlation to construct a more informative prior in an automatic and systematic manner.

\begin{figure}[!t]
    \centerline{\includegraphics[width=1\linewidth]{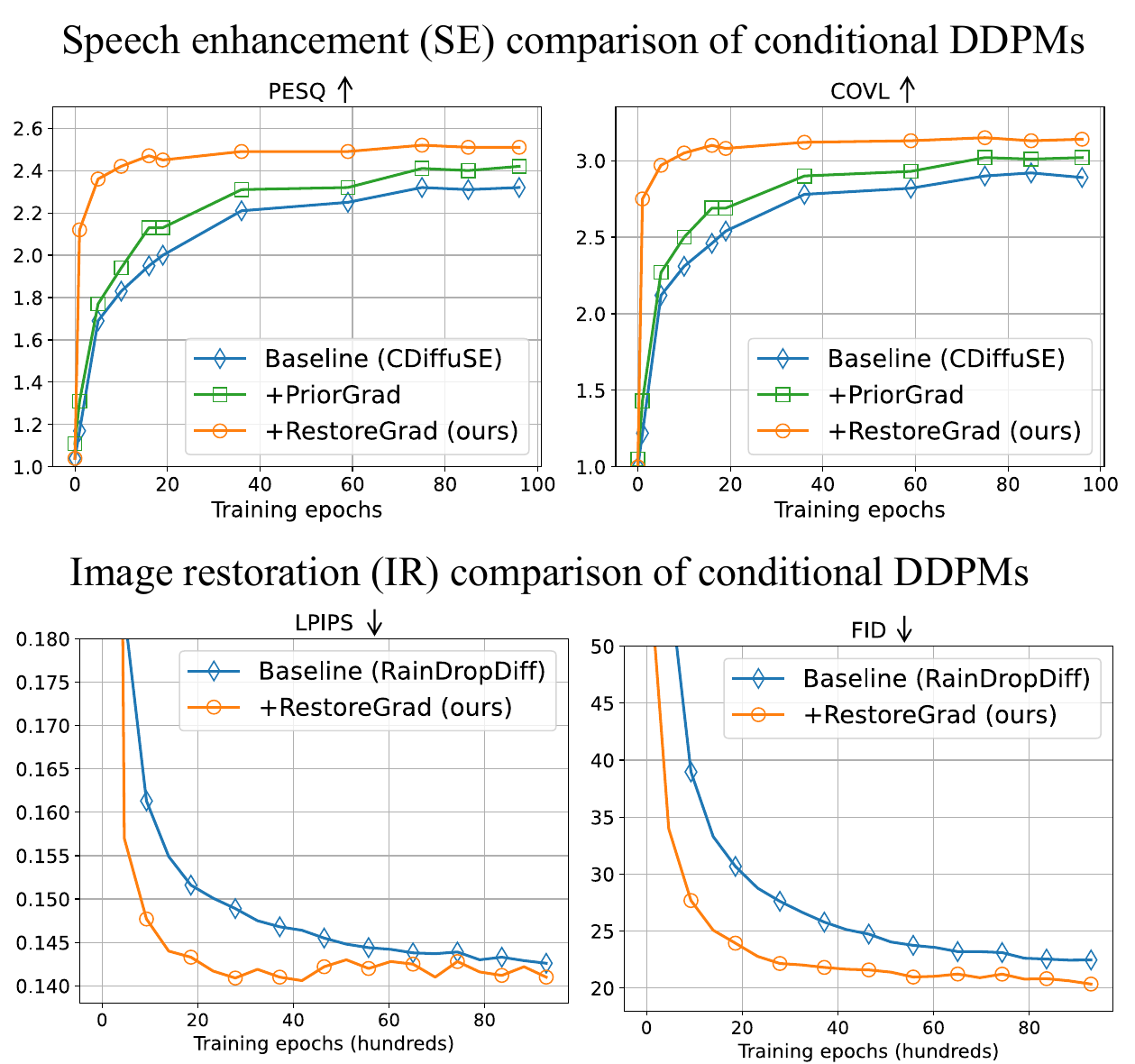}}
    \vspace{-0.15cm}
    \caption{Model learning performance. (Top) In the speech domain, RestoreGrad outperforms PriorGrad \citep{lee2021priorgrad}, a recently proposed improvement to the baseline DDPM (CDiffuSE \citep{lu2022conditional}) which leverages handcrafted priors. (Bottom) In the image domain, RestoreGrad provides a paradigm to improve the DDPM baseline (RainDropDiff \citep{ozdenizci2023restoring}) when there is no existing recipe for obtaining better priors.} 
\label{fig: training_curve}
\end{figure}

\begin{figure}[!t]  
    \centerline{\includegraphics[width=1\linewidth]{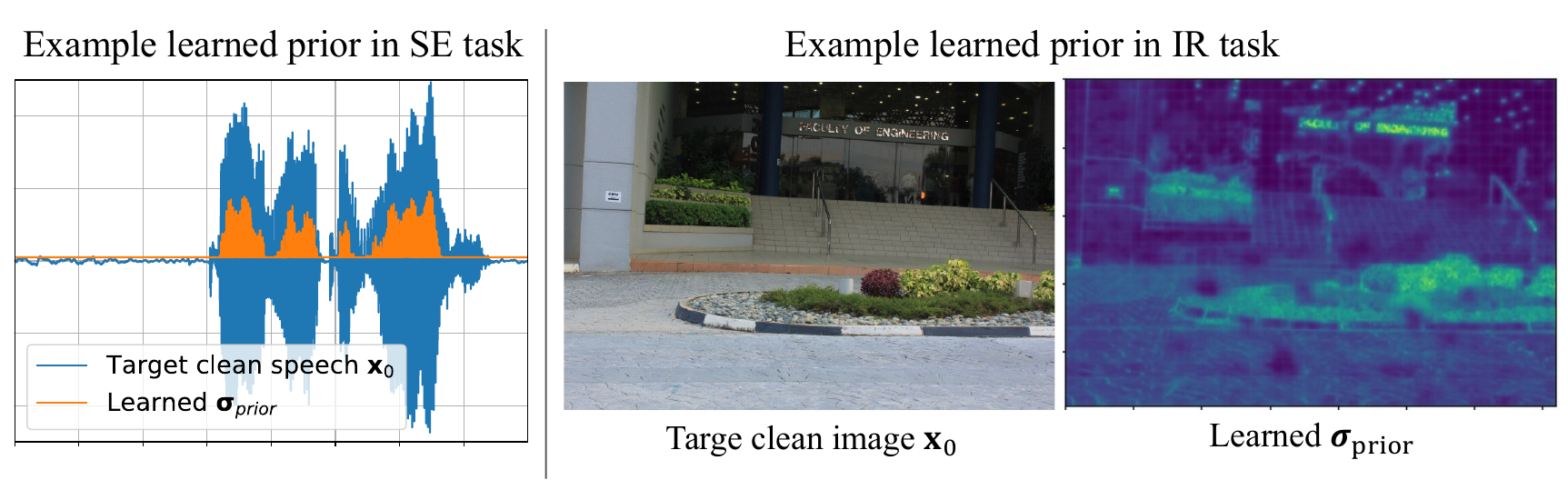}}
    \vspace{-0.15cm}
    \caption{Visualization of the learned prior distribution. Here, the distribution of the prior is modeled as: $p_\psi(\boldsymbol{\epsilon}|\mathbf{y})\coloneq\mathcal{N}(\boldsymbol{\epsilon}; \mathbf{0},\mbox{diag}\{\boldsymbol{\sigma}^2_{\text{prior}}(\mathbf{y};\psi)\})$, where $\boldsymbol{\sigma}_{\text{prior}}$ is estimated by the prior encoder $\psi$ with input $\mathbf{y}$. It appears that $\boldsymbol{\sigma}_{\text{prior}}$ follows the level variation of the speech waveform (in SE) and preserves the structure of the original image (in IR). This indicates that an informative prior approximating the data distribution has been obtained, leading to improved efficiency of the diffusion process.} 
\label{fig: learned priors}
\end{figure}

To explore the idea, we introduce \textit{RestoreGrad}, a new paradigm for improving conditional DDPM by learning the prior distribution in tandem with the diffusion model, focusing on signal restoration applications. We apply RestoreGrad to speech enhancement (SE) and image restoration (IR) tasks to demonstrate its generality for signals of different nature. For SE, we compare with PriorGrad \citep{lee2021priorgrad} which provides guidance on handcrafting suitable priors in the speech domain. For IR, we show that RestoreGrad serves as a promising solution for improving the baseline DDPM even in a domain that lacks such a recipe for engineering the prior. As shown in Figure \ref{fig: training_curve}, models trained using RestoreGrad are more data and compute-efficient than the baseline DDPMs and PriorGrad; they converge faster to achieve higher quality of the restored signal. Further shown in Figure \ref{fig: learned priors}, the learned prior is more informative as it better correlates with the desired signal than an isotropic covariance, potentially simplifying the diffusion trajectory for improved model efficiency. 

Our main contributions are summarized as follows:
\begin{itemize}[leftmargin=*]
\vspace{-0.25cm}
    \item We study the problem of learning the prior distribution \textit{jointly} with the conditional DDPM for signal restoration applications, aiming at providing a more systematic, \textit{learning-based} treatment to address the inefficiency incurred by existing selections of the prior distribution. 
    \vspace{-0.05cm}
    \item We propose a new framework called RestoreGrad that learns the prior in conjuncture with the DDPM model through a \textit{prior encoder}, by exploiting the correlation between the targe signal and input degraded signal encoded by an auxiliary \textit{posterior encoder}, for improved model efficiency. 
    Our \textit{two-encoder} learning framework is established based on a novel evidence lower bound (ELBO) that seamlessly integrates the DDPM into the variational autoencoder (VAE) \citep{kingma2013auto} to harness the advantages of both methodologies.
    \vspace{-0.05cm}
    \item Experiments demonstrate that the proposed paradigm is quite general and can benefit both training and sampling of DDPMs, achieving considerable improvements with lightweight encoders in high quality signal restoration tasks of various modalities including images and audio.
\end{itemize}

\section{Background on DDPMs}

\subsection{Forward Process}
DDPMs \citep{ho2020denoising,sohl2015deep} slowly corrupt the training data using Gaussian noise in the forward process. Let $q_{\text{data}}(\mathbf{x}_0)$ be the data density of the original data $\mathbf{x}_0$. The forward process is a fixed Markov Chain that sequentially corrupts
the data $\mathbf{x}_0 \sim q_{\text{data}}(\mathbf{x}_0)$ in $T$ diffusion steps, by injecting Gaussian noise according to a variance schedule $\{\beta_t\}_{t=1}^{T}\in(0,1)$:
\begin{equation}
    q(\mathbf{x}_{1:T}|\mathbf{x}_0)\coloneq\prod_{t=1}^Tq(\mathbf{x}_t|\mathbf{x}_{t-1}),
\label{eq: ddpm forward}
\end{equation}
where $q(\mathbf{x}_t|\mathbf{x}_{t-1})\coloneq\mathcal{N}(\mathbf{x}_t;\sqrt{1-\beta_t}\mathbf{x}_{t-1}, \beta_t\mathbf{I})$ is the transition probability at step $t$. It allows the direct sampling of $\mathbf{x}_t$ according to $q(\mathbf{x}_t|\mathbf{x}_0)=\mathcal{N}(\mathbf{x}_t;\sqrt{\bar{\alpha}_t}\mathbf{x}_0,\sqrt{1-\bar{\alpha}_t}\mathbf{I})$, where $\bar{\alpha}_t\coloneq\prod_{i=1}^t\alpha_i$ with $\alpha_t\coloneq 1-\beta_t$; i.e., we can sample $\mathbf{x}_t=\sqrt{\bar{\alpha}_t}\mathbf{x}_0+\sqrt{1-\bar{\alpha}_t}\boldsymbol{\epsilon}$, where $\boldsymbol{\epsilon}\sim\mathcal{N}(\mathbf{0},\mathbf{I})$. A notable assumption is that with a carefully designed variance schedule $\beta_t$ and large enough $T$, such that $\bar{\alpha}_T$ is sufficiently small, $q(\mathbf{x}_T|\mathbf{x}_0)$ converges to $\mathcal{N}(\mathbf{x}_T;\mathbf{0},\mathbf{I})$ so that the distribution of $\mathbf{x}_T$ is well approximated by the standard Gaussian.

\subsection{Reverse Process}
One can generate new data samples from $q_{\text{data}}(\mathbf{x}_0)$ by reversing the predefined forward process utilizing the same functional form. That is, we can progressively transform a noise $\mathbf{x}_T\sim p(\mathbf{x}_T)=\mathcal{N}(\mathbf{x}_T; \mathbf{0}, \mathbf{I})$ back into the data by approximating the reverse of the forward transition probability. This process is defined by the joint distribution $p_\theta(\mathbf{x}_{0:T})$ of a Markov Chain with learned transitions:
\begin{equation}
    p_\theta(\mathbf{x}_{0:T})\coloneq p(\mathbf{x}_T)\prod_{t=1}^Tp_\theta(\mathbf{x}_{t-1}|\mathbf{x}_t),
\label{eq: ddpm reverse}
\end{equation}
where $p_\theta(\mathbf{x}_{t-1}|\mathbf{x}_t)\coloneq\mathcal{N}(\mathbf{x}_{t-1};\boldsymbol{\mu}_\theta(\mathbf{x}_t,t), \boldsymbol{\Sigma}_\theta(\mathbf{x}_t,t))$ is the reverse transition probability parameterized by a network $\theta$. 

\subsection{DDPM Learning Framework}
Ideally, we would train the model $\theta$ with a maximum likelihood objective such that the probability assigned by the model $p_\theta(\mathbf{x}_0)$ to each training example is as large as possible, which is unfortunately intractable \citep{croitoru2023diffusion}. To circumvent such difficulty, DDPMs \citep{ho2020denoising} instead maximize an ELBO of the data log-likelihood, by introducing a sequence of hidden variables $\mathbf{x}_{1:T}$ and the approximate variational distribution $q(\mathbf{x}_{1:T}|\mathbf{x}_0)$: 
\begin{equation}
    \log p(\mathbf{x}_0)\geq\mathbb{E}_{q(\mathbf{x}_{1:T}|\mathbf{x}_0)}\left[\log\frac{p_\theta({\mathbf{x}_{0:T}})}{q(\mathbf{x}_{1:T}|\mathbf{x}_0)}\right].
\label{eq: ddpm log likelihood}
\end{equation}
With the above parametric modeling of the forward and reverse processes, the ELBO (\ref{eq: ddpm log likelihood}) suggests training the network $\theta$ such that, at each time step $t$, $p_\theta(\mathbf{x}_{t-1}|\mathbf{x}_t)$ is as close as possible to the true forward process posterior conditioned on $\mathbf{x}_0$ \citep{luo2022understanding,croitoru2023diffusion}, i.e., $q(\mathbf{x}_{t-1}|\mathbf{x}_t,\mathbf{x}_0)=\mathcal{N}(\mathbf{x}_{t-1};\tilde{\boldsymbol{\mu}}_t(\mathbf{x}_t,\mathbf{x}_0),\tilde{\beta}_t\mathbf{I})$,
where $\tilde{\beta}_t\coloneq\frac{1-\bar{\alpha}_{t-1}}{1-\bar{\alpha}_t}\beta_t$ and  $\tilde{\boldsymbol{\mu}}_t(\mathbf{x}_t,\mathbf{x}_0)\coloneq\frac{\sqrt{\bar{\alpha}_{t-1}}\beta_t}{1-\bar{\alpha}_t}\mathbf{x}_0+\frac{\sqrt{\alpha_t}(1-\bar{\alpha}_{t-1})}{1-\bar{\alpha}_t}\mathbf{x}_t$.

Based on using a fixed covariance $\boldsymbol{\Sigma}_\theta(\mathbf{x}_t,t)=\sigma_t^2\mathbf{I}$ (e.g., $\sigma_t^2=\tilde{\beta}_t$) as in \citet{ho2020denoising}, maximizing (\ref{eq: ddpm log likelihood}) corresponds to training a network $\boldsymbol{\mu}_\theta(\mathbf{x}_t,t)$ that predicts $\tilde{\boldsymbol{\mu}}_t(\mathbf{x}_t,\mathbf{x}_0)$. Alternatively, \citet{ho2020denoising} suggested the following reparameterization to rewrite the mean as a function of noise: 
\begin{equation}
       \boldsymbol{\mu}_\theta(\mathbf{x}_t,t)=\frac{1}{\sqrt{\alpha_t}}\left(\mathbf{x}_t-\frac{\beta_t}{\sqrt{1-\bar{\alpha}_t}}\boldsymbol{\epsilon}_\theta(\mathbf{x}_t,t)\right).
\label{eq: mean}
\end{equation}
They train a network $\boldsymbol{\epsilon}_\theta(\mathbf{x}_t,t)$ to predict the real noise $\boldsymbol{\epsilon}\sim\mathcal{N}(\mathbf{0},\mathbf{I})$ and use (\ref{eq: mean}) to compute the mean. Practically it is carried out by minimizing a simplified training objective: 
\begin{equation}
    \mathcal{L}_{\text{simple}}(\theta)\coloneq\mathbb{E}_{\mathbf{x}_0,\boldsymbol{\epsilon},t}\left[\norm{\boldsymbol{\epsilon}-\boldsymbol{\epsilon}_\theta(\mathbf{x}_t, t)}^2\right],
\label{eq: ddpm loss simple}
\end{equation}
which measures, for a random time step $t\sim\mathcal{U}(\{1,\dots,T\})$, the distance between the actual noise and estimated noise.

\subsection{Signal Restoration by Conditional DDPMs}
Signal restoration is concerned with recovering the original signals from their degraded observations, which are of paramount importance in reality but remaining challenging, as noises are ubiquitous and may be strong enough to cause significant degradation of the signal quality. Recently, adoption of deep generative models \citep{kingma2013auto,goodfellow2014generative,ho2020denoising} for signal restoration tasks has considerably increased due to their remarkable capabilities of generating missing components in the data, with conditional DDPMs \citep{croitoru2023diffusion,cao2024survey} demonstrating substantial promise. More formally, let $\mathbf{y}$ denote the degraded observation of the clean signal $\mathbf{x}_0$. Recovering $\mathbf{x}_0$ given $\mathbf{y}$ by a model $\theta$ can be cast as maximizing the conditional likelihood of data $p_\theta(\mathbf{x}_0|\mathbf{y})$. The problem is in general intractable, but can be approximated by using a DDPM conditioned on $\mathbf{y}$. The main idea is to learn a diffusion model $\theta$ with $\mathbf{y}$ provided as a conditioner in the reverse process (\ref{eq: ddpm reverse}):
\begin{equation}
    p_\theta(\mathbf{x}_{0:T}|\mathbf{y})\coloneq p(\mathbf{x}_T)\prod_{t=1}^Tp_\theta(\mathbf{x}_{t-1}|\mathbf{x}_t,\mathbf{y}),
\label{eq: cddpm reverse}
\end{equation}
where $p_\theta(\mathbf{x}_{t-1}|\mathbf{x}_t,\mathbf{y})\coloneq\mathcal{N}(\mathbf{x}_{t-1};\boldsymbol{\mu}_\theta(\mathbf{x}_t,\mathbf{y},t), \sigma_t^2\mathbf{I})$ assuming a fixed covariance. In practice, a noise estimator network $\boldsymbol{\epsilon}_\theta(\mathbf{x}_t, \mathbf{y}, t)$ is adopted to predict the mean, following the practice in \citet{ho2020denoising}.
\section{Proposed Method: Integrating DDPM and VAE for Learnable Diffusion Prior}

We start with the conditional VAE \citep{sohn2015learning} formulation to maximize the conditional data log-likelihood, $\log p(\mathbf{x}_0|\mathbf{y})=\log\int p(\mathbf{x}_0,\boldsymbol{\epsilon}|\mathbf{y}) d\boldsymbol{\epsilon}$, where $\boldsymbol{\epsilon}$ is an introduced latent variable. To avoid intractable integral, in VAEs an ELBO is utilized as the surrogate objective by introducing an approximate posterior $q(\boldsymbol{\epsilon}|\mathbf{x}_0,\mathbf{y})$ \citep{harvey2022conditional}:
\begin{equation}
\begin{aligned}
    \log p(\mathbf{x}_0|\mathbf{y})\geq&\underbrace{\mathbb{E}_{q(\boldsymbol{\epsilon}|\mathbf{x}_0,\mathbf{y})}\left[\log p(\mathbf{x}_0|\mathbf{y},\boldsymbol{\epsilon})\right]}_{\text{reconstruction term}} \\ 
    &- \underbrace{D_{\text{KL}}\left(q(\boldsymbol{\epsilon}|\mathbf{x}_0,\mathbf{y}) || p(\boldsymbol{\epsilon}|\mathbf{y})\right)}_{\text{prior matching term}}.
\label{eq: vae original elbo}
\end{aligned}
\end{equation}
The \textit{reconstruction} and \textit{prior matching} terms are typically realized by an encoder-decoder architecture with $\boldsymbol{\epsilon}$ being the bottleneck representation sampled from the latent distribution. VAEs generally benefit from learnable latent spaces for good modeling efficiency. However, their generative performance often lags behind DDPMs that employ an iterative, more sophisticated decoding (reconstruction) process.

In this work, our aim is to embrace the best of both worlds, i.e., \textit{remarkable generative power (DDPM) and modeling efficiency (VAE) to achieve improved output signal quality and training/sampling efficiency simultaneously.} 

\vspace{0.2cm}

\begin{proposition}[Incorporation of diffusion process into VAE]
    By introducing a sequence of hidden variables $\mathbf{x}_{1:T}$, under the setup of conditional diffusion models where the Markov Chain assumption is employed on the forward process $q(\mathbf{x}_{1:T}|\mathbf{x}_0)\coloneq\prod_{t=1}^Tq(\mathbf{x}_t|\mathbf{x}_{t-1})$ and the reverse process $p_\theta(\mathbf{x}_{0:T}|\mathbf{y})\coloneq p(\mathbf{x}_T)\prod_{t=1}^Tp_\theta(\mathbf{x}_{t-1}|\mathbf{x}_t,\mathbf{y})$ parameterized by a DDPM $\theta$, and assuming that $\mathbf{x}_T=\boldsymbol{\epsilon}$ (i.e., the latent noise of DDPM samples from the VAE latent distribution), we have the lower bound on $\log p(\mathbf{x}_0|\mathbf{y},\boldsymbol{\epsilon})$ in the reconstruction term of the VAE (\ref{eq: vae original elbo}) as:
    \begin{equation}
        \log p(\mathbf{x}_0|\mathbf{y},\boldsymbol{\epsilon})\geq\mathbb{E}_{q(\mathbf{x}_{1:T}|\mathbf{x}_0)}\left[\log\frac{p_\theta({\mathbf{x}_{0:T}}|\mathbf{y})}{q(\mathbf{x}_{1:T}|\mathbf{x}_0)}\right],
    \label{eq: cddpm lower bound}
    \end{equation}
    which is the ELBO of the conditional DDPM (i.e., the conditional version of (\ref{eq: ddpm log likelihood})). 
\label{proposition: 1}
\end{proposition}
The proof (based on Markov Chain property) is provided in Appendix \ref{sec: appendix proof prop 1}. Note that the assumption $\mathbf{x}_T=\boldsymbol{\epsilon}$ follows the standard DDPM to sample $\mathbf{x}_T$ from the distribution of the prior noise $\boldsymbol{\epsilon}$, practically achieved by using a large enough $T$ and a carefully designed variance schedule $\{\beta_t\}_{t=1}^T$. In our case, we adopt the same assumption to enable sampling $\mathbf{x}_T$ from the latent space of VAE. We interpret Proposition \ref{proposition: 1} as a seamless integration of DDPM into the VAE framework to achieve improved generative (decoding) capabilities. 

Having incorporated the DDPM as the decoder, we now discuss the encoding part, i.e., the prior matching term in (\ref{eq: vae original elbo}). A straightforward design could be using a network $\psi$ (\textit{Prior Net}) to parameterize the prior distribution as $p_\psi(\boldsymbol{\epsilon}|\mathbf{y})$, while assuming the posterior to be a fixed form of distribution like the standard Gaussian. However, this may in turn discard any useful information inherent between $\mathbf{x}_0$ and $\mathbf{y}$. To take advantage of the adequate correlation present in signal restoration settings, we propose to also parameterize the posterior distribution with another network $\phi$ (\textit{Posterior Net}), to incorporate richer information about the target signal distribution into the learning of the prior. Together with (\ref{eq: cddpm lower bound}), we introduce the \textbf{new lower bound} of the conditional data log-likelihood:
\begin{equation}
\begin{aligned}
    \log p(\mathbf{x}_0|\mathbf{y})\geq&\mathbb{E}_{q_\phi(\boldsymbol{\epsilon}|\mathbf{x}_0,\mathbf{y})}\Biggr[\underbrace{\mathbb{E}_{q(\mathbf{x}_{1:T}|\mathbf{x}_0)}\left[\log\frac{p_{\theta}(\mathbf{x}_{0:T}|\mathbf{y})}{q(\mathbf{x}_{1:T}|\mathbf{x}_0)}\right]}_{\text{conditional DDPM}}\Biggr] \\
    &- D_{\text{KL}}\bigr(\underbrace{q_{\phi}(\boldsymbol{\epsilon}|\mathbf{x}_0,\mathbf{y})}_{\text{Posterior Net}} || \underbrace{p_{\psi}(\boldsymbol{\epsilon}|\mathbf{y})}_{\text{Prior Net}}\bigr).
\label{eq: vae elbo}
\end{aligned}
\end{equation}
Based on the assumption $\mathbf{x}_T=\boldsymbol{\epsilon}$, the conditional DDPM samples the latent noise $\mathbf{x}_T$ from the distribution of $\boldsymbol{\epsilon}$ which is jointly estimated by the two encoders, $\phi$ and $\psi$. The two-encoder design is inspired by \citet{kohl2018probabilistic} for image segmentation with traditional U-Nets. Here, we adopt the idea in the context of DDPM for signal restoration, which is effective as it incorporates posterior information exploiting the correlation between clean and degraded signals. Based on (\ref{eq: vae elbo}), we introduce the training objective of RestoreGrad:

\begin{figure*}[!t]  
    \centerline{\includegraphics[width=1\linewidth]{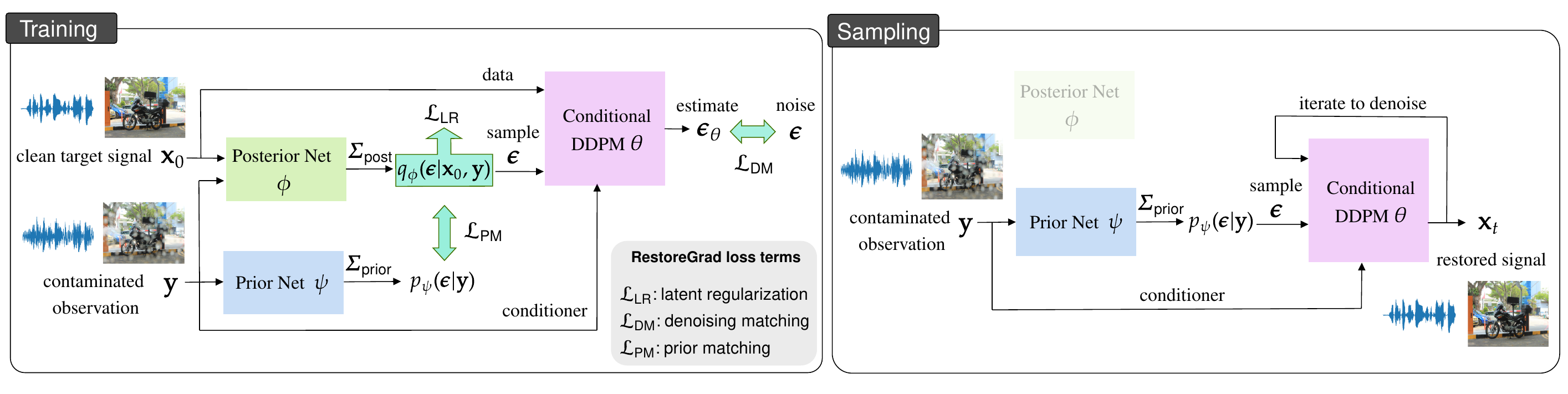}}
    \vspace{-0.5cm}
    \caption{Proposed RestoreGrad. During training, the conditional DDPM $\theta$, Prior Net $\psi$, and Posterior Net $\phi$ are jointly optimized by (\ref{eq: elbo of restoregrad}). During inference, the DDPM $\theta$ samples the latent noise $\boldsymbol{\epsilon}$ from the jointly learned prior distribution to synthesize the clean signal. (Summary of the algorithm details is presented in Appendix \ref{sec: algorithms}.)} 
\label{fig: restoregrad}
\vspace{-0.15cm}
\end{figure*} 

\vspace{0.2cm}

\begin{proposition}[RestoreGrad]
    Assume the prior and posterior distributions are both zero-mean Gaussian, parameterized as $p_{\psi}(\boldsymbol{\epsilon}|\mathbf{y})=\mathcal{N}(\boldsymbol{\epsilon};\mathbf{0},\boldsymbol{\Sigma}_{\text{prior}}(\mathbf{y};\psi))$ and $q_{\phi}(\boldsymbol{\epsilon}|\mathbf{x}_0, \mathbf{y})=\mathcal{N}(\boldsymbol{\epsilon};\mathbf{0},\boldsymbol{\Sigma}_{\text{post}}(\mathbf{x}_0,\mathbf{y};\phi))$, respectively, where the covariances are estimated by the Prior Net $\psi$ (taking $\mathbf{y}$ as input) and Posterior Net $\phi$ (taking both $\mathbf{x}_0$ and $\mathbf{y}$ as input). Let us simply use $\boldsymbol{\Sigma}_{\text{prior}}$ and $\boldsymbol{\Sigma}_{\text{post}}$ hereafter to refer to $\boldsymbol{\Sigma}_{\text{prior}}(\mathbf{y};\psi)$ and $\boldsymbol{\Sigma}_{\text{post}}(\mathbf{x}_0,\mathbf{y};\phi)$ for concise notation. 
    Then, with the direct sampling property in the forward path $\mathbf{x}_t=\sqrt{\bar{\alpha}_t}\mathbf{x}_0+\sqrt{1-\bar{\alpha}_t}\boldsymbol{\epsilon}$ at arbitrary timestep $t$ where $\boldsymbol{\epsilon}\sim q_{\phi}(\boldsymbol{\epsilon}|\mathbf{x}_0,\mathbf{y})$, and assuming the reverse process has the same covariance as the true forward process posterior conditioned on $\mathbf{x}_0$, by utilizing the conditional DDPM $\boldsymbol{\epsilon}_\theta(\mathbf{x}_t,\mathbf{y},t)$ as the noise estimator of the true noise $\boldsymbol{\epsilon}$, we have the modified ELBO, $-\mathcal{L}(\theta,\phi,\psi)$, associated with (\ref{eq: vae elbo}): 
    \begin{equation}
    \begin{aligned}
        \mathcal{L}(\theta,\phi,\psi) & = \underbrace{\frac{\bar{\alpha}_T}{2}\mathbb{E}_{\mathbf{x}_0}\norm{\mathbf{x}_0}^2_{\boldsymbol{\Sigma}^{-1}_{\text{post}}}+\frac{1}{2}\log\abs{\boldsymbol{\Sigma}_{\text{post}}}}_{\text{Latent Regularization (LR) terms}} \\
        +&\underbrace{\sum_{t=1}^{T}\gamma_t\mathbb{E}_{(\mathbf{x}_0,\mathbf{y}),\boldsymbol{\epsilon}\sim \mathcal{N}(\mathbf{0},\boldsymbol{\Sigma}_{\text{post}})}\norm{\boldsymbol{\epsilon}-\boldsymbol{\epsilon}_\theta(\mathbf{x}_t,\mathbf{y},t)}^2_{\boldsymbol{\Sigma}^{-1}_{\text{post}}}}_{\text{Denoising Matching (DM) terms}} \\
        +&\underbrace{\frac{1}{2}\bigr(\log\frac{\abs{\boldsymbol{\Sigma}_{\text{prior}}}}{\abs{\boldsymbol{\Sigma}_{\text{post}}}}+\text{tr}(\boldsymbol{\Sigma}_{\text{prior}}^{-1}\boldsymbol{\Sigma}_{\text{post}})\bigr)}_{\text{Prior Matching (PM) terms}} + \, \text{C},
    \end{aligned}
    \label{eq: elbo for restoregrad}
    \end{equation}
    where $\gamma_t= \begin{cases} \frac{\beta_t^2}{2\sigma_t^2\alpha_t(1-\bar{\alpha}_t)}, & t>1 \\ \frac{1}{2\alpha_1}, & t=1\end{cases}$ are weighting factors, $\norm{\mathbf{x}}^2_{\boldsymbol{\Sigma}^{-1}}=\mathbf{x}^T\boldsymbol{\Sigma}^{-1}\mathbf{x}$, $\sigma_t^2=\frac{1-\bar{\alpha}_{t-1}}{1-\bar{\alpha_t}}\beta_t$ and $C$ is some constant not depending on learnable parameters $\theta$, $\phi$, $\psi$.
\label{proposition: 2}
\end{proposition}

The derivation (see Appendix \ref{sec: appendix proof prop 2}) is based on combining VAE and the results in \citet{lee2021priorgrad}. \textit{Notably, we join the conditional DDPM with the posterior/prior encoders and optimize all modules at once, by connecting the DDPM prior space with the latent space estimated by the encoders.} To this end, the sampling $\boldsymbol{\epsilon}\sim q_{\phi}(\boldsymbol{\epsilon}|\mathbf{x}_0, \mathbf{y})$ is performed by the standard reparameterization trick as in VAEs, unlocking end-to-end training via gradient descent on the loss terms:
\begin{itemize}[leftmargin=*]
\vspace{-0.2cm}
    \item \underline{\textit{Latent Regularization (LR)} terms}: to help learn a reasonable latent space; e.g., minimizing $\log\abs{\boldsymbol{\Sigma}_{\text{post}}}$ avoids $\boldsymbol{\Sigma}_{\text{post}}$ from becoming arbitrary large due to the presence of its inverse in the weighted norms.
    \vspace{-0.1cm}
    \item \underline{\textit{Denoising Matching (DM)} terms}: responsible for training the DDPM to predict the true noise.
    \vspace{-0.1cm}
    \item \underline{\textit{Prior Matching (PM)} terms}: to shape a desirable latent space by aligning the prior and posterior distributions. Note that we model the distributions as zero-mean, based on that signals can be properly normalized. 
\end{itemize}

\subsection{Training of RestoreGrad} 
With the the conditional DDPM $\theta$, Prior Net $\psi$, and Posterior Net $\phi$ defined in Proposition \ref{proposition: 2}, optimization can be performed to learn the model parameters of $\theta,\psi,\phi$ based on the modified ELBO. The RestoreGrad framework jointly trains the three neural network modules by minimizing (\ref{eq: elbo for restoregrad}) as depicted in Figure \ref{fig: restoregrad}. Following existing DDPM literature, we approximate the objective by dropping the weighting constant $\gamma_t$ of the DM terms, leading to the simplified loss:
\begin{equation}
\begin{small}
\begin{aligned}
    \min_{\theta,\phi,\psi} \,\,\, \eta&\bigr(\underbrace{\bar{\alpha}_T\norm{\mathbf{x}_0}^2_{\boldsymbol{\Sigma}^{-1}_{\text{post}}}+\log\abs{\boldsymbol{\Sigma}_{\text{post}}}}_{\mathcal{L}_{\text{LR}}}\bigr)+\underbrace{\norm{\boldsymbol{\epsilon}-\boldsymbol{\epsilon}_\theta(\mathbf{x}_t,\mathbf{y},t)}^2_{\boldsymbol{\Sigma}^{-1}_{\text{post}}}}_{\mathcal{L}_{\text{DM}}} \\
    +&\lambda \underbrace{\bigr(\log\frac{\abs{\boldsymbol{\Sigma}_{\text{prior}}}}{\abs{\boldsymbol{\Sigma}_{\text{post}}}}+\text{tr}(\boldsymbol{\Sigma}_{\text{prior}}^{-1}\boldsymbol{\Sigma}_{\text{post}})\bigr)}_{\mathcal{L}_{\text{PM}}},
\end{aligned}
\end{small}
\label{eq: elbo of restoregrad}
\end{equation}
where we approximate the expectations by randomly sampling $(\mathbf{x}_0,\mathbf{y})\sim q_{\text{data}}(\mathbf{x}_0,\mathbf{y})$ and $\boldsymbol{\epsilon}\sim \mathcal{N}(\mathbf{0},\boldsymbol{\Sigma}_{\text{post}})$, and the summation over $t$ by sampling $t\sim \mathcal{U}(\{1,\dots,T\})$ (exploiting the independency due to Markov assumption \citep{nichol2021improved}) in each training iteration. We have also introduced $\eta>0$ for the LR terms and $\lambda>0$ for PM terms, to exert flexible control of the learned latent space.

\subsection{Sampling of RestoreGrad} 
In applications that RestoreGrad is mainly concerned with, the target signal $\mathbf{x}_0$ is not available in inference time. As in Figure \ref{fig: restoregrad}, the conditional DDPM then samples $\boldsymbol{\epsilon}\sim p_{\psi}(\boldsymbol{\epsilon}|\mathbf{y})=\mathcal{N}(\mathbf{0},\boldsymbol{\Sigma}_{\text{prior}})$ from the Prior Net instead; the Posterior Net is no longer needed.

\subsection{The Role of Posterior Information} 
In the training stage of RestoreGrad, the latent code $\boldsymbol{\epsilon}$ samples from the posterior $q_\phi(\boldsymbol{\epsilon}|\mathbf{x}_0,\mathbf{y})$ which exploits both the ground truth signal $\mathbf{x}_0$ and conditioner $\mathbf{y}$. It is thus more advantageous than existing works on adaptive priors (e.g., PriorGrad \citep{lee2021priorgrad}) that only utilize the conditioner $\mathbf{y}$. To observe the benefits brought by the posterior information, we can make comparison with a variant of RestoreGrad where the Posterior Net is excluded in training:
\begin{equation}
\small
    \min_{\theta,\psi} \,\,\, \eta\bigr(\bar{\alpha}_T\norm{\mathbf{x}_0}^2_{\boldsymbol{\Sigma}^{-1}_{\text{prior}}}+\log\abs{\boldsymbol{\Sigma}_{\text{prior}}}\bigr)+\norm{\boldsymbol{\epsilon}-\boldsymbol{\epsilon}_\theta(\mathbf{x}_t,\mathbf{y},t)}^2_{\boldsymbol{\Sigma}^{-1}_{\text{prior}}},
\label{eq: elbo of restoregrad no post net}
\end{equation}
which basically removes the Posterior Net $\phi$ and only trains the Prior Net $\psi$ and DDPM $\theta$. Interestingly, our experimental results show that RestoreGrad indeed performs better with the Posterior Net than without it in model training.

\section{Experiments}
\label{sec: exp}

\subsection{Application to Speech Enhancement (SE)}

\subsubsection{Experimental Setup}

\noindent\textbf{Dataset:} We validate performance on the benchmark SE dataset \textit{VoiceBank+DEMAND} \citep{valentini2016investigating}, consisting of clean speech clips collected from the \textit{VoiceBank} corpus \citep{veaux2013voice}, mixed with ten types of noise profiles from the \textit{DEMAND} database \citep{thiemann2013diverse}. Specifically, the training utterances from VoiceBank are artificially contaminated with the noise samples from DEMAND at 0, 5, 10, and 15 dB signal-to-noise ratio (SNR) levels, amounting to 11,572 utterances. The testing utterances are mixed with different noise samples at 2.5, 7.5, 12.5, and 17.5 dB  SNR levels, amounting to 824 utterances. 

\noindent\textbf{Evaluation Metrics:} We consider: \textbf{PESQ:} Perceptual Evaluation of Speech Quality \citep{itu862}. \textbf{SI-SNR:} Scale-Invariant SNR \citep{le2019sdr}. \textbf{SSNR}: Segmental SNR \cite{hu2007evaluation}. \textbf{CSIG, CBAK, COVL:} Mean-opinion-score predictors of signal distortion, background-noise intrusiveness, and overall signal quality, respectively \citep{hu2007evaluation}. 

\noindent\textbf{Models:} The following models are compared:
\begin{itemize}[leftmargin=*]
\vspace{-0.15cm}
    \item \textbf{Baseline DDPM}: We adopt the \textit{CDiffuSE (Base)} model from \citet{lu2022conditional}, which is based on DiffWave \citep{kong2020diffwave} with 4.28M learnable parameters. 
    \item \textbf{PriorGrad}: We implement the PriorGrad \citep{lee2021priorgrad} on top of CDiffuSE by changing the prior distribution from $\mathcal{N}(\mathbf{0},\mathbf{I})$ to $\mathcal{N}(\mathbf{0},\boldsymbol{\Sigma}_{\text{y}})$, where $\boldsymbol{\Sigma}_{\text{y}}$ is the covariance of the data-dependent prior computed based on the conditioner $\mathbf{y}$, using the rule-based estimation approach for the application to vocoder in \citet{lee2021priorgrad}.
    \item \textbf{RestoreGrad}: We incorporate Prior Net and Posterior Net on top of CDiffuSE. Both modules adopt the ResNet-20 architect \citep{he2016deep}, suitably modified to 1-D convolutions for waveform processing, each with only 93K learnable parameters (only 2\% of the CDiffuSE model).
\end{itemize}

\noindent\textbf{Configurations:} 
We adopted the basic configurations same as in \citet{lu2022conditional}. The waveforms were processed at 16kHz sampling rate. The number of forward diffusion steps was $T=50$. The variance schedule was $\beta_t\in[10^{-4}, 0.035]$, linearly spaced. The batch size was 16. The fast sampling scheme in \citet{kong2020diffwave} was used in the reverse process with 6 steps to reduce inference complexity, with the 6-step inference variance schedule $\beta^{\text{infer}}_t=[10^{-4}, 10^{-3}, 0.01, 0.05, 0.2, 0.35]$. Adam optimizer \citep{kingma2014adam} was utilized with a learning rate of $2\times 10^{-4}$. We set $\eta=0.1$ and $\lambda=0.5$ for (\ref{eq: elbo of restoregrad}).

\subsubsection{Results}
\noindent\textbf{Improved Model Convergence:} 
As shown in Figure \ref{fig: training_curve} (test set performance), RestoreGrad shows better convergence behavior over PriorGrad (using handcrafted prior) and CDiffuSE (using standard Gaussian prior). For example, \textit{PriorGrad reaches 2.4 in PESQ at 96 epochs, whereas RestoreGrad reaches it in (roughly) 10 epochs, indicating a 10$\times$ speed-up.} The results suggest that jointly learning the prior distribution can be beneficial for conditional DDPMs.

\begin{figure}[!t]
    \centering
    \includegraphics[width=1\linewidth]{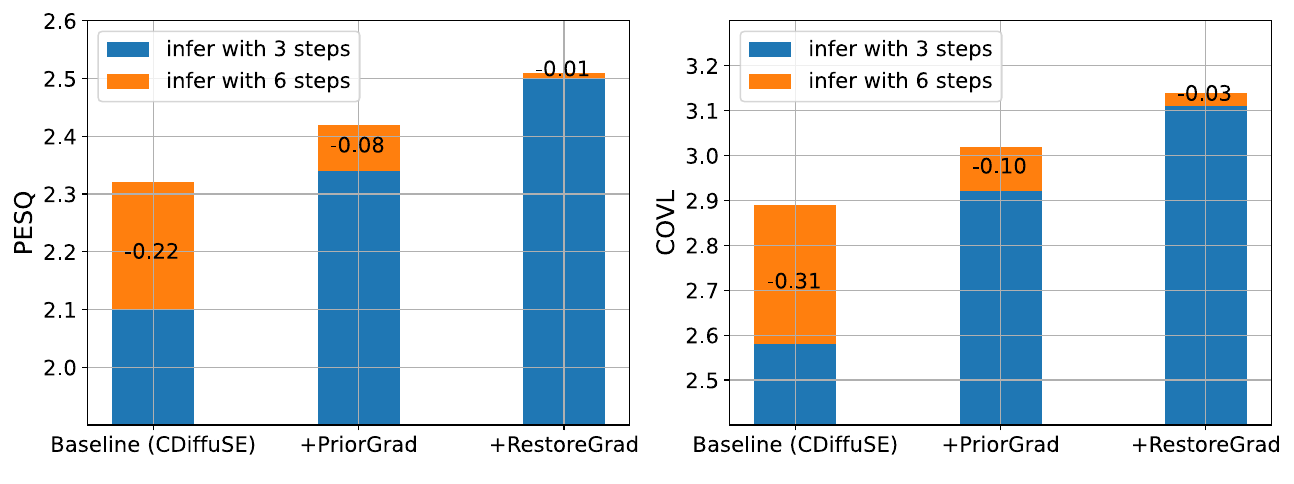}
    \vspace{-0.75cm}
    \caption{Robustness to the reduction in reverse sampling time steps for inference.} 
\label{fig: tolerance to reduced interence sampling steps}
\vspace{-0.cm}
\end{figure}

\noindent\textbf{Robustness to Reduced Reverse Steps in Inference:} 
RestoreGrad can potentially reduce the inference complexity. In Figure \ref{fig: tolerance to reduced interence sampling steps}, we show how the trained diffusion models withstand the reduction in the number of inference steps. In each model, we trained the network for 96 epochs and then inferenced with 3 reverse steps to compare with the originally adopted 6-step scheme in \citet{lu2022conditional}. The noise schedule for the 3-step scheme was $\beta^{\text{infer}}_t=[0.05, 0.2, 0.35]$, a subset of the 6-step schedule that resulted in best performance. We can see that the baseline DDPM is most sensitive to the step reduction, while PriorGrad shows certain resistance as leveraging a closer-to-data prior distribution. \textit{Finally, RestoreGrad barely degrades with reduced sampling steps, echoing that a better prior has been obtained as it recovers higher fidelity signal even in fewer reverse steps}. 

\begin{figure}[!t]
    \centering
    \includegraphics[width=0.8\linewidth]{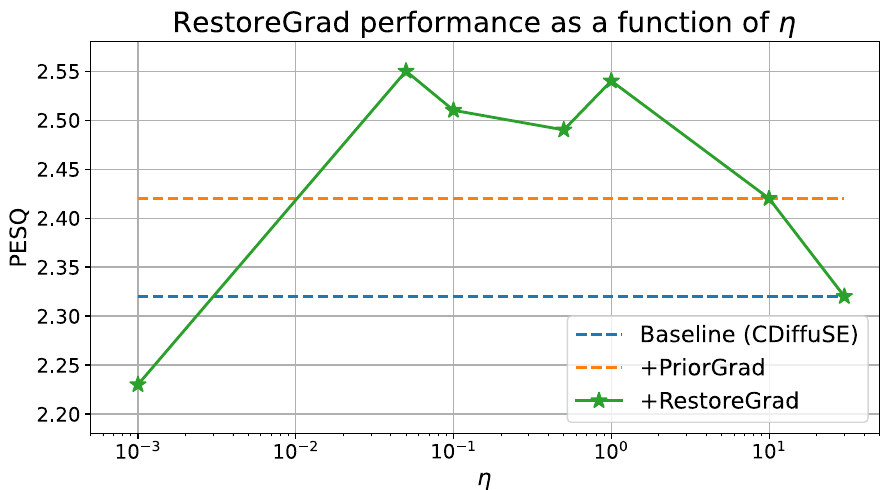}
    \vspace{-0.3cm}
    \caption{Effect of latent regularization weight $\eta$ for $\mathcal{L}_{\text{LR}}$ on SE. } 
\label{fig: eta effect}
\end{figure}

\noindent\textbf{Effect of $\eta$:} 
An important factor in our prior learning scheme is the regularization weight $\eta$ for $\mathcal{L}_{\text{LR}}$ of the training loss. An appropriate value of $\eta$ should be large enough to properly regularize the learned latent space for avoiding instability, while not adversely affecting signal reconstruction performance. It is thus interesting to see how the performance varies with the choice of $\eta$. \textit{Empirically, we found the overall SE performance not to be very sensitive to the value of $\eta$ across a wide range,} as shown in Figure \ref{fig: eta effect}: Roughly in the range of $[10^{-2}, 10]$ of the $\eta$ value we see that RestoreGrad gives better results over both PriorGrad and CDiffuSE.

\noindent\textbf{Comparison to Fully-Trained CDiffuSE:} 
We present in Table \ref{table: se sota comp 1} more detailed comparison of RestoreGrad with the baseline CDiffuSE. Here, the scores of CDiffuSE were directly taken from the results reported in \citet{lu2022conditional} where the model has been fully trained for 445 epochs. For PriorGrad and RestoreGrad we report the mean$\pm$std computed based on results of 10 independent samplings. \textit{We can see that with RestoreGrad applied, the SE model can achieve better performance over the baseline CDiffuSE by only training for 96 epochs (4.6 times lesser) in all the metrics}. In addition, halving the number of reverse steps in inference time still maintains better performance than the fully-trained CDiffuSE and also the PriorGrad. 

\begin{table}[!t]
\centering
\begin{small}
\setlength{\tabcolsep}{2pt} 
\caption{Comparison with the fully-trained CDiffuSE model performance reported in \citet{lu2022conditional}.}
\vspace{0.15cm}
\label{table: se sota comp 1}
\resizebox{1\linewidth}{!}{%
\begin{NiceTabular}{lccccccc}
\toprule 
 \multirow{2}{*}{Methods} & \multirow{1}{*}{\# train} & \multirow{1}{*}{\# infer} & \multirow{2}{*}{PESQ $\uparrow$} & \multirow{2}{*}{CSIG $\uparrow$} & \multirow{2}{*}{CBAK $\uparrow$}& \multirow{2}{*}{COVL $\uparrow$} & \multirow{2}{*}{SI-SNR $\uparrow$} \\
 & \multirow{1}{*}{epochs} & \multirow{1}{*}{steps} & & & & & \\
 \midrule
 CDiffuSE & 445 & 6 & 2.44 & 3.66 & 2.83 & 3.03 & - \\
 \midrule
 + PriorGrad & 96 & 6 & 2.42$\pm$3e-3 & 3.67$\pm$2e-3 & 2.93$\pm$1e-3 & 3.03$\pm$2e-3 & 14.21$\pm$2e-3 \\
 \midrule
 \multirow{1}{*}{+ RestoreGrad}  & \multirow{2}{*}{96} & 6 & \textbf{2.51}$\pm$6e-4 & \textbf{3.80}$\pm$4e-4 & \textbf{3.00}$\pm$3e-4 & \textbf{3.14}$\pm$5e-4 & \textbf{14.74}$\pm$3e-4 \\
 \multirow{1}{*}{\quad\quad(ours)} &  & 3 & \underline{2.50}$\pm$3e-4 & \underline{3.75}$\pm$2e-4 & \underline{2.99}$\pm$2e-4 & \underline{3.11}$\pm$3e-4 & \underline{14.65}$\pm$2e-4 \\
 \bottomrule
\end{NiceTabular}
}
\\
\vspace{0.1cm}
\scriptsize
*Bold text for best and underlined text for second best values.
\end{small}
\vspace{-0.1cm}
\end{table}

\noindent\textbf{Signal Quality and Encoder Complexity Trade-Offs:} We further present results using three different model sizes (24K, 93K, 370K) for the Prior and Posterior Nets (encoders) in Table \ref{table: se quality complexity trade-off}, along with latency and GPU memory usage (presented as the ratio of encoder to DDPM). The results clearly show that the restored speech quality improves with increasing encoder size. \textit{This indicates that there is a trade-off between the restoration signal quality and encoder model complexity}. Notably, \textit{the latency and memory usage of the encoder modules are relatively small compared to the DDPM decoding ($<$2.6\% latency and $<$18.2\% memory usage of the DDPM processing}), suggesting that RestoreGrad is capable of achieving improved performance without incurring considerable increase in complexity compared to the adopted DDPM model.

\begin{table}[!t]
\centering
\begin{small}
\setlength{\tabcolsep}{4pt} 
\caption{SE comparison of RestoreGrad models using encoder modules of different sizes and the corresponding latency and GPU memory usage  (measured on one NVIDIA Tesla V100 GPU) presented as the ratio of encoder to DDPM.}
\vspace{0.15cm}
\label{table: se quality complexity trade-off}
\resizebox{\linewidth}{!}{%
\begin{tabular}{lcccc|cc}
\toprule
  Encoder size & PESQ$\uparrow$ & COVL$\uparrow$ & SSNR$\uparrow$ & SI-SNR$\uparrow$ & Proc. Time & Memory \\
\midrule
    Tiny (24K) & 2.48 & 3.11  & 5.10 & 13.74 & 1.9\% & 6.5\%  \\
    Base (93K) & 2.51 & 3.14  & 5.92 &  14.74 & 2.2\% & 10.3\% \\
    Large (370K) & 2.54 & 3.16 & 6.15 & 15.01 & 2.6\% & 18.2\% \\
\bottomrule
\end{tabular}
}
\end{small}
\vspace{-0.1cm}
\end{table}

\textbf{Posterior Net Helps:} Finally, we validate the benefits brought by employing Posterior Net in the training phase by comparing with the RestoreGrad models trained without Posterior Net as (\ref{eq: elbo of restoregrad no post net}) for some $\eta$. For fairness, all models were trained with 96 epochs, inferred with 6 steps. In Table \ref{table: se no post net}, we observe that \textit{RestoreGrad achieves better results with Posterior Net than without it, indicating the benefits of being informed of the target $\mathbf{x}_0$ by utilizing the Posterior Net}. We also observe that without regularizing the latent space (i.e., with $\eta=0$) it could lead to training divergence.

\begin{table}[!t]
\centering
\begin{small}
\setlength{\tabcolsep}{1.2pt} 
\caption{Performance of RestoreGrad models trained with and without using Posterior Net.}
\vspace{0.1cm}
\label{table: se no post net}
\resizebox{0.98\linewidth}{!}{%
\begin{NiceTabular}{lcccc}
\toprule
  SE models & PESQ$\uparrow$ & COVL$\uparrow$ & SSNR$\uparrow$ & SI-SNR$\uparrow$ \\
\midrule
    CDiffuSE (trained for 96 epochs) & 2.32 & 2.89 & 3.94 & 11.84 \\
    + PriorGrad & 2.42 & 3.03 & 5.53 & 14.21 \\
    + RestoreGrad & \textbf{2.51} & \textbf{3.14} & \textbf{5.92} & \textbf{14.74} \\
\midrule
    + RestoreGrad w/o Posterior Net ($\eta=0$) & --- & training & diverged & --- \\
    + RestoreGrad w/o Posterior Net ($\eta=0.01$) & 2.47 & 3.08 & 4.96 &11.22 \\
    + RestoreGrad w/o Posterior Net ($\eta=1$) & 2.48 & 3.12 & 5.11 & 13.29 \\
\bottomrule
\end{NiceTabular}
}
\\
\vspace{0.1cm}
\scriptsize
*Best values in bold.
\end{small}
\vspace{-0.1cm}
\end{table}

\subsection{Application to Image Restoration (IR)}
\label{sec: ir on weathers}

\subsubsection{Experimental Setup}
\noindent\textbf{Dataset:} Following \citet{ozdenizci2023restoring}, we consider the IR task of recovering clean images from their degraded versions contaminated by synthesized noises corresponding to different weather conditions. Two datasets are considered, where one is a weather-specific dataset called \textit{RainDrop} \citep{qian2018attentive} and the other is a multi-weather dataset named \textit{AllWeather} \citep{valanarasu2022transweather}. The RainDrop dataset consists of images captured with raindrops on the camera sensor which obstruct the view. It has 861 training images and a test set of 58 images dedicated for quantitative evaluations. The AllWeather dataset is a curated training dataset from \citet{valanarasu2022transweather}, which has 18,069 samples composed of subsets of training images from \textit{Snow100K} \citep{liu2018desnownet}, \textit{Outdoor-Rain} \citep{li2019heavy} and \textit{RainDrop} \citep{qian2018attentive}, in order to create a balanced training set across three weather conditions. 

\noindent\textbf{Evaluation Metrics:} Quantitative evaluations of restored images are performed via Peak Signal-to-Noise Ratio (\textbf{PSNR}) \citep{huynh2008scope}, Structural SIMilarity (\textbf{SSIM}) \citep{wang2004image}, Learned Perceptual Image Patch Similarity (\textbf{LPIPS}) \citep{zhang2018unreasonable}, and Fréchet Inception Distance (\textbf{FID}) \citep{heusel2017gans}.

\noindent\textbf{Models:} The following IR models are compared:
\begin{itemize}[leftmargin=*]
\vspace{-0.15cm}
    \item \textbf{Baseline DDPMs}: We consider the \textit{$\text{RainDropDiff}_{64}$} and \textit{$\text{WeatherDiff}_{64}$} in \citet{ozdenizci2023restoring} trained on the RainDrop and AllWeather datasets, respectively, as baseline DDPMs. Our work is based on the implementation provided by \citet{ozdenizci2023restoring}. 
    \vspace{-0.45cm}
    \item \textbf{RestoreGrad}: We incorporate the encoder modules, Prior Net and Posterior Net, on top of the baseline DDPMs. Both encoder modules adopt the ResNet-20 architect \citep{he2016deep} with only 0.27M learnable parameters, significantly smaller ($<0.3\%$) than the baseline DDPM models.
\end{itemize}

\vspace{-0.01cm}
\noindent\textbf{Configurations:} 
We used Adam optimizer with a learning rate of $2\times 10^{-5}$. An exponential moving average with a weight of 0.999 was applied. We used $T=1000$ and linear noise schedule $\beta_t\in[10^{-4}, 0.02]$, same as \citet{ozdenizci2023restoring}. A batch size of 4 was used.

\begin{table*}[!t]
\centering
\begin{small}
\setlength{\tabcolsep}{1pt} 
\caption{Comparison with existing IR models. The multi-weather (MW) models were trained on the AllWeather training set \citep{valanarasu2022transweather} and tested on three different weather types: Snow100K-L \citep{liu2018desnownet}, Outdoor-Rain \citep{li2019heavy}, and RainDrop \citep{qian2018attentive}. Several weather-specific (WS) models that were trained on individual weather types are also presented for reference.}
\vspace{0.3cm}
\label{table: ir sota comp allweather}
\resizebox{\linewidth}{!}{%
\begin{NiceTabular}{l|lcc||lcc||lcc}
\toprule 
 \multirow{2}{*}{Type} & \multirow{2}{*}{Methods} & \multicolumn{2}{c||}{Snow100K-L} & \multirow{2}{*}{Methods} & \multicolumn{2}{c||}{Outdoor-Rain} & \multirow{2}{*}{Methods} & \multicolumn{2}{c}{RainDrop}  \\
 \cmidrule(lr){3-4}
 \cmidrule(lr){6-7}
 \cmidrule(lr){9-10}
 & & \multirow{1}{*}{PSNR $\uparrow$} & \multirow{1}{*}{SSIM $\uparrow$} & & \multirow{1}{*}{PSNR $\uparrow$} & \multirow{1}{*}{SSIM $\uparrow$} & & \multirow{1}{*}{PSNR $\uparrow$} & \multirow{1}{*}{SSIM $\uparrow$} \\
 \midrule
 \multirow{5}{*}{WS} & RESCAN \citep{li2018recurrent} &  26.08 & 0.8108 & HRGAN \citep{li2019heavy} & 21.56 &   0.8550 & AttentiveGAN \citep{qian2018attentive} & 31.59 & 0.9170 \\
 & DesnowNet \citep{liu2018desnownet} &   27.17 & 0.8983 & PCNet \citep{jiang2021rain} &  26.19 &  0.9015 & RaindropAttn \citep{quan2019deep} & 31.44 &  0.9263 \\
 & DDMSNet \citep{zhang2021deep} &   28.85 & 0.8772 & MPRNet \citep{zamir2021multi} &  28.03 & 0.9192 & IDT \citep{xiao2022image} & 31.87 &  0.9313 \\
 & SnowDiff \citep{ozdenizci2023restoring} &  30.43 &  0.9145 & RainHazeDiff \citep{ozdenizci2023restoring} &  28.38 & 0.9320 & RainDropDiff \citep{ozdenizci2023restoring} & 32.29&  0.9422 \\
 & DTPM \citep{ye2024learning} & \underline{30.92} & \underline{0.9174} & DTPM \citep{ye2024learning} & \textbf{30.99} & 0.9340 & DTPM \citep{ye2024learning} & \textbf{32.72} & \textbf{0.9440} \\  
 \midrule
 \multirow{5}{*}{MW} & All-in-One \citep{li2020all} &  28.33 & 0.8820 & All-in-One \citep{li2020all} & 24.71 &  0.8980 & All-in-One \citep{li2020all} & 31.12 & 0.9268 \\
 & TransWeather \citep{valanarasu2022transweather} &  29.31 & 0.8879 & TransWeather \citep{valanarasu2022transweather} & 28.83 & 0.9000 & TransWeather \citep{valanarasu2022transweather} & 30.17 & 0.9157 \\
   \cmidrule(lr){2-4}
   \cmidrule(lr){5-7}
   \cmidrule(lr){8-10}
   & WeatherDiff \citep{ozdenizci2023restoring} & 30.09 & 0.9041 & WeatherDiff \citep{ozdenizci2023restoring} & 29.64 & 0.9312 & WeatherDiff \citep{ozdenizci2023restoring} & 30.71 & 0.9312 \\
 & + RestoreGrad (ours) & 30.82 & 0.9159 & + RestoreGrad (ours) & \underline{30.83} & \underline{0.9411} & + RestoreGrad (ours) & 31.78 & 0.9394 \\
 & + RestoreGrad (ours) -- trained longer & \textbf{31.16} & \textbf{0.9175} & + RestoreGrad (ours) -- trained longer & 30.70 & \textbf{0.9418} & + RestoreGrad (ours) -- trained longer & \underline{32.26} & \underline{0.9414} \\
 \bottomrule
\end{NiceTabular}
}
\\
\vspace{0.1cm}
\scriptsize
*Bold text for best and underlined text for second best values.
\end{small}
\vspace{0.4cm}
\end{table*}

\begin{figure*}[!tp]
    \centering
    \includegraphics[width=\linewidth]{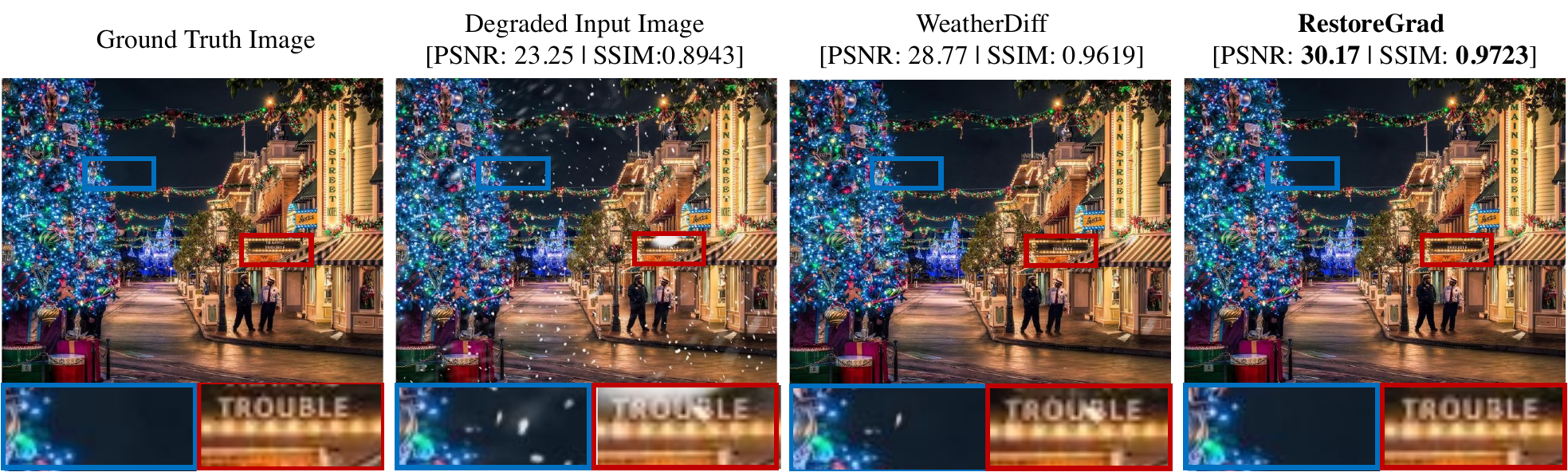}
    \vspace{-0.55cm}
    \caption{Image restoration examples using a test image taken from the Snow100K-L test set. We provide more examples, including other degradations (\textbf{desnowing}, \textbf{deraining}, \textbf{raindrop removal} and \textbf{deblurring}) in Appendix \ref{sec: additional results on IR appendix}.} 
\label{fig: ir example}
\vspace{0.45cm}
\end{figure*}

\subsubsection{Results}
\noindent\textbf{Model Convergence:} As presented in Figure \ref{fig: training_curve} (test set performance), RestoreGrad demonstrates faster convergence and better restored image quality over the baseline DDPM (RainDropDiff). For example, \textit{RainDropDiff reaches 0.143 in LPIPS at 9.2K epochs, while RestoreGrad reaches it in 1.8K epochs only, indicating a 5$\times$ speed-up} due to the effectiveness of the joint prior learning scheme.

\noindent\textbf{Comparison with Existing IR Models:}
We compare our method with existing models of the multi-weather (MW) type in Table \ref{table: ir sota comp allweather}, including All-in-One \citep{li2020all} and TransWeather \citep{valanarasu2022transweather} in addition to WeatherDiff \citep{ozdenizci2023restoring}. The models were all trained on the AllWeather dataset and the performance numbers were taken from \citet{ozdenizci2023restoring}, where the WeatherDiff was trained for 1,775 epochs and inferenced with 25 steps. \textit{Our RestoreGrad was trained for only 887 epochs (2$\times$ fewer) and inferenced with 10 steps to already achieve the best performance in the MW category}. In addition, when trained for more epochs (1,551 epochs, which is 1.14$\times$ fewer than WeatherDiff), RestoreGrad achieves further improvements as shown in the last row the table. On the other hand, our method, while trained on multi-weather data, achieves a comparable performance to the recently proposed Diffusion Texture Prior Model (DTPM) \cite{ye2024learning} individually trained on the three weather-specific (WS) datasets. \textit{Notably, our method does not require a pretraining stage on a large dataset of high-quality images, unlike DTPM which is pretrained on 55,000 images samples.} This suggests the generality and effectiveness of our method to improve baseline DDPM models.

\noindent\textbf{IR Example:} Figure \ref{fig: ir example} presents examples of restored images by the models. It can be seen that RestoreGrad is able to better recover the original image, \textit{especially in regions of the blue and red boxes where the baseline WeatherDiff fails to remove the snow obstructions}. The higher PSNR and SSIM scores of RestoreGrad also reflect the improvements.

\noindent\textbf{Other IR Tasks:}
Our method demonstrates the advantages of learnable priors in image \textit{deblurring} and \textit{super-resolution} tasks, suggesting its \textit{generality}. See Appendix \ref{sec: additional results on IR appendix}.

\subsection{Generalization to Out-of-Distribution (OOD) and Realistic Data}

We have so far evaluated the models on in-domain scenarios with synthetic noisy data where RestoreGrad has shown substantial improvements. A natural question is whether the demonstrated improvements have actually come at the expense of the model's generalizability to unseen or realistic data. To address the concern, we evaluate the IR models on two additional datasets,\textit{RainDS-Real} from \citet{quan2021removing} and \textit{Snow100K-Real} from \citet{liu2018desnownet} consisting of real-world images, using the reference-free Natural Image Quality Evaluator (NIQE) metric \citep{mittal2012making} (a lower score indicates better quality). In Table \ref{table: ir sota comp realistic} we see that RestoreGrad is able to perform on par with or better than WeatherDiff and the non-generative model of TransWeather. For OOD testing, we evaluate the SE models on the \textit{CHiME-3} dataset \citep{barker2017third} unseen during model training. Table \ref{table: se sota comp ood} compares RestoreGrad with CDiffuSE also trained for 96 epochs, DOSE \citep{tai2024dose}, and two discriminative SE models, Demucs \citep{defossez2020real} and WaveCRN \citep{hsieh2020wavecrn}. We can see that RestoreGrad is able to perform equally well as the CDiffuSE while outperforming DOSE and the non-generative SE models. The results show that RestoreGrad is capable of improving in-domain performance while maintaining desirable generalizability of generative models.

\begin{table}[!t]
\centering
\begin{small}
\setlength{\tabcolsep}{4pt} 
\caption{Evaluation on realistic image datasets of the IR models trained on synthetic images of AllWeather training set.}
\vspace{0.2cm}
\label{table: ir sota comp realistic}
\centering
\resizebox{1\columnwidth}{!}{%
\begin{NiceTabular}{lccc}
\toprule 
 \multirow{2}{*}{Methods} & \multirow{2}{*}{Gen. } & RainDS-Real & Snow-Real \\
 \cmidrule(lr){3-3}
 \cmidrule(lr){4-4}
 &  & NIQE $\downarrow$ & NIQE $\downarrow$ \\
 \midrule
 TransWeather \citep{valanarasu2022transweather} & N & 4.005 & 3.161 \\
 \midrule
 WeatherDiff \citep{ozdenizci2023restoring} & Y & \underline{3.050} & \textbf{2.985} \\
 + RestoreGrad (ours) & Y & \textbf{2.556} & \underline{3.015}  \\
 \bottomrule
\end{NiceTabular}
}\\
\vspace{0.1cm}
\scriptsize
*Bold text for best and underlined text for second best values. The column ``Gen.'' indicates if the model is generative (Y) or not (N).
\end{small}
\vspace{-0.2cm}
\end{table}

\begin{table}[!t]
\centering
\begin{small}
\setlength{\tabcolsep}{3pt} 
\caption{Evaluation of SE models on CHiME-3 test set, where the models were trained on VoiceBank+DEMAND training set.}
\vspace{0.2cm}
\label{table: se sota comp ood}
\resizebox{1\columnwidth}{!}{%
\begin{NiceTabular}{lcccccc}
\toprule 
 Methods & Gen. & PESQ$\uparrow$ & CSIG$\uparrow$ & CBAK$\uparrow$ & COVL$\uparrow$ & SI-SNR$\uparrow$ \\
\midrule
 Unprocessed & - & 1.27 & 2.61 & 1.93 & 1.88 & 7.51 \\
 \midrule
 Demucs \citep{defossez2020real} & N & 1.38 & 2.50 & 2.08 & 1.88 & - \\
 WaveCRN \citep{hsieh2020wavecrn} & N & 1.43 & 2.53 & 2.03 & 1.91 & - \\
 DOSE \citep{tai2024dose} & Y &  1.52 & 2.71 & \textbf{2.15} & 2.06 & -\\
 \midrule
 CDiffuSE \citep{lu2022conditional} & Y & \textbf{1.55} & \underline{2.87} & 2.09 & \underline{2.15} & \underline{7.67}\\
 + RestoreGrad (ours) & Y &\underline{1.54} & \textbf{2.88} & \underline{2.14} & \textbf{2.16} & \textbf{8.45} \\
 \bottomrule
\end{NiceTabular}
}
\\
\vspace{0.1cm}
\scriptsize
*Bold text for best and underlined text for second best values. The column ``Gen.'' indicates if the model is generative (Y) or not (N).
\end{small}
\vspace{-0.2cm}
\end{table}
\section{Related Work}

\noindent\textbf{Diffusion Efficiency Improvements:} 
\citet{das2023image} utilized the shortest path between two Gaussians and \citet{song2020denoising} generalized DDPMs via a class of non-Markovian diffusion processes to reduce the number of diffusion steps. \citet{nichol2021improved} introduced a few simple modifications to improve the log-likelihood. \citet{pandey2022diffusevae, pandey2021vaes} used DDPMs to refine VAE-generated samples. \citet{rombach2022high} performed the diffusion process in the lower dimensional latent space of an autoencoder to achieve high-resolution image synthesis, and \citet{liu2023audioldm} studied using such latent diffusion models for audio. \citet{popov2021grad} explored using a text encoder to extract better representations for continuous-time diffusion-based text-to-speech generation. More recently, \citet{nielsendiffenc} explored using a time-dependent image encoder to parameterize the mean of the diffusion process. Orthogonal to the above, PriorGrad \citep{lee2021priorgrad} and follow-up work \citep{koizumi22_interspeech} studied utilizing informative prior extracted from the conditioner data for improving learning efficiency. \textit{However, they become sub-optimal when the conditioner are degraded versions of the target data, posing challenges in applications like signal restoration tasks.}

\noindent\textbf{Diffusion-Based Signal Restoration:}
Built on top of the diffusion models for audio generation, e.g., \citet{kong2020diffwave,chen2020wavegrad,leng2022binauralgrad}, many SE models have been proposed. The pioneering work of \citet{lu2022conditional} introduced conditional DDPMs to the SE task and demonstrated the potential. Other works \citep{serra2022universal,welker2022speech,richter2023speech,yen2023cold,lemercier2023storm,tai2024dose} have also attempted to improve SE by exploiting diffusion models. In the vision domain, diffusion models have demonstrated impressive performance for IR tasks \citep{li2023diffusion,zhu2023denoising,huang2024wavedm,luo2023refusion,xia2023diffir,fei2023generative,hurault2022gradient,liu20232,chung2024direct,chungdiffusion,zhoudenoising,xiaodreamclean,zheng2024diffusion,liu2024residual,ye2024learning}, and a comprehensive view of recent advancements is provided by \citet{he2025diffusion}. A notable IR work is by \citet{ozdenizci2023restoring} that achieved impressive performance on several benchmark datasets for restoring vision in adverse weather conditions. \textit{Despite showing promising results, existing works have not fully exploited prior information about the data as they mostly settle on standard Gaussian priors.} 
\vspace{-0.05cm}
\section{Conclusion}
\label{sec: conclusion}
\vspace{-0.05cm}
We investigated the potential of learning the prior distribution in tandem with the conditional DDPM for improved efficiency. We demonstrated the advantages of RestoreGrad that leverages learning-based priors, providing a more systematic way of estimating the prior than existing selections. A limitation of the current work is that it focuses on signal restoration applications, where we suitably assume a zero-mean Gaussian prior and only learn its covariance. In the future, it can be interesting to explore using a more generic prior form and extend the idea to other applications.

\section*{Impact Statement}

This paper presents work whose goal is to advance the field of Machine Learning. There are many potential societal consequences of our work, none of which we feel must be specifically highlighted here.


\bibliography{example_paper}

\begin{thebibliography}{94}
\providecommand{\natexlab}[1]{#1}
\providecommand{\url}[1]{\texttt{#1}}
\expandafter\ifx\csname urlstyle\endcsname\relax
  \providecommand{\doi}[1]{doi: #1}\else
  \providecommand{\doi}{doi: \begingroup \urlstyle{rm}\Url}\fi

\bibitem[Abd El-Fattah et~al.(2008)Abd El-Fattah, Dessouky, Diab, and Abd El-Samie]{abd2008speech}
Abd El-Fattah, M., Dessouky, M.~I., Diab, S., and Abd El-Samie, F.
\newblock Speech enhancement using an adaptive {W}iener filtering approach.
\newblock \emph{Progress in Electromagnetics Research M}, 4:\penalty0 167--184, 2008.

\bibitem[Agustsson \& Timofte(2017)Agustsson and Timofte]{Agustsson_2017_CVPR_Workshops}
Agustsson, E. and Timofte, R.
\newblock {NTIRE 2017 Challenge on Single Image Super-Resolution: Dataset and Study}.
\newblock In \emph{IEEE Conference on Computer Vision and Pattern Recognition (CVPR) Workshops}, 2017.

\bibitem[Barker et~al.(2017)Barker, Marxer, Vincent, and Watanabe]{barker2017third}
Barker, J., Marxer, R., Vincent, E., and Watanabe, S.
\newblock The third {‘CHiME’}speech separation and recognition challenge: Analysis and outcomes.
\newblock \emph{Computer Speech \& Language}, 46:\penalty0 605--626, 2017.

\bibitem[Benita et~al.(2024)Benita, Elad, and Keshet]{benitadiffar}
Benita, R., Elad, M., and Keshet, J.
\newblock {DiffAR}: Denoising diffusion autoregressive model for raw speech waveform generation.
\newblock In \emph{International Conference on Learning Representations (ICLR)}, 2024.

\bibitem[Blau \& Michaeli(2018)Blau and Michaeli]{blau2018perception}
Blau, Y. and Michaeli, T.
\newblock The perception-distortion tradeoff.
\newblock In \emph{IEEE/CVF Conference on Computer Vision and Pattern Recognition (CVPR)}, pp.\  6228--6237, 2018.

\bibitem[Cao et~al.(2024)Cao, Tan, Gao, Xu, Chen, Heng, and Li]{cao2024survey}
Cao, H., Tan, C., Gao, Z., Xu, Y., Chen, G., Heng, P.-A., and Li, S.~Z.
\newblock A survey on generative diffusion models.
\newblock \emph{IEEE Transactions on Knowledge and Data Engineering}, 2024.

\bibitem[Chen et~al.(2020)Chen, Zhang, Zen, Weiss, Norouzi, and Chan]{chen2020wavegrad}
Chen, N., Zhang, Y., Zen, H., Weiss, R.~J., Norouzi, M., and Chan, W.
\newblock {WaveGrad}: Estimating gradients for waveform generation.
\newblock In \emph{International Conference on Learning Representations (ICLR)}, 2020.

\bibitem[Chung et~al.(2023{\natexlab{a}})Chung, Kim, Mccann, Klasky, and Ye]{chungdiffusion}
Chung, H., Kim, J., Mccann, M.~T., Klasky, M.~L., and Ye, J.~C.
\newblock Diffusion posterior sampling for general noisy inverse problems.
\newblock In \emph{International Conference on Learning Representations (ICLR)}, 2023{\natexlab{a}}.

\bibitem[Chung et~al.(2023{\natexlab{b}})Chung, Kim, and Ye]{chung2024direct}
Chung, H., Kim, J., and Ye, J.~C.
\newblock Direct diffusion bridge using data consistency for inverse problems.
\newblock In \emph{Advances in Neural Information Processing Systems (NeurIPS)}, 2023{\natexlab{b}}.

\bibitem[Croitoru et~al.(2023)Croitoru, Hondru, Ionescu, and Shah]{croitoru2023diffusion}
Croitoru, F.-A., Hondru, V., Ionescu, R.~T., and Shah, M.
\newblock Diffusion models in vision: A survey.
\newblock \emph{IEEE Transactions on Pattern Analysis and Machine Intelligence}, 2023.

\bibitem[Das et~al.(2023)Das, Fotiadis, Batra, Nabiei, Liao, Vakili, Shiu, and Bernacchia]{das2023image}
Das, A., Fotiadis, S., Batra, A., Nabiei, F., Liao, F., Vakili, S., Shiu, D.-s., and Bernacchia, A.
\newblock Image generation with shortest path diffusion.
\newblock In \emph{International Conference on Machine Learning (ICML)}, pp.\  7009--7024, 2023.

\bibitem[Defossez et~al.(2020)Defossez, Synnaeve, and Adi]{defossez2020real}
Defossez, A., Synnaeve, G., and Adi, Y.
\newblock Real time speech enhancement in the waveform domain.
\newblock In \emph{Annual Conference of the International Speech Communication Association (Interspeech)}, pp.\  3291--3295, 2020.

\bibitem[Fei et~al.(2023)Fei, Lyu, Pan, Zhang, Yang, Luo, Zhang, and Dai]{fei2023generative}
Fei, B., Lyu, Z., Pan, L., Zhang, J., Yang, W., Luo, T., Zhang, B., and Dai, B.
\newblock Generative diffusion prior for unified image restoration and enhancement.
\newblock In \emph{IEEE/CVF Conference on Computer Vision and Pattern Recognition (CVPR)}, pp.\  9935--9946, 2023.

\bibitem[Freirich et~al.(2021)Freirich, Michaeli, and Meir]{freirich2021theory}
Freirich, D., Michaeli, T., and Meir, R.
\newblock A theory of the distortion-perception tradeoff in wasserstein space.
\newblock In \emph{Advances in Neural Information Processing Systems (NeurIPS)}, pp.\  25661--25672, 2021.

\bibitem[Goodfellow et~al.(2014)Goodfellow, Pouget-Abadie, Mirza, Xu, Warde-Farley, Ozair, Courville, and Bengio]{goodfellow2014generative}
Goodfellow, I., Pouget-Abadie, J., Mirza, M., Xu, B., Warde-Farley, D., Ozair, S., Courville, A., and Bengio, Y.
\newblock Generative adversarial nets.
\newblock In \emph{Advances in Neural Information Processing Systems (NIPS)}, 2014.

\bibitem[Gulati et~al.(2020)Gulati, Qin, Chiu, Parmar, Zhang, Yu, Han, Wang, Zhang, Wu, and Pang]{gulati2020conformer}
Gulati, A., Qin, J., Chiu, C.-C., Parmar, N., Zhang, Y., Yu, J., Han, W., Wang, S., Zhang, Z., Wu, Y., and Pang, R.
\newblock Conformer: Convolution-augmented transformer for speech recognition.
\newblock In \emph{Annual Conference of the International Speech Communication Association (Interspeech)}, pp.\  5036--5040, 2020.

\bibitem[Harvey et~al.(2022)Harvey, Naderiparizi, and Wood]{harvey2022conditional}
Harvey, W., Naderiparizi, S., and Wood, F.
\newblock Conditional image generation by conditioning variational auto-encoders.
\newblock In \emph{International Conference on Learning Representations (ICLR)}, 2022.

\bibitem[He et~al.(2025)He, Shen, Fang, Xiao, Tang, Zhang, Zuo, Guo, and Li]{he2025diffusion}
He, C., Shen, Y., Fang, C., Xiao, F., Tang, L., Zhang, Y., Zuo, W., Guo, Z., and Li, X.
\newblock Diffusion models in low-level vision: A survey.
\newblock \emph{IEEE Transactions on Pattern Analysis and Machine Intelligence}, 2025.

\bibitem[He et~al.(2016)He, Zhang, Ren, and Sun]{he2016deep}
He, K., Zhang, X., Ren, S., and Sun, J.
\newblock Deep residual learning for image recognition.
\newblock In \emph{IEEE Conference on Computer Vision and Pattern Recognition (CVPR)}, pp.\  770--778, 2016.

\bibitem[Heusel et~al.(2017)Heusel, Ramsauer, Unterthiner, Nessler, and Hochreiter]{heusel2017gans}
Heusel, M., Ramsauer, H., Unterthiner, T., Nessler, B., and Hochreiter, S.
\newblock {GANs} trained by a two time-scale update rule converge to a local {N}ash equilibrium.
\newblock In \emph{Advances in Neural Information Processing Systems (NIPS)}, 2017.

\bibitem[Ho et~al.(2020)Ho, Jain, and Abbeel]{ho2020denoising}
Ho, J., Jain, A., and Abbeel, P.
\newblock Denoising diffusion probabilistic models.
\newblock In \emph{Advances in Neural Information Processing Systems (NeurIPS)}, pp.\  6840--6851, 2020.

\bibitem[Hsieh et~al.(2020)Hsieh, Wang, Lu, and Tsao]{hsieh2020wavecrn}
Hsieh, T.-A., Wang, H.-M., Lu, X., and Tsao, Y.
\newblock Wavecrn: An efficient convolutional recurrent neural network for end-to-end speech enhancement.
\newblock \emph{IEEE Signal Processing Letters}, 27:\penalty0 2149--2153, 2020.

\bibitem[Hu \& Loizou(2007)Hu and Loizou]{hu2007evaluation}
Hu, Y. and Loizou, P.~C.
\newblock Evaluation of objective quality measures for speech enhancement.
\newblock \emph{IEEE Transactions on Audio, Speech, and Language Processing}, 16\penalty0 (1):\penalty0 229--238, 2007.

\bibitem[Huang et~al.(2024)Huang, Huang, Liu, Yan, Dong, Lyu, Chen, and Chen]{huang2024wavedm}
Huang, Y., Huang, J., Liu, J., Yan, M., Dong, Y., Lyu, J., Chen, C., and Chen, S.
\newblock Wave{DM}: Wavelet-based diffusion models for image restoration.
\newblock \emph{IEEE Transactions on Multimedia}, 2024.

\bibitem[Hurault et~al.(2022)Hurault, Leclaire, and Papadakis]{hurault2022gradient}
Hurault, S., Leclaire, A., and Papadakis, N.
\newblock Gradient step denoiser for convergent plug-and-play.
\newblock In \emph{International Conference on Learning Representations (ICLR)}, 2022.

\bibitem[Huynh-Thu \& Ghanbari(2008)Huynh-Thu and Ghanbari]{huynh2008scope}
Huynh-Thu, Q. and Ghanbari, M.
\newblock Scope of validity of psnr in image/video quality assessment.
\newblock \emph{Electronics letters}, 44\penalty0 (13):\penalty0 800--801, 2008.

\bibitem[{ITU-T Rec. P.862.2}(2005)]{itu862}
{ITU-T Rec. P.862.2}.
\newblock Wideband extension to recommendation {P}.862 for the assessment of wideband telephone networks and speech codecs.
\newblock \emph{International Telecommunication Union}, 2005.

\bibitem[Jiang et~al.(2021)Jiang, Wang, Yi, Chen, Wang, Wang, Jiang, and Lin]{jiang2021rain}
Jiang, K., Wang, Z., Yi, P., Chen, C., Wang, Z., Wang, X., Jiang, J., and Lin, C.-W.
\newblock Rain-free and residue hand-in-hand: A progressive coupled network for real-time image deraining.
\newblock \emph{IEEE Transactions on Image Processing}, 30:\penalty0 7404--7418, 2021.

\bibitem[Kingma \& Ba(2014)Kingma and Ba]{kingma2014adam}
Kingma, D.~P. and Ba, J.
\newblock Adam: A method for stochastic optimization.
\newblock In \emph{International Conference on Learning Representations (ICLR)}, 2014.

\bibitem[Kingma \& Welling(2014)Kingma and Welling]{kingma2013auto}
Kingma, D.~P. and Welling, M.
\newblock Auto-encoding variational bayes.
\newblock In \emph{International Conference on Learning Representations (ICLR)}, 2014.

\bibitem[Kohl et~al.(2018)Kohl, Romera-Paredes, Meyer, De~Fauw, Ledsam, Maier-Hein, Eslami, Jimenez~Rezende, and Ronneberger]{kohl2018probabilistic}
Kohl, S., Romera-Paredes, B., Meyer, C., De~Fauw, J., Ledsam, J.~R., Maier-Hein, K., Eslami, S., Jimenez~Rezende, D., and Ronneberger, O.
\newblock A probabilistic {U-Net} for segmentation of ambiguous images.
\newblock In \emph{Advances in Neural Information Processing Systems (NIPS)}, 2018.

\bibitem[Koizumi et~al.(2022)Koizumi, Zen, Yatabe, Chen, and Bacchiani]{koizumi22_interspeech}
Koizumi, Y., Zen, H., Yatabe, K., Chen, N., and Bacchiani, M.
\newblock {{SpecGrad}: Diffusion probabilistic model based neural vocoder with adaptive noise spectral shaping}.
\newblock In \emph{Annual Conference of the International Speech Communication Association (Interspeech)}, pp.\  803--807, 2022.

\bibitem[Kong et~al.(2021)Kong, Ping, Huang, Zhao, and Catanzaro]{kong2020diffwave}
Kong, Z., Ping, W., Huang, J., Zhao, K., and Catanzaro, B.
\newblock {DiffWave}: A versatile diffusion model for audio synthesis.
\newblock In \emph{International Conference on Learning Representations (ICLR)}, 2021.

\bibitem[Kupyn et~al.(2019)Kupyn, Martyniuk, Wu, and Wang]{kupyn2019deblurgan}
Kupyn, O., Martyniuk, T., Wu, J., and Wang, Z.
\newblock {DeblurGAN}-v2: Deblurring (orders-of-magnitude) faster and better.
\newblock In \emph{IEEE/CVF Conference on Computer Vision and Pattern Recognition (CVPR)}, pp.\  8878--8887, 2019.

\bibitem[Le~Roux et~al.(2019)Le~Roux, Wisdom, Erdogan, and Hershey]{le2019sdr}
Le~Roux, J., Wisdom, S., Erdogan, H., and Hershey, J.~R.
\newblock {SDR}--half-baked or well done?
\newblock In \emph{IEEE International Conference on Acoustics, Speech and Signal Processing (ICASSP)}, pp.\  626--630, 2019.

\bibitem[Lee et~al.(2022)Lee, Kim, Shin, Tan, Liu, Meng, Qin, Chen, Yoon, and Liu]{lee2021priorgrad}
Lee, S.-g., Kim, H., Shin, C., Tan, X., Liu, C., Meng, Q., Qin, T., Chen, W., Yoon, S., and Liu, T.-Y.
\newblock {PriorGrad}: Improving conditional denoising diffusion models with data-dependent adaptive prior.
\newblock In \emph{International Conference on Learning Representations (ICLR)}, 2022.

\bibitem[Lemercier et~al.(2023)Lemercier, Richter, Welker, and Gerkmann]{lemercier2023storm}
Lemercier, J.-M., Richter, J., Welker, S., and Gerkmann, T.
\newblock {StoRM}: A diffusion-based stochastic regeneration model for speech enhancement and dereverberation.
\newblock \emph{IEEE/ACM Transactions on Audio, Speech, and Language Processing}, 2023.

\bibitem[Leng et~al.(2022)Leng, Chen, Guo, Liu, Chen, Tan, Mandic, He, Li, Qin, zhao, and Liu]{leng2022binauralgrad}
Leng, Y., Chen, Z., Guo, J., Liu, H., Chen, J., Tan, X., Mandic, D., He, L., Li, X., Qin, T., zhao, s., and Liu, T.-Y.
\newblock {BinauralGrad}: A two-stage conditional diffusion probabilistic model for binaural audio synthesis.
\newblock In \emph{Advances in Neural Information Processing Systems (NeurIPS)}, pp.\  23689--23700, 2022.

\bibitem[Li et~al.(2019)Li, Cheong, and Tan]{li2019heavy}
Li, R., Cheong, L.-F., and Tan, R.~T.
\newblock Heavy rain image restoration: Integrating physics model and conditional adversarial learning.
\newblock In \emph{IEEE/CVF Conference on Computer Vision and Pattern Recognition (CVPR)}, pp.\  1633--1642, 2019.

\bibitem[Li et~al.(2020)Li, Tan, and Cheong]{li2020all}
Li, R., Tan, R.~T., and Cheong, L.-F.
\newblock All in one bad weather removal using architectural search.
\newblock In \emph{IEEE/CVF Conference on Computer Vision and Pattern Recognition (CVPR)}, pp.\  3175--3185, 2020.

\bibitem[Li et~al.(2018)Li, Wu, Lin, Liu, and Zha]{li2018recurrent}
Li, X., Wu, J., Lin, Z., Liu, H., and Zha, H.
\newblock Recurrent squeeze-and-excitation context aggregation net for single image deraining.
\newblock In \emph{Proceedings of the European Conference on Computer Vision (ECCV)}, pp.\  254--269, 2018.

\bibitem[Li et~al.(2023)Li, Ren, Jin, Lan, Wang, Zeng, Wang, and Chen]{li2023diffusion}
Li, X., Ren, Y., Jin, X., Lan, C., Wang, X., Zeng, W., Wang, X., and Chen, Z.
\newblock Diffusion models for image restoration and enhancement--{A} comprehensive survey.
\newblock \emph{arXiv preprint arXiv:2308.09388}, 2023.

\bibitem[Liu et~al.(2023{\natexlab{a}})Liu, Vahdat, Huang, Theodorou, Nie, and Anandkumar]{liu20232}
Liu, G.-H., Vahdat, A., Huang, D.-A., Theodorou, E., Nie, W., and Anandkumar, A.
\newblock $\text{I}^2\text{SB}$: Image-to-image {S}chr{\"o}dinger bridge.
\newblock In \emph{International Conference on Machine Learning (ICML)}, pp.\  22042--22062, 2023{\natexlab{a}}.

\bibitem[Liu et~al.(2023{\natexlab{b}})Liu, Chen, Yuan, Mei, Liu, Mandic, Wang, and Plumbley]{liu2023audioldm}
Liu, H., Chen, Z., Yuan, Y., Mei, X., Liu, X., Mandic, D., Wang, W., and Plumbley, M.~D.
\newblock Audio{LDM}: Text-to-audio generation with latent diffusion models.
\newblock In \emph{International Conference on Machine Learning (ICML)}, 2023{\natexlab{b}}.

\bibitem[Liu et~al.(2024)Liu, Wang, Fan, Wang, Tang, and Qu]{liu2024residual}
Liu, J., Wang, Q., Fan, H., Wang, Y., Tang, Y., and Qu, L.
\newblock Residual denoising diffusion models.
\newblock In \emph{Proceedings of the IEEE/CVF Conference on Computer Vision and Pattern Recognition (CVPR)}, pp.\  2773--2783, 2024.

\bibitem[Liu et~al.(2019)Liu, Suganuma, Sun, and Okatani]{liu2019dual}
Liu, X., Suganuma, M., Sun, Z., and Okatani, T.
\newblock Dual residual networks leveraging the potential of paired operations for image restoration.
\newblock In \emph{IEEE/CVF Conference on Computer Vision and Pattern Recognition (CVPR)}, pp.\  7007--7016, 2019.

\bibitem[Liu et~al.(2018)Liu, Jaw, Huang, and Hwang]{liu2018desnownet}
Liu, Y.-F., Jaw, D.-W., Huang, S.-C., and Hwang, J.-N.
\newblock {DesnowNet}: Context-aware deep network for snow removal.
\newblock \emph{IEEE Transactions on Image Processing}, 27\penalty0 (6):\penalty0 3064--3073, 2018.

\bibitem[Lu et~al.(2021)Lu, Tsao, and Watanabe]{lu2021study}
Lu, Y.-J., Tsao, Y., and Watanabe, S.
\newblock A study on speech enhancement based on diffusion probabilistic model.
\newblock In \emph{Asia-Pacific Signal and Information Processing Association Annual Summit and Conference (APSIPA ASC)}, pp.\  659--666, 2021.

\bibitem[Lu et~al.(2022)Lu, Wang, Watanabe, Richard, Yu, and Tsao]{lu2022conditional}
Lu, Y.-J., Wang, Z.-Q., Watanabe, S., Richard, A., Yu, C., and Tsao, Y.
\newblock Conditional diffusion probabilistic model for speech enhancement.
\newblock In \emph{IEEE International Conference on Acoustics, Speech and Signal Processing (ICASSP)}, pp.\  7402--7406, 2022.

\bibitem[Luo(2022)]{luo2022understanding}
Luo, C.
\newblock Understanding diffusion models: A unified perspective.
\newblock \emph{arXiv preprint arXiv:2208.11970}, 2022.

\bibitem[Luo et~al.(2023)Luo, Gustafsson, Zhao, Sj{\"o}lund, and Sch{\"o}n]{luo2023refusion}
Luo, Z., Gustafsson, F.~K., Zhao, Z., Sj{\"o}lund, J., and Sch{\"o}n, T.~B.
\newblock Refusion: Enabling large-size realistic image restoration with latent-space diffusion models.
\newblock In \emph{IEEE/CVF Conference on Computer Vision and Pattern Recognition (CVPR)}, pp.\  1680--1691, 2023.

\bibitem[Majumdar et~al.(2021)Majumdar, Balam, Hrinchuk, Lavrukhin, Noroozi, and Ginsburg]{majumdar2021citrinet}
Majumdar, S., Balam, J., Hrinchuk, O., Lavrukhin, V., Noroozi, V., and Ginsburg, B.
\newblock Citrinet: Closing the gap between non-autoregressive and autoregressive end-to-end models for automatic speech recognition.
\newblock \emph{arXiv preprint arXiv:2104.01721}, 2021.

\bibitem[Mittal et~al.(2012)Mittal, Soundararajan, and Bovik]{mittal2012making}
Mittal, A., Soundararajan, R., and Bovik, A.~C.
\newblock Making a “completely blind” image quality analyzer.
\newblock \emph{IEEE Signal Processing Letters}, 20\penalty0 (3):\penalty0 209--212, 2012.

\bibitem[Nichol \& Dhariwal(2021)Nichol and Dhariwal]{nichol2021improved}
Nichol, A.~Q. and Dhariwal, P.
\newblock Improved denoising diffusion probabilistic models.
\newblock In \emph{International Conference on Machine Learning (ICML)}, pp.\  8162--8171, 2021.

\bibitem[Nielsen et~al.(2024)Nielsen, Christensen, Dittadi, and Winther]{nielsendiffenc}
Nielsen, B. M.~G., Christensen, A., Dittadi, A., and Winther, O.
\newblock {DiffEnc}: Variational diffusion with a learned encoder.
\newblock In \emph{International Conference on Learning Representations (ICLR)}, 2024.

\bibitem[{\"O}zdenizci \& Legenstein(2023){\"O}zdenizci and Legenstein]{ozdenizci2023restoring}
{\"O}zdenizci, O. and Legenstein, R.
\newblock Restoring vision in adverse weather conditions with patch-based denoising diffusion models.
\newblock \emph{IEEE Transactions on Pattern Analysis and Machine Intelligence}, 2023.

\bibitem[Pandey et~al.(2021)Pandey, Mukherjee, Rai, and Kumar]{pandey2021vaes}
Pandey, K., Mukherjee, A., Rai, P., and Kumar, A.
\newblock {VAEs} meet diffusion models: Efficient and high-fidelity generation.
\newblock In \emph{NeurIPS 2021 Workshop on Deep Generative Models and Downstream Applications}, 2021.

\bibitem[Pandey et~al.(2022)Pandey, Mukherjee, Rai, and Kumar]{pandey2022diffusevae}
Pandey, K., Mukherjee, A., Rai, P., and Kumar, A.
\newblock Diffuse{VAE}: Efficient, controllable and high-fidelity generation from low-dimensional latents.
\newblock \emph{Transactions on Machine Learning Research}, 2022.

\bibitem[Pascual et~al.(2017)Pascual, Bonafonte, and Serra]{pascual2017segan}
Pascual, S., Bonafonte, A., and Serra, J.
\newblock {SEGAN}: Speech enhancement generative adversarial network.
\newblock In \emph{Annual Conference of the International Speech Communication Association (Interspeech)}, pp.\  3642--3646, 2017.

\bibitem[Phan et~al.(2020)Phan, McLoughlin, Pham, Ch{\'e}n, Koch, De~Vos, and Mertins]{phan2020improving}
Phan, H., McLoughlin, I.~V., Pham, L., Ch{\'e}n, O.~Y., Koch, P., De~Vos, M., and Mertins, A.
\newblock Improving gans for speech enhancement.
\newblock \emph{IEEE Signal Processing Letters}, 27:\penalty0 1700--1704, 2020.

\bibitem[Popov et~al.(2021)Popov, Vovk, Gogoryan, Sadekova, and Kudinov]{popov2021grad}
Popov, V., Vovk, I., Gogoryan, V., Sadekova, T., and Kudinov, M.
\newblock {Grad-TTS}: A diffusion probabilistic model for text-to-speech.
\newblock In \emph{International Conference on Machine Learning (ICML)}, pp.\  8599--8608, 2021.

\bibitem[Qian et~al.(2018)Qian, Tan, Yang, Su, and Liu]{qian2018attentive}
Qian, R., Tan, R.~T., Yang, W., Su, J., and Liu, J.
\newblock Attentive generative adversarial network for raindrop removal from a single image.
\newblock In \emph{IEEE/CVF Conference on Computer Vision and Pattern Recognition (CVPR)}, pp.\  2482--2491, 2018.

\bibitem[Quan et~al.(2021)Quan, Yu, Liang, and Yang]{quan2021removing}
Quan, R., Yu, X., Liang, Y., and Yang, Y.
\newblock Removing raindrops and rain streaks in one go.
\newblock In \emph{IEEE/CVF Conference on Computer Vision and Pattern Recognition (CVPR)}, pp.\  9147--9156, 2021.

\bibitem[Quan et~al.(2019)Quan, Deng, Chen, and Ji]{quan2019deep}
Quan, Y., Deng, S., Chen, Y., and Ji, H.
\newblock Deep learning for seeing through window with raindrops.
\newblock In \emph{IEEE/CVF International Conference on Computer Vision (ICCV)}, pp.\  2463--2471, 2019.

\bibitem[Richter et~al.(2023)Richter, Welker, Lemercier, Lay, and Gerkmann]{richter2023speech}
Richter, J., Welker, S., Lemercier, J.-M., Lay, B., and Gerkmann, T.
\newblock Speech enhancement and dereverberation with diffusion-based generative models.
\newblock \emph{IEEE/ACM Transactions on Audio, Speech, and Language Processing}, 2023.

\bibitem[Rim et~al.(2020)Rim, Lee, Won, and Cho]{rim_2020_ECCV}
Rim, J., Lee, H., Won, J., and Cho, S.
\newblock Real-world blur dataset for learning and benchmarking deblurring algorithms.
\newblock In \emph{European Conference on Computer Vision (ECCV)}, 2020.

\bibitem[Rombach et~al.(2022)Rombach, Blattmann, Lorenz, Esser, and Ommer]{rombach2022high}
Rombach, R., Blattmann, A., Lorenz, D., Esser, P., and Ommer, B.
\newblock High-resolution image synthesis with latent diffusion models.
\newblock In \emph{IEEE/CVF Conference on Computer Vision and Pattern Recognition (CVPR)}, pp.\  10684--10695, 2022.

\bibitem[Serr{\`a} et~al.(2022)Serr{\`a}, Pascual, Pons, Araz, and Scaini]{serra2022universal}
Serr{\`a}, J., Pascual, S., Pons, J., Araz, R.~O., and Scaini, D.
\newblock Universal speech enhancement with score-based diffusion.
\newblock \emph{arXiv preprint arXiv:2206.03065}, 2022.

\bibitem[Sohl-Dickstein et~al.(2015)Sohl-Dickstein, Weiss, Maheswaranathan, and Ganguli]{sohl2015deep}
Sohl-Dickstein, J., Weiss, E., Maheswaranathan, N., and Ganguli, S.
\newblock Deep unsupervised learning using nonequilibrium thermodynamics.
\newblock In \emph{International Conference on Machine Learning (ICML)}, pp.\  2256--2265, 2015.

\bibitem[Sohn et~al.(2015)Sohn, Lee, and Yan]{sohn2015learning}
Sohn, K., Lee, H., and Yan, X.
\newblock Learning structured output representation using deep conditional generative models.
\newblock In \emph{Advances in Neural Information Processing Systems (NIPS)}, 2015.

\bibitem[Song et~al.(2020)Song, Meng, and Ermon]{song2020denoising}
Song, J., Meng, C., and Ermon, S.
\newblock Denoising diffusion implicit models.
\newblock In \emph{International Conference on Learning Representations (ICLR)}, 2020.

\bibitem[Stoica et~al.(2005)Stoica, Moses, et~al.]{stoica2005spectral}
Stoica, P., Moses, R.~L., et~al.
\newblock \emph{Spectral analysis of signals}, volume 452.
\newblock Pearson Prentice Hall Upper Saddle River, NJ, 2005.

\bibitem[Strauss \& Edler(2021)Strauss and Edler]{strauss2021flow}
Strauss, M. and Edler, B.
\newblock A flow-based neural network for time domain speech enhancement.
\newblock In \emph{IEEE International Conference on Acoustics, Speech and Signal Processing (ICASSP)}, pp.\  5754--5758, 2021.

\bibitem[Tai et~al.(2023{\natexlab{a}})Tai, Lei, Zhou, Trajcevski, and Zhong]{tai2024dose}
Tai, W., Lei, Y., Zhou, F., Trajcevski, G., and Zhong, T.
\newblock {DOSE}: Diffusion dropout with adaptive prior for speech enhancement.
\newblock In \emph{Advances in Neural Information Processing Systems (NeurIPS)}, 2023{\natexlab{a}}.

\bibitem[Tai et~al.(2023{\natexlab{b}})Tai, Zhou, Trajcevski, and Zhong]{tai2023revisiting}
Tai, W., Zhou, F., Trajcevski, G., and Zhong, T.
\newblock Revisiting denoising diffusion probabilistic models for speech enhancement: Condition collapse, efficiency and refinement.
\newblock In \emph{AAAI Conference on Artificial Intelligence (AAAI)}, volume~37, pp.\  13627--13635, 2023{\natexlab{b}}.

\bibitem[Tao et~al.(2018)Tao, Gao, Shen, Wang, and Jia]{tao2018scale}
Tao, X., Gao, H., Shen, X., Wang, J., and Jia, J.
\newblock Scale-recurrent network for deep image deblurring.
\newblock In \emph{IEEE/CVF Conference on Computer Vision and Pattern Recognition (CVPR)}, pp.\  8174--8182, 2018.

\bibitem[Thiemann et~al.(2013)Thiemann, Ito, and Vincent]{thiemann2013diverse}
Thiemann, J., Ito, N., and Vincent, E.
\newblock The diverse environments multi-channel acoustic noise database ({DEMAND}): A database of multichannel environmental noise recordings.
\newblock In \emph{Proceedings of Meetings on Acoustics}, 2013.

\bibitem[Timofte et~al.(2017)Timofte, Agustsson, Van~Gool, Yang, Zhang, Lim, et~al.]{Timofte_2017_CVPR_Workshops}
Timofte, R., Agustsson, E., Van~Gool, L., Yang, M.-H., Zhang, L., Lim, B., et~al.
\newblock {NTIRE 2017 Challenge on Single Image Super-Resolution: Methods and Results}.
\newblock In \emph{IEEE Conference on Computer Vision and Pattern Recognition (CVPR) Workshops}, 2017.

\bibitem[Valanarasu et~al.(2022)Valanarasu, Yasarla, and Patel]{valanarasu2022transweather}
Valanarasu, J. M.~J., Yasarla, R., and Patel, V.~M.
\newblock {TransWeather}: Transformer-based restoration of images degraded by adverse weather conditions.
\newblock In \emph{IEEE/CVF Conference on Computer Vision and Pattern Recognition (CVPR)}, pp.\  2353--2363, 2022.

\bibitem[Valentini-Botinhao et~al.(2016)Valentini-Botinhao, Wang, Takaki, and Yamagishi]{valentini2016investigating}
Valentini-Botinhao, C., Wang, X., Takaki, S., and Yamagishi, J.
\newblock Investigating {RNN}-based speech enhancement methods for noise-robust text-to-speech.
\newblock In \emph{ISCA Workshop on Speech Synthesis Workshop (SSW)}, pp.\  146--152, 2016.

\bibitem[Veaux et~al.(2013)Veaux, Yamagishi, and King]{veaux2013voice}
Veaux, C., Yamagishi, J., and King, S.
\newblock The voice bank corpus: Design, collection and data analysis of a large regional accent speech database.
\newblock In \emph{International Conference Oriental COCOSDA held jointly with 2013 Conference on Asian Spoken Language Research and Evaluation (O-COCOSDA/CASLRE)}, 2013.

\bibitem[Wang et~al.(2004)Wang, Bovik, Sheikh, and Simoncelli]{wang2004image}
Wang, Z., Bovik, A.~C., Sheikh, H.~R., and Simoncelli, E.~P.
\newblock Image quality assessment: from error visibility to structural similarity.
\newblock \emph{IEEE ransactions on Image Processing}, 13\penalty0 (4):\penalty0 600--612, 2004.

\bibitem[Welker et~al.(2022)Welker, Richter, and Gerkmann]{welker2022speech}
Welker, S., Richter, J., and Gerkmann, T.
\newblock Speech enhancement with score-based generative models in the complex {STFT} domain.
\newblock In \emph{Annual Conference of the International Speech Communication Association (Interspeech)}, pp.\  2928--2932, 2022.

\bibitem[Xia et~al.(2023)Xia, Zhang, Wang, Wang, Wu, Tian, Yang, and Van~Gool]{xia2023diffir}
Xia, B., Zhang, Y., Wang, S., Wang, Y., Wu, X., Tian, Y., Yang, W., and Van~Gool, L.
\newblock Diff{IR}: Efficient diffusion model for image restoration.
\newblock In \emph{IEEE/CVF International Conference on Computer Vision (ICCV)}, pp.\  13095--13105, 2023.

\bibitem[Xiao et~al.(2022)Xiao, Fu, Liu, Wu, and Zha]{xiao2022image}
Xiao, J., Fu, X., Liu, A., Wu, F., and Zha, Z.-J.
\newblock Image de-raining transformer.
\newblock \emph{IEEE Transactions on Pattern Analysis and Machine Intelligence}, 2022.

\bibitem[Xiao et~al.(2024)Xiao, Feng, Zhang, Liu, Yang, Zhu, Fu, Zhu, Liu, and Zha]{xiaodreamclean}
Xiao, J., Feng, R., Zhang, H., Liu, Z., Yang, Z., Zhu, Y., Fu, X., Zhu, K., Liu, Y., and Zha, Z.-J.
\newblock {DreamClean}: Restoring clean image using deep diffusion prior.
\newblock In \emph{International Conference on Learning Representations (ICLR)}, 2024.

\bibitem[Ye et~al.(2024)Ye, Chen, Chai, Xing, Qin, Lin, and Zhu]{ye2024learning}
Ye, T., Chen, S., Chai, W., Xing, Z., Qin, J., Lin, G., and Zhu, L.
\newblock Learning diffusion texture priors for image restoration.
\newblock In \emph{Proceedings of the IEEE/CVF Conference on Computer Vision and Pattern Recognition (CVPR)}, pp.\  2524--2534, 2024.

\bibitem[Yen et~al.(2023)Yen, Germain, Wichern, and Le~Roux]{yen2023cold}
Yen, H., Germain, F.~G., Wichern, G., and Le~Roux, J.
\newblock Cold diffusion for speech enhancement.
\newblock In \emph{IEEE International Conference on Acoustics, Speech and Signal Processing (ICASSP)}, 2023.

\bibitem[Zamir et~al.(2021)Zamir, Arora, Khan, Hayat, Khan, Yang, and Shao]{zamir2021multi}
Zamir, S.~W., Arora, A., Khan, S., Hayat, M., Khan, F.~S., Yang, M.-H., and Shao, L.
\newblock Multi-stage progressive image restoration.
\newblock In \emph{Proceedings of the IEEE/CVF Conference on Computer Vision and Pattern Recognition (CVPR)}, pp.\  14821--14831, 2021.

\bibitem[Zhang et~al.(2021)Zhang, Li, Yu, Luo, and Li]{zhang2021deep}
Zhang, K., Li, R., Yu, Y., Luo, W., and Li, C.
\newblock Deep dense multi-scale network for snow removal using semantic and depth priors.
\newblock \emph{IEEE Transactions on Image Processing}, 30:\penalty0 7419--7431, 2021.

\bibitem[Zhang et~al.(2018)Zhang, Isola, Efros, Shechtman, and Wang]{zhang2018unreasonable}
Zhang, R., Isola, P., Efros, A.~A., Shechtman, E., and Wang, O.
\newblock The unreasonable effectiveness of deep features as a perceptual metric.
\newblock In \emph{IEEE/CVF Conference on Computer Vision and Pattern Recognition (CVPR)}, pp.\  586--595, 2018.

\bibitem[Zheng et~al.(2025)Zheng, He, Chen, Bao, and Zhu]{zheng2024diffusion}
Zheng, K., He, G., Chen, J., Bao, F., and Zhu, J.
\newblock Diffusion bridge implicit models.
\newblock In \emph{International Conference on Learning Representations (ICLR)}, 2025.

\bibitem[Zhou et~al.(2024)Zhou, Lou, Khanna, and Ermon]{zhoudenoising}
Zhou, L., Lou, A., Khanna, S., and Ermon, S.
\newblock Denoising diffusion bridge models.
\newblock In \emph{International Conference on Learning Representations (ICLR)}, 2024.

\bibitem[Zhu et~al.(2023)Zhu, Zhang, Liang, Cao, Wen, Timofte, and Van~Gool]{zhu2023denoising}
Zhu, Y., Zhang, K., Liang, J., Cao, J., Wen, B., Timofte, R., and Van~Gool, L.
\newblock Denoising diffusion models for plug-and-play image restoration.
\newblock In \emph{IEEE/CVF Conference on Computer Vision and Pattern Recognition (CVPR)}, pp.\  1219--1229, 2023.

\end{thebibliography}
\bibliographystyle{icml2025}

\appendix
\onecolumn
\section{Proof of Proposition 3.1}
\label{sec: appendix proof prop 1}
\noindent\textbf{Proposition 3.1} 
\textit{
    (Incorporation of diffusion process into VAE)\textbf{.}
    By introducing a sequence of hidden variables $\mathbf{x}_{1:T}$, under the setup of conditional diffusion models where the Markov Chain assumption is employed on the forward process $q(\mathbf{x}_{1:T}|\mathbf{x}_0)\coloneq\prod_{t=1}^Tq(\mathbf{x}_t|\mathbf{x}_{t-1})$ and the reverse process $p_\theta(\mathbf{x}_{0:T}|\mathbf{y})\coloneq p(\mathbf{x}_T)\prod_{t=1}^Tp_\theta(\mathbf{x}_{t-1}|\mathbf{x}_t,\mathbf{y})$ parameterized by a DDPM $\theta$, and assuming that $\mathbf{x}_T=\boldsymbol{\epsilon}$ (i.e., the latent noise of DDPM samples from the VAE latent distribution), we have the lower bound on $\log p(\mathbf{x}_0|\mathbf{y},\boldsymbol{\epsilon})$ in the reconstruction term of the VAE (\ref{eq: vae original elbo}) as:
    \begin{equation*}
        \log p(\mathbf{x}_0|\mathbf{y},\boldsymbol{\epsilon})\geq\mathbb{E}_{q(\mathbf{x}_{1:T}|\mathbf{x}_0)}\left[\log\frac{p_\theta({\mathbf{x}_{0:T}}|\mathbf{y})}{q(\mathbf{x}_{1:T}|\mathbf{x}_0)}\right],
    \end{equation*}
    which is the ELBO of the conditional DDPM (i.e., the conditional version of (\ref{eq: ddpm log likelihood})).
}

\vspace{0.25cm}

\textit{Proof:}

We lower bound the conditional data log-likelihood $\log p(\mathbf{x}_0|\mathbf{y},\boldsymbol{\epsilon})$ by utilizing a sequence of hidden latent representations $\mathbf{x}_{1:T}$ and the approximate variational distribution $q(\mathbf{x}_{1:T}|\mathbf{x}_0)$:
\begin{equation}
\begin{aligned}
    \log p(\mathbf{x}_0|\mathbf{y},\boldsymbol{\epsilon})=&\log\int p_\theta(\mathbf{x}_{0:T}|\mathbf{y},\boldsymbol{\epsilon})d\mathbf{x}_{1:T} \\
    =&\log\int \frac{p_\theta(\mathbf{x}_{0:T}|\mathbf{y},\boldsymbol{\epsilon})q(\mathbf{x}_{1:T}|\mathbf{x}_0)}{q(\mathbf{x}_{1:T}|\mathbf{x}_0)}d\mathbf{x}_{1:T} \\
    =&\log\mathbb{E}_{q(\mathbf{x}_{1:T}|\mathbf{x}_0)}\left[\frac{p_\theta({\mathbf{x}_{0:T}}|\mathbf{y},\boldsymbol{\epsilon})}{q(\mathbf{x}_{1:T}|\mathbf{x}_0)}\right] \\
    \geq&\mathbb{E}_{q(\mathbf{x}_{1:T}|\mathbf{x}_0)}\left[\log\frac{p_\theta({\mathbf{x}_{0:T}}|\mathbf{y},\boldsymbol{\epsilon})}{q(\mathbf{x}_{1:T}|\mathbf{x}_0)}\right],
\label{eq: cddpm lower bound appendix}
\end{aligned}
\end{equation}
where the inequality is obtained by applying Jensen’s inequality.

The reverse diffusion process incorporating $\boldsymbol{\epsilon}$ is given as:
\begin{equation}
\begin{aligned}
    p_\theta(\mathbf{x}_{0:T}|\mathbf{y},\boldsymbol{\epsilon})\coloneq & p(\mathbf{x}_T)\prod_{t=1}^Tp_\theta(\mathbf{x}_{t-1}|\mathbf{x}_t,\mathbf{y},\boldsymbol{\epsilon}) \\
    =&p(\mathbf{x}_T)\prod_{t=1}^Tp_\theta(\mathbf{x}_{t-1}|\mathbf{x}_t,\mathbf{y},\mathbf{x}_T) \\
    =&p(\mathbf{x}_T)\prod_{t=1}^Tp_\theta(\mathbf{x}_{t-1}|\mathbf{x}_t,\mathbf{y})\eqcolon p_\theta(\mathbf{x}_{0:T}|\mathbf{y}),
\end{aligned}
\end{equation}
by using the assumption that the diffusion prior samples from the VAE latent space, i.e., $\mathbf{x}_T=\boldsymbol{\epsilon}$, and utilizing the Markov Chain property on the reverse transition probability to remove the $\mathbf{x}_T$ as a condition. This indicates that:
\begin{equation}
    \mathbb{E}_{q(\mathbf{x}_{1:T}|\mathbf{x}_0)}\left[\log\frac{p_\theta({\mathbf{x}_{0:T}}|\mathbf{y},\boldsymbol{\epsilon})}{q(\mathbf{x}_{1:T}|\mathbf{x}_0)}\right]=\mathbb{E}_{q(\mathbf{x}_{1:T}|\mathbf{x}_0)}\left[\log\frac{p_\theta({\mathbf{x}_{0:T}}|\mathbf{y})}{q(\mathbf{x}_{1:T}|\mathbf{x}_0)}\right].
\label{eq: cddpm lower bound appendix 2} 
\end{equation}

Together, (\ref{eq: cddpm lower bound appendix}) and (\ref{eq: cddpm lower bound appendix 2}) lead to the result in Proposition \ref{proposition: 1}.

\section{Derivation of Proposition 3.2}
\label{sec: appendix proof prop 2}
\noindent\textbf{Proposition 3.2} (RestoreGrad)\textbf{.}\textit{
Assume the prior and posterior distributions are both zero-mean Gaussian, parameterized as $p_{\psi}(\boldsymbol{\epsilon}|\mathbf{y})=\mathcal{N}(\boldsymbol{\epsilon};\mathbf{0},\boldsymbol{\Sigma}_{\text{prior}}(\mathbf{y};\psi))$ and $q_{\phi}(\boldsymbol{\epsilon}|\mathbf{x}_0, \mathbf{y})=\mathcal{N}(\boldsymbol{\epsilon};\mathbf{0},\boldsymbol{\Sigma}_{\text{post}}(\mathbf{x}_0,\mathbf{y};\phi))$, respectively, where the covariances are estimated by the Prior Net $\psi$ (taking $\mathbf{y}$ as input) and Posterior Net $\phi$ (taking both $\mathbf{x}_0$ and $\mathbf{y}$ as input). Let us simply use $\boldsymbol{\Sigma}_{\text{prior}}$ and $\boldsymbol{\Sigma}_{\text{post}}$ hereafter to refer to $\boldsymbol{\Sigma}_{\text{prior}}(\mathbf{y};\psi)$ and $\boldsymbol{\Sigma}_{\text{post}}(\mathbf{x}_0,\mathbf{y};\phi)$ for concise notation. 
Then, with the direct sampling property in the forward path $\mathbf{x}_t=\sqrt{\bar{\alpha}_t}\mathbf{x}_0+\sqrt{1-\bar{\alpha}_t}\boldsymbol{\epsilon}$ at arbitrary timestep $t$ where $\boldsymbol{\epsilon}\sim q_{\phi}(\boldsymbol{\epsilon}|\mathbf{x}_0,\mathbf{y})$, and assuming the reverse process has the same covariance as the true forward process posterior conditioned on $\mathbf{x}_0$, by utilizing the conditional DDPM $\boldsymbol{\epsilon}_\theta(\mathbf{x}_t,\mathbf{y},t)$ as the noise estimator of the true noise $\boldsymbol{\epsilon}$, we have the modified ELBO, $-\mathcal{L}(\theta,\phi,\psi)$, associated with (\ref{eq: vae elbo}): 
\begin{equation*}
\begin{aligned}
    \mathcal{L}(\theta,\phi,\psi) = & \underbrace{\frac{\bar{\alpha}_T}{2}\mathbb{E}_{\mathbf{x}_0}\norm{\mathbf{x}_0}^2_{\boldsymbol{\Sigma}^{-1}_{\text{post}}}+\frac{1}{2}\log\abs{\boldsymbol{\Sigma}_{\text{post}}}}_{\text{Latent Regularization (LR) terms}}
    +\underbrace{\sum_{t=1}^{T}\gamma_t\mathbb{E}_{(\mathbf{x}_0,\mathbf{y}),\boldsymbol{\epsilon}\sim \mathcal{N}(\mathbf{0},\boldsymbol{\Sigma}_{\text{post}})}\norm{\boldsymbol{\epsilon}-\boldsymbol{\epsilon}_\theta(\mathbf{x}_t,\mathbf{y},t)}^2_{\boldsymbol{\Sigma}^{-1}_{\text{post}}}}_{\text{Denoising Matching (DM) terms}} \\
    &+\underbrace{\frac{1}{2}\bigr(\log\frac{\abs{\boldsymbol{\Sigma}_{\text{prior}}}}{\abs{\boldsymbol{\Sigma}_{\text{post}}}}+\text{tr}(\boldsymbol{\Sigma}_{\text{prior}}^{-1}\boldsymbol{\Sigma}_{\text{post}})\bigr)}_{\text{Prior Matching (PM) terms}} + \, \text{C},
\end{aligned}
\end{equation*}
where $\gamma_t= \begin{cases} \frac{\beta_t^2}{2\sigma_t^2\alpha_t(1-\bar{\alpha}_t)}, & t>1 \\ \frac{1}{2\alpha_1}, & t=1\end{cases}$ are weighting factors, $\norm{\mathbf{x}}^2_{\boldsymbol{\Sigma}^{-1}}=\mathbf{x}^T\boldsymbol{\Sigma}^{-1}\mathbf{x}$, $\sigma_t^2=\frac{1-\bar{\alpha}_{t-1}}{1-\bar{\alpha_t}}\beta_t$ and $C$ is some constant not depending on learnable parameters $\theta$, $\phi$, $\psi$.some constant not depending on the learnable parameters $\theta$, $\phi$, and $\psi$.
}

\vspace{0.25cm}

\textit{Derivation:}

Recall our proposed lower bound in (\ref{eq: vae elbo}) to incorporate the conditional DDPM into the VAE framework is given as:
\begin{equation}
    \mathbb{E}_{q_{\phi}(\boldsymbol{\epsilon}|\mathbf{x}_0,\mathbf{y})}\Biggr[\underbrace{\mathbb{E}_{q(\mathbf{x}_{1:T}|\mathbf{x}_0)}\left[\frac{p_{\theta}(\mathbf{x}_{0:T}|\mathbf{y})}{q(\mathbf{x}_{1:T}|\mathbf{x}_0)}\right]}_{-\mathcal{L}(\theta,\phi)}\Biggr] - D_{\text{KL}}\bigr(q_{\phi}(\boldsymbol{\epsilon}|\mathbf{x}_0,\mathbf{y}) || p_{\psi}(\boldsymbol{\epsilon}|\mathbf{y})\bigr).
\label{eq: rg elbo appendix}
\end{equation}

Note that as assumed in standard DDPMs, the forward diffusion process gradually corrupts the data distribution into the prior distribution, which can be achieved by carefully designing the variance schedule for the forward pass, i.e., $\{\beta_t\}_{t=1}^T$, such that $\mathbf{x}_T\to\boldsymbol{\epsilon}$ (as a result of $\bar{\alpha}_T\to0$). More specifically, the $q(\mathbf{x}_T|\mathbf{x}_0)$ of the forward diffusion process converges in distribution to the approximate posterior $q_{\phi}(\boldsymbol{\epsilon}|\mathbf{x}_0,\mathbf{y})$ from the posterior encoder $\phi$. Then, the term $\mathcal{L}(\theta,\phi)$ in (\ref{eq: rg elbo appendix}) suggests training a conditional diffusion model $\theta$ to reverse the diffusion trajectory from the estimated distribution of $\boldsymbol{\epsilon}$ given by the posterior encoder $\phi$ back to the target data distribution of $\mathbf{x}_0$. 

The $-$ELBO $\mathcal{L}(\theta, \phi)$ can be shown to be expanded as \citep{luo2022understanding,sohl2015deep,ho2020denoising}:
\begin{equation}
\begin{aligned}
    \mathcal{L}(\theta,\phi)\coloneq&-\mathbb{E}_{q(\mathbf{x}_{1:T}|\mathbf{x}_0)}\left[\log\frac{p_\theta({\mathbf{x}_{0:T}}|\mathbf{y})}{q(\mathbf{x}_{1:T}|\mathbf{x}_0)}\right] \\
    =&-\mathbb{E}_{q(\mathbf{x}_{1:T}|\mathbf{x}_0)}\left[\log\frac{p(\mathbf{x}_T)\prod_{t=1}^Tp_\theta(\mathbf{x}_{t-1}|\mathbf{x}_t,\mathbf{y})}{\prod_{t=1}^T q(\mathbf{x}_t|\mathbf{x}_{t-1})}\right] \\
    =&-\mathbb{E}_{q(\mathbf{x}_{1:T}|\mathbf{x}_0)}\left[\log\frac{p(\mathbf{x}_T)p_\theta(\mathbf{x}_0|\mathbf{x}_1,\mathbf{y})\prod_{t=2}^T p_\theta(\mathbf{x}_{t-1}|\mathbf{x}_t,\mathbf{y})}{q(\mathbf{x}_1|\mathbf{x}_0)\prod_{t=2}^T q(\mathbf{x}_t|\mathbf{x}_{t-1})}\right] \\
    =&-\mathbb{E}_{q(\mathbf{x}_{1:T}|\mathbf{x}_0)}\left[\log p_\theta(\mathbf{x}_0|\mathbf{x}_1,\mathbf{y})\right] -\mathbb{E}_{q(\mathbf{x}_{1:T}|\mathbf{x}_0)}\left[\log\frac{p(\mathbf{x}_T)}{q(\mathbf{x}_T|\mathbf{x}_0)}\right] 
    -\sum_{t=2}^T\mathbb{E}_{q(\mathbf{x}_{1:T}|\mathbf{x}_0)}\left[\log\frac{p_\theta(\mathbf{x}_{t-1}|\mathbf{x}_t,\mathbf{y})}{q(\mathbf{x}_{t-1}|\mathbf{x}_t,\mathbf{x}_0)}\right] \\
    =&-\mathbb{E}_{q(\mathbf{x}_1|\mathbf{x}_0)}\left[\log p_\theta(\mathbf{x}_0|\mathbf{x}_1,\mathbf{y})\right] -\mathbb{E}_{q(\mathbf{x}_T|\mathbf{x}_0)}\left[\log\frac{p(\mathbf{x}_T)}{q(\mathbf{x}_T|\mathbf{x}_0)}\right] 
    -\sum_{t=2}^T\mathbb{E}_{q(\mathbf{x}_t,\mathbf{x}_{t-1}|\mathbf{x}_0)}\left[\log\frac{p_\theta(\mathbf{x}_{t-1}|\mathbf{x}_t,\mathbf{y})}{q(\mathbf{x}_{t-1}|\mathbf{x}_t,\mathbf{x}_0)}\right] \\
    =&\mathcal{L}_0+\mathcal{L}_T+\sum_{t=2}^T\mathcal{L}_{t-1},
\label{eq: ddpm elbo appendix}
\end{aligned}
\end{equation}
where
\begin{align}
    \mathcal{L}_0&\coloneq -\mathbb{E}_{q(\mathbf{x}_1|\mathbf{x}_0)}\left[\log p_\theta(\mathbf{x}_0|\mathbf{x}_1,\mathbf{y})\right], \\
    \mathcal{L}_{t-1}&\coloneq \mathbb{E}_{q(\mathbf{x}_t|\mathbf{x}_0)}\left[\mathcal{D}_{\text{KL}}\left(q(\mathbf{x}_{t-1}|\mathbf{x}_t,\mathbf{x}_0)||p_\theta(\mathbf{x}_{t-1}|\mathbf{x}_t,\mathbf{y})\right)\right], \\    
    \mathcal{L}_T&\coloneq \mathcal{D}_{\text{KL}}\left(q(\mathbf{x}_T|\mathbf{x}_0)||p(\mathbf{x}_T)\right).
\label{eq: ddpm elbo appendix next}
\end{align}

According to \citet{lee2021priorgrad}, the terms of the loss function for training the noise estimator network $\theta$ of the conditional DDPM for an arbitrary $\boldsymbol{\epsilon}\sim\mathcal{N}(\mathbf{0},\boldsymbol{\Sigma})$ can be explicitly written as: 

\begin{equation}
\begin{aligned}
    \mathcal{L}_0=&\frac{1}{2}\log\left((2\pi\beta_1)^d\abs{\boldsymbol{\Sigma}}\right)+\frac{1}{2\alpha_1}
    \mathbb{E}_{\mathbf{x}_0,\boldsymbol{\epsilon}\sim\mathcal{N}(\mathbf{0},\boldsymbol{\Sigma})}
    \norm{\boldsymbol{\epsilon}-\boldsymbol{\epsilon}_\theta(\mathbf{x}_1,\mathbf{y},1)}^2_{\boldsymbol{\Sigma}^{-1}}, \\\\
    \mathcal{L}_{t-1}=&\frac{\beta_t}{2\alpha_t(1-\bar{\alpha}_{t-1})}\mathbb{E}_{\mathbf{x}_0,\boldsymbol{\epsilon}\sim\mathcal{N}(\mathbf{0},\boldsymbol{\Sigma})}
    \norm{\boldsymbol{\epsilon}-\boldsymbol{\epsilon}_\theta(\mathbf{x}_t,\mathbf{y},t)}^2_{\boldsymbol{\Sigma}^{-1}} \\
    =&\frac{\beta_t^2}{2\sigma_t^2\alpha_t(1-\bar{\alpha}_t)}\mathbb{E}_{\mathbf{x}_0,\boldsymbol{\epsilon}\sim\mathcal{N}(\mathbf{0},\boldsymbol{\Sigma})}
    \norm{\boldsymbol{\epsilon}-\boldsymbol{\epsilon}_\theta(\mathbf{x}_t,\mathbf{y},t)}^2_{\boldsymbol{\Sigma}^{-1}}, \\\\
    \mathcal{L}_T=&\frac{\bar{\alpha}_T}{2}\mathbb{E}_{\mathbf{x}_0}\norm{\mathbf{x}_0}^2_{\boldsymbol{\Sigma}^{-1}}-\frac{d}{2}(\bar{\alpha}_T+\log(1-\bar{\alpha}_T)),
\label{eq: loss terms priorgrad}
\end{aligned}
\end{equation}
with $\bar{\alpha}_t\coloneq\prod_{i=1}^t\alpha_i$ and $\alpha_t\coloneq 1-\beta_t$ for $t=1,\dots,T$ where $\{\beta_t\}_{t=1}^{T}$ is the noise variance schedule as a hyperparameter, $d$ is the parameter freedom and $\sigma_t^2=\frac{1-\bar{\alpha}_{t-1}}{1-\bar{\alpha_t}}\beta_t$.

In our case, we have assumed modeling of the posterior distribution where the $\boldsymbol{\epsilon}$ is sampled from as the zero-mean Gaussian $\boldsymbol{\epsilon}\sim\mathcal{N}(\mathbf{0},\boldsymbol{\Sigma}_{\text{post}})$ where the covariance $\boldsymbol{\Sigma}_{\text{post}}\coloneq\boldsymbol{\Sigma}_{\text{post}}(\mathbf{x}_0,\mathbf{y};\phi)$ is estimated by the Posterior Net $\phi$, taking both the ground truth data $\mathbf{x}_0$ and the conditioner $\mathbf{y}$ as input.
By directly plugging in $\boldsymbol{\Sigma}=\boldsymbol{\Sigma}_{\text{post}}$ for each term in (\ref{eq: loss terms priorgrad}), we obtain:
\begin{equation}
    \mathcal{L}(\theta,\phi)=\frac{\bar{\alpha}_T}{2}\mathbb{E}_{\mathbf{x}_0}\norm{\mathbf{x}_0}^2_{\boldsymbol{\Sigma}^{-1}_{\text{post}}}+\frac{1}{2}\log\abs{\boldsymbol{\Sigma}_{\text{post}}} 
    +\sum_{t=1}^{T}\gamma_t\mathbb{E}_{(\mathbf{x}_0,\mathbf{y}),\boldsymbol{\epsilon}\sim\mathcal{N}(\mathbf{0},\boldsymbol{\Sigma}_{\text{post}})}\norm{\boldsymbol{\epsilon}-\boldsymbol{\epsilon}_\theta(\underbrace{\sqrt{\bar{\alpha}_t}\mathbf{x}_0+\sqrt{1-\bar{\alpha}_t}\boldsymbol{\epsilon}}_{\mathbf{x}_t},\mathbf{y},t)}^2_{\boldsymbol{\Sigma}^{-1}_{\text{post}}}+\textit{C},
\label{eq: prop 2 ddpm}
\end{equation}
where
\begin{equation*}
    \gamma_t= 
        \begin{cases}
        \frac{\beta_t^2}{2\sigma_t^2\alpha_t(1-\bar{\alpha}_t)}, & t>1 \\
        \frac{1}{2\alpha_1}, & t=1
        \end{cases}
\end{equation*}
and $C$ is some constant not depending on the learnable parameters.

For the prior matching term in (\ref{eq: rg elbo appendix}), we can utilize the analytic form of the KL divergence between two Gaussians which leads to:
\begin{equation}
    D_{\text{KL}}\bigr(q_{\phi}(\boldsymbol{\epsilon}|\mathbf{x}_0,\mathbf{y}) || p_{\psi}(\boldsymbol{\epsilon}|\mathbf{y})\bigr)=\frac{1}{2}\bigr(\log\frac{\abs{\boldsymbol{\Sigma}_{\text{prior}}}}{\abs{\boldsymbol{\Sigma}_{\text{post}}}}+\text{tr}(\boldsymbol{\Sigma}_{\text{prior}}^{-1}\boldsymbol{\Sigma}_{\text{post}})\bigr),
\label{eq: prop 2 kl div}
\end{equation}
where the covariances $\boldsymbol{\Sigma}_{\text{prior}}\coloneq\boldsymbol{\Sigma}_{\text{prior}}(\mathbf{y};\psi)$ and $\boldsymbol{\Sigma}_{\text{post}}\coloneq\boldsymbol{\Sigma}_{\text{post}}(\mathbf{x}_0,\mathbf{y};\phi)$.

Combining (\ref{eq: prop 2 ddpm}) and (\ref{eq: prop 2 kl div}), we have obtained the $-$ELBO $\mathcal{L}(\theta,\phi,\psi)$ of Proposition \ref{proposition: 2}.

\section{Implementation Details}
\label{sec: appendix exp}

\subsection{Algorithms}
\label{sec: algorithms}
\begin{minipage}{0.53\textwidth}
\begin{algorithm}[H]
    \footnotesize
    \caption{Training of RestoreGrad}
    \For{$i=0,1,2...,N_{\text{iter}}$}
    {
        Sample $(\mathbf{x}_0,\mathbf{y})\sim q_{\text{data}}(\mathbf{x}_0,\mathbf{y})$ \\
        $\boldsymbol{\Sigma}_{\text{prior}}\gets$Prior Net($\mathbf{y};\psi$) \\
        $\boldsymbol{\Sigma}_{\text{post}}\gets$Posterior Net($\mathbf{x}_0,\mathbf{y};\phi$) \\
        Sample $\boldsymbol{\epsilon}\sim \mathcal{N}(\mathbf{0},\boldsymbol{\Sigma}_{\text{post}})$ and $t\sim \mathcal{U}(\{1,\dots,T\})$ \\
        $\mathbf{x}_t=\sqrt{\bar{\alpha}_t}\mathbf{x}_0+\sqrt{1-\bar{\alpha}_t}\boldsymbol{\epsilon}$ \\
        $\mathcal{L}_{\text{LR}}=\bar{\alpha}_T||\mathbf{x}_0||^2_{\boldsymbol{\Sigma}^{-1}_{\text{post}}}+\log\abs{\boldsymbol{\Sigma}_{\text{post}}}$ \\
        $\mathcal{L}_{\text{DM}}=||\boldsymbol{\epsilon}-\boldsymbol{\epsilon}_\theta(\mathbf{x}_t,\mathbf{y},t)||^2_{\boldsymbol{\Sigma}^{-1}_{\text{post}}}$ \\
        $\mathcal{L}_{\text{PM}}=\log\frac{\abs{\boldsymbol{\Sigma}_{\text{prior}}}}{\abs{\boldsymbol{\Sigma}_{\text{post}}}}+\text{tr}(\boldsymbol{\Sigma}_{\text{prior}}^{-1}\boldsymbol{\Sigma}_{\text{post}})$ \\
        Update $\theta,\psi,\phi$ with $\nabla_{\theta,\psi,\phi} \,\, \eta\mathcal{L}_{\text{LR}}+\mathcal{L}_{\text{DM}}+\lambda\mathcal{L}_{\text{PM}}$ 
    }
\label{algo: training}
\end{algorithm}
\end{minipage}
\hfill
\begin{minipage}{0.44\textwidth}
\begin{algorithm}[H]
    \footnotesize
    \caption{Sampling of RestoreGrad}
    $\boldsymbol{\Sigma}_{\text{prior}}\gets$Prior Net($\mathbf{y};\psi$) \\
    Sample $\mathbf{x}_T\sim \mathcal{N}(\mathbf{0},\boldsymbol{\Sigma}_{\text{prior}})$ \\
    \For{$t=T,T-1,...,1$}
    {
        \eIf {$t>0$}{Sample $\boldsymbol{\epsilon}\sim\mathcal{N}(\mathbf{0},\boldsymbol{\Sigma}_{\text{prior}})$}{$\boldsymbol{\epsilon}=0$}
        $\mathbf{x}_{t-1}=\frac{1}{\sqrt{\alpha_t}}\left(\mathbf{x}_t-\frac{1-\alpha_t}{\sqrt{1-\bar{\alpha}}_t}\boldsymbol{\epsilon}_{\theta}(\mathbf{x}_t,\mathbf{y},t)\right)+\sigma_t\boldsymbol{\epsilon}$
        
    }
    \Return $\mathbf{x}_0$
\label{algo: sampling}
\end{algorithm}
\end{minipage}

\subsection{Experiments on Speech Enhancement (SE)}
\subsubsection{Dataset}
We used the VoiceBank+DEMAND dataset \citep{valentini2016investigating} with the same experimental setup as in previous work \citep{pascual2017segan,phan2020improving,strauss2021flow,lu2022conditional} to perform a direct comparison. The clean speech and noise recordings were provided from the VoiceBank corpus \citep{veaux2013voice} and the Diverse Environments Multichannel Acoustic Noise Database (DEMAND) \citep{thiemann2013diverse}, respectively, each recorded with sampling rate of 48kHz. Noisy speech inputs used for training were composed by mixing the two datasets with four signal-to-noise ratio (SNR) settings from \{0, 5, 10, 15\} dB, using 10 types of noise (2 artificially generated + 8 real recorded from the DEMAND dataset) and 28 speakers from the Voice Bank corpus. The test set inputs were made with four SNR settings different from the training set, i.e., \{2.5, 7.5, 12.5, 17.5\} dB, using the remaining 5 noise types from DEMAND and 2 speakers from the VoiceBank corpus. There are totally 11,527 utterances for training and 824 for testing. Note that the speaker and noise classes were uniquely selected for the training and test sets. The dataset is publicly available at: \url{https://datashare.ed.ac.uk/handle/10283/2826}. In our experiments, the audio steams were resampled to 16kHz sampling rate.

\subsubsection{Model Architecture}
\noindent\textbf{Baseline DDPM-based SE Model:} The baseline SE model considered in this work, i.e., CDiffuSE \citep{lu2022conditional}, performs enhancement in the time domain. We utilized the CDiffuSE base model, which has approximately 4.28M learnable parameters, from the implementation at: \url{https://github.com/neillu23/CDiffuSE}. The model is implemented based on DiffWave \citep{kong2020diffwave}, a versatile diffusion probabilistic model for conditional and unconditional waveform generation. The basic model structure of CDiffuSE is similar to that of DiffWave. However, since the target task is SE, CDiffuSE uses the noisy spectral features as the conditioner, rather than the clean Mel-spectral features used in DiffWave utilized for vocoders. After the reverse process is completed, the enhanced waveform further combine the observed noisy signal $\mathbf{y}$ with the ratio 0.2 to recover the high frequency speech in the final enhanced waveform, as suggested in \citet{abd2008speech,defossez2020real}. 

\noindent\textbf{PriorGrad:} We implemented the PriorGrad \citep{lee2021priorgrad} on top of the CDiffuSE model by using a data-dependent prior $\mathcal{N}(\mathbf{0},\boldsymbol{\Sigma}_{\text{y}})$, where $\boldsymbol{\Sigma}_{\text{y}}$ is the covariance of the prior distribution computed based on using the mel-spectrogram of the noisy input $\mathbf{y}$. Following the application to vocoder in \citet{lee2021priorgrad}, we leveraged a normalized frame-level energy of the mel-spectrogram for acquiring data-dependent prior, exploiting the fact that the spectral energy contains an exact correlation to the waveform variance (by Parseval’s theorem \citep{stoica2005spectral}). More specifically, we computed the frame-level energy by taking the square root of the sum of $\exp(\mathbf{Y})$ over the frequency axis for each time frame, where $\mathbf{Y}$ is the mel-spectrogram of the noisy input $\mathbf{y}$ from the training data. We then normalized the frame-level energy to a range of $(0, 1]$ to acquire the data-dependent diagonal variance $\boldsymbol{\Sigma}_{Y}$. Then we upsampled $\boldsymbol{\Sigma}_{Y}$ in the frame level to $\boldsymbol{\Sigma}_{y}$ in the waveform-level using the given hop length of computing the mel-spectrogram. We imposed the minimum standard deviation of the prior to 0.1 through clipping to ensure numerical stability during training, as suggested in \citet{lee2021priorgrad}.

\noindent\textbf{Prior Net and Posterior Net for RestoreGrad:} The additional encoder modules for the RestoreGrad adopt the ResNet-20 architect \citep{he2016deep} using the implementation from: \url{https://github.com/akamaster/pytorch_resnet_cifar10}. We suitably modified the original 2-D convolutions in ResNet-20 to 1-D convolutions for waveform processing. The modified ResNet-20 model has only 93K learnable parameters (only 2\% of the size of CDiffuSE model). The Prior Net takes the noisy speech waveform $\mathbf{y}$ as input, while the Posterior Net takes both the clean and noisy waveforms, $\mathbf{x}_0$ and $\mathbf{y}$, as input, which are concatenated along the channel dimension. We employed the exponential nonlinearity at the network output for estimating the variances of the prior and posterior distributions.

\subsubsection{Optimization and Inference}
We used the same configurations of CDiffuSE (Base) \citep{lu2022conditional} for optimizing all the models, where the batch size was 16, the Adam optimizer was used with a learning rate of $2\times 10^{-4}$, and the diffusion steps $T=50$ with linearly spaced $\beta_t\in[10^{-4}, 0.035]$. For RestoreGrad, we imposed the minimum standard deviation $\sigma_{\text{min}}=0.1$ by adding it to the output of the Prior Net and Posterior Net to ensure stability during training. The fast sampling scheme in \citet{kong2020diffwave} was used in the reverse processes with 6-step inference schedule $\beta^{\text{infer}}_t=[10^{-4}, 10^{-3}, 0.01, 0.05, 0.2, 0.35]$. The models were trained on one NVIDIA Tesla V100 GPU of 32 GB CUDA memory and finished training for 96 epochs in 1 day.

\subsubsection{Evaluation Metrics}
\noindent\textbf{PESQ:} a speech quality measure using the wide-band version recommended in ITU-T P.862.2 \citep{itu862}. It basically models the mean opinion scores (MOS) that cover a scale from 1 (bad) to 5 (excellent). We used the Python-based PESQ implementation from: \url{https://github.com/ludlows/python-pesq}.

\noindent\textbf{SI-SNR:} a variant of the conventional SNR measure taking into account the scale-invariance of audio signals. The SI-SDR is a more robust and meaningful metric than the traditional SNR for measuring speech quality. A higher SI-SNR score indicates better perceptual speech quality. We adopted the SI-SNR implementation from: \url{https://lightning.ai/docs/torchmetrics/stable/audio/scale_invariant_signal_noise_ratio.html}.

\noindent\textbf{SSNR:} an SNR measure, instead of working on the whole signal, that calculates the average of the SNR values of short segments (segment length $=30$ msec, $75\%$ overlap, $\text{SNR}_{\text{min}}=-10$ dB, $\text{SNR}_{\text{max}}=35$ dB). We use the Python-based SSNR implementation from: \url{https://github.com/schmiph2/pysepm}.

\noindent\textbf{CSIG:} The mean opinion score (MOS) prediction of the signal distortion (from 1 to 5, the higher the better) \citep{hu2007evaluation}. We used the implementation from: \url{https://github.com/schmiph2/pysepm}.

\noindent\textbf{CBAK:} MOS prediction of the intrusiveness of background noises (from 1 to 5, the higher the better) \citep{hu2007evaluation}. We used the implementation from: \url{https://github.com/schmiph2/pysepm}.

\noindent\textbf{COVL:} MOS prediction of the overall effect (from 1 to 5, the higher the better) \citep{hu2007evaluation}. We used the implementation from: \url{https://github.com/schmiph2/pysepm}.

\subsection{Experiments on Image Restoration (IR)}

\subsubsection{Datasets}
We used three standard benchmark image restoration datasets considering adverse weather conditions of snow, heavy rain with haze, and raindrops on the camera sensor, following \citet{ozdenizci2023restoring}. 

\noindent\textbf{Snow100K \citep{liu2018desnownet}:} a dataset for evaluation of image desnowing models. The images are split into approximately equal sizes of three Snow100K-S/M/L sub-test sets (with 16,611/16,588/16,801 samples, respectively), indicating the synthetic snow strength imposed via snowflake sizes (light/mid/heavy). We used the Snow100K-L sub-test set for evaluation. The dataset can be downloaded from: \url{https://sites.google.com/view/yunfuliu/desnownet}.

\noindent\textbf{Outdoor-Rain \citep{li2019heavy}:} a dataset of simultaneous rain and
fog which exploits a physics-based generative model to simulate not only dense synthetic rain streaks, but also incorporating more realistic scene views, constructing an inverse
problem of simultaneous image deraining and dehazing. We used the test set, denoted in \citet{li2019heavy} as Test1, which is of size 750 for quantitative evaluations. The dataset can be accessed at: \url{https://github.com/liruoteng/HeavyRainRemoval}.

\noindent\textbf{RainDrop \citep{qian2018attentive}:} a dataset of images with raindrops
introducing artifacts on the camera sensor and obstructing the view. It consists of 861 training images with synthetic raindrops, and a test set of 58 images dedicated for quantitative evaluations, denoted in \citet{qian2018attentive} as RainDrop-A. The dataset is provided at: \url{https://github.com/rui1996/DeRaindrop}.

In addition, we also used the composite dataset for multi-weather IR model training:

\noindent\textbf{AllWeather \citep{valanarasu2022transweather}:} is a dataset of 18,069 samples composed of subsets of training images from the training sets of the three datasets above, in order to create a balanced training set across three weather conditions with a similar approach to \citet{li2020all}. The dataset is publicly available at: \url{https://github.com/jeya-maria-jose/TransWeather}.

\subsubsection{Model Architecture}
\label{appendix: model arch of ir for weather}
\noindent\textbf{Baseline DDPM-based IR Models:} The baseline IR models considered in this work, i.e., the RainDropDiff and WeatherDiff from \citet{ozdenizci2023restoring}, perform patch-based diffusive restoration of the images. The models perform diffusion process at the patch level, where overlapping $p\times p$ patches are taken as input. When sampling, all $p\times p$ patches extracted from the image with a hop size $r$ are processed by the DDPM model, utilizing the mean estimated noise based sampling updates for the overlapping pixels to synthesize the clean image. In this work, we considered $p=64$ and $r=16$, which correspond to the RainDropDiff$_{64}$and WeatherDiff$_{64}$ models (with 110M and 82 M learnable parameters, respectively) provided by the authors at: \url{https://github.com/IGITUGraz/WeatherDiffusion}.

\noindent\textbf{Prior Net and Posterior Net for RestoreGrad:} The additional encoder modules for the RestoreGrad adopt the ResNet-20 architect \citep{he2016deep} using the implementation from: \url{https://github.com/akamaster/pytorch_resnet_cifar10}. The ResNet-20 model has 0.27M learnable parameters, which is less than 0.3\% of the size of RainDropDiff and WeatherDiff. The Prior Net takes the noisy image $\mathbf{y}$ as input, while the Posterior Net takes both the clean and noisy images, $\mathbf{x}_0$ and $\mathbf{y}$, as input, which are concatenated along the channel dimension. We employed the exponential nonlinearity at the network output for estimating the variances of the prior and posterior distributions.

\subsubsection{Optimization and Inference}
\label{appendix: opt and infer of ir for weather}
We used the same configurations of \citet{ozdenizci2023restoring} for optimizing all the models, except the batch size was 4 instead of 16 due to GPU memory limitation. The Adam optimizer with a fixed learning rate of $2\times 10^{-5}$ was used for training models without weight decay, and an exponential moving average with a weight of 0.999 was applied during parameter updates. The number of diffusion steps was $T=1000$ and the noise schedule was $\beta_t\in[10^{-4}, 0.02]$, linearly spaced. For inference, we used the sampling scheme with 10 timesteps for each model that we trained on our own. We did not use the deterministic implicit sampling scheme as in \citet{ozdenizci2023restoring} for our RestoreGrad-based DDPM models as we found using the normal stochastic sampling scheme actually works better. The models were trained on 2 NVIDIA Tesla V100 GPU of 32 GB CUDA memory and finished training for 9,261 epochs on the RainDrop dataset in 12 days and 887 epochs on the AllWeather dataset in 21 days.

\subsubsection{Evaluation Metrics}
\noindent\textbf{PSNR:} a non-linear full-reference metric that compares the pixel values of the original reference image to the values of the degraded image based on the mean squared error \citep{huynh2008scope}. A higher PSNR indicates better reconstruction quality of images in terms of distortion. PSNR can be calculated for the different color spaces. We followed \citet{ozdenizci2023restoring} to compute PSNR based on the luminance channel Y of the YCbCr color space. We used the implementation form \url{https://github.com/JingyunLiang/SwinIR} for PSNR calculation.

\noindent\textbf{SSIM:} a non-linear full-reference metric compares the luminance, contrast and structure of the original and degraded image \citep{wang2004image}. It provides a value from 0 to 1,  the closer the score is to 1, the more similar the degraded image is to the reference image. We followed \citet{ozdenizci2023restoring} to compute SSIM based on the luminance channel Y of the YCbCr color space. We used the implementation form \url{https://github.com/JingyunLiang/SwinIR} for SSIM.

\textbf{LPIPS} \citep{zhang2018unreasonable} and \textbf{FID} \citep{heusel2017gans}: to provide better quantification of perceptual quality over the traditional distortion measures of PSNR and SSIM \citep{blau2018perception,freirich2021theory}. For the LPIPS we used the implementation from \url{https://github.com/richzhang/PerceptualSimilarity}, and for FID we used the  implementation from \url{https://github.com/chaofengc/IQA-PyTorch}. In both metrics, a lower score indicates better perceptual quality of the restored image.

\subsection{Experiments on Generalization to OOD and Realistic Data}

\subsubsection{Datasets}
The additional datasets considered for experiments on realistic data for the IR task are:

\noindent\textbf{RainDS-Real \citep{qian2018attentive}:} is the raindrop removal test subset of the RainDS dataset presented in  \citet{qian2018attentive}. It consists of 98 real-world captured raindrop obstructed images. The dataset is publicly available at: \url{https://github.com/Songforrr/RainDS_CCN}.

\noindent\textbf{Snow100K-Real \citep{liu2018desnownet}:} is the subset of the Snow100K dataset \citep{liu2018desnownet} that consists of 1,329 realistic snowy images for testing real-world restoration cases. The dataset can be accessed at: \url{https://sites.google.com/view/yunfuliu/desnownet}.

The additional dataset considered for experiments on OOD data of the SE task is:

\noindent\textbf{CHiME-3 \citep{barker2017third}:} is a 6-channel microphone recording of talkers speaking in a noisy environment, sampled at 16 kHz. It consists of 7,138 and 1,320 simulated utterances for training and testing, respectively, which are generated by artificially mixing clean speech data with noisy backgrounds of four types, i.e. cafe, bus, street, and pedestrian area. In this paper, we only take the 5-th channel recordings for the experiments. The dataset information can be found at: \url{https://www.chimechallenge.org/challenges/chime3/data}.

\subsubsection{Evaluation Metrics}
The additional evaluation metric used in the corresponding section is:

\noindent\textbf{NIQE:} is a reference-free quality assessment of real-world restoration performance introduced by \citet{mittal2012making} which measures the naturalness of a given image without using any reference image. A lower NIQE score indicates better perceptual image quality. We used the NIQE implementation from: \url{https://github.com/chaofengc/IQA-PyTorch}.

\subsection{Applications to Other IR tasks}

\subsubsection{Datasets}

The datasets considered for experiments on image deblurring and super-resolution tasks are:

\noindent\textbf{RealBlur \citep{rim_2020_ECCV}:} a large-scale dataset of real-world blurred images and ground truth sharp images for learning and benchmarking single image deblurring methods. The images were captured both in the camera raw and JPEG formats, leading to two datasets: \textit{RealBlur-R} from the raw images and \textit{RealBlur-J} from the JPEG images. Each training set consists of 3,758 image pairs and each test set consists of 980 image pairs. The dataset can be downloaded from: \url{https://cg.postech.ac.kr/research/realblur/}.

\noindent\textbf{DIV2K \citep{Agustsson_2017_CVPR_Workshops,Timofte_2017_CVPR_Workshops}:} a dataset of 2K resolution high quality images collected from the Internet as part of the NTIRE 2017 super-resolution challenge. There are 800, 100, and 100 images for training, validation, and testing, respectively. The dataset provides $\times2$, $\times3$, and $\times4$ downscaled images with bicubic and unknown downgrading operations. The dataset can be downloaded from: \url{https://data.vision.ee.ethz.ch/cvl/DIV2K/}.

\subsubsection{Model Architecture}
The baseline conditional DDPM (cDDPM) implements the same architecture as the patch-based denoising diffusion model of WeatherDiff \citep{ozdenizci2023restoring}. The Prior Net and Posterior Net of RestoreGrad also adopt the same ResNet models as in the IR experiments under adverse weather conditions. For more details please refer to Appendix \ref{appendix: model arch of ir for weather}.

\subsubsection{Optimization and Inference}
The models were optimized and inferenced using the same configurations and settings as given in Appendix \ref{appendix: opt and infer of ir for weather} for the IR experiments under adverse weather conditions. The models were trained on 2 NVIDIA Tesla V100 GPU of 32 GB CUDA memory and finished training for 853 epochs on the RealBlur-\{R,J\} dataset each in 5 days and 2000 epochs on the DIV2K-\{$\times2$,$\times4$\} dataset each in 3 days.

\section{Additional Experimental Results}
\label{sec: additional results}

\vspace{0.2cm}

\subsection{Additional Results on SE}
\label{sec: additional results se}

\vspace{0.2cm}

\begin{figure}[!t]
    \centering
    \includegraphics[width=0.85\linewidth]{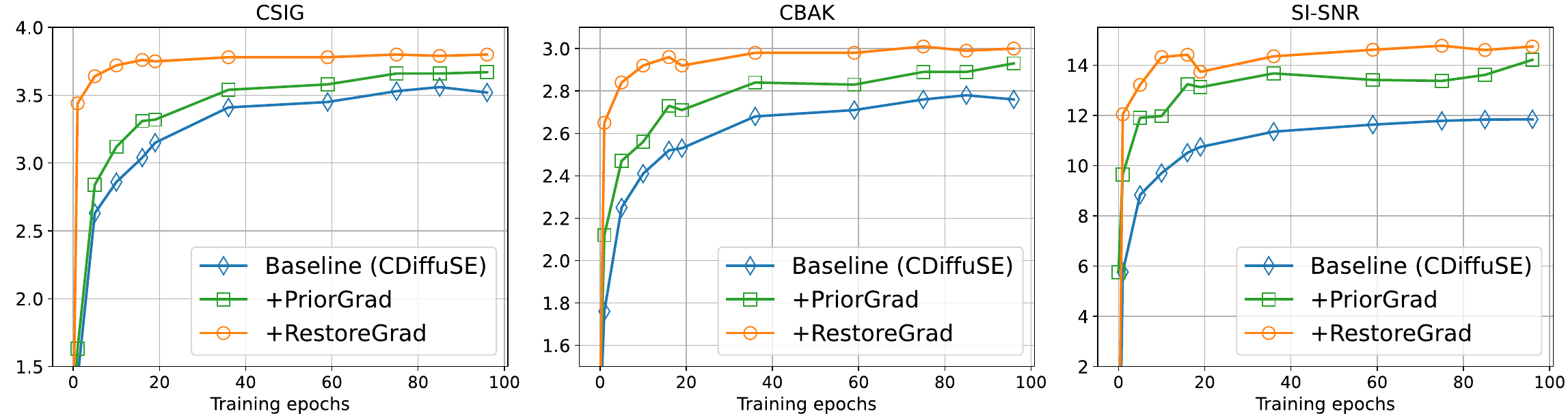}
    \vspace{-0.2cm}
    \caption{Model learning performance in terms of CSIG, CBAK, and SI-SNR metrics. Improved training behavior of RestoreGrad over CDiffuSE and PriorGrad is observed among all metrics.} 
\label{fig: training curves other}
\end{figure}

\noindent\textbf{Model Learning Performance in Terms of Other Metrics:}
In addition to the results evaluated by PESQ and COVL in Figure \ref{fig: training_curve}, we provide the learning curves in terms of the CSIG, CBAK, and SI-SNR metrics in Figure \ref{fig: training curves other}, to further support the advantages of RestoreGrad over the baseline DDPM and PriorGrad for improved training behavior and efficiency.

\vspace{0.2cm}

\begin{figure}[!t]
    \centering
    \includegraphics[width=0.85\linewidth]{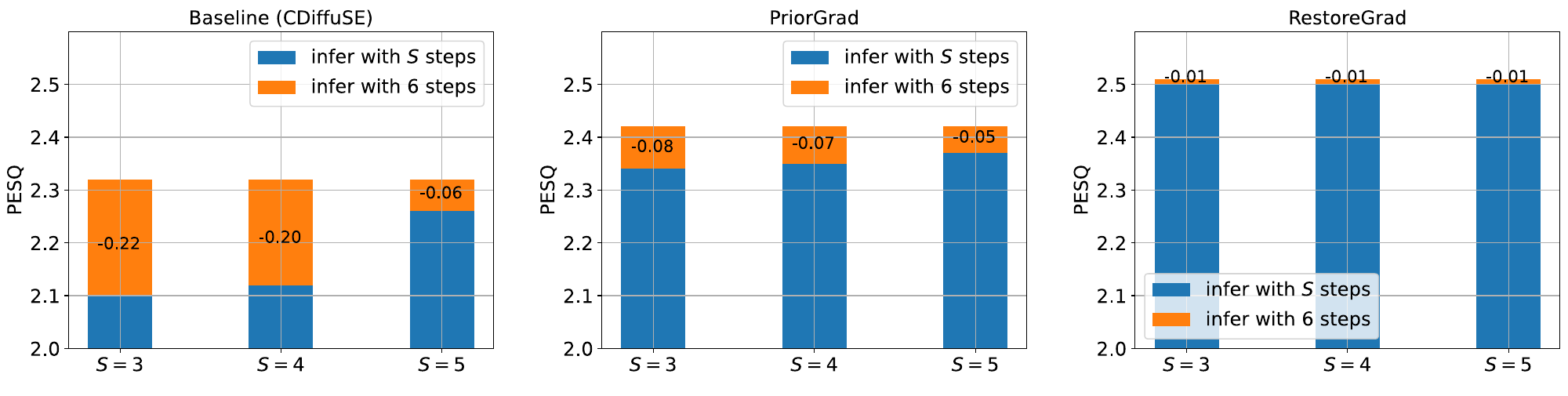}
    \vspace{-0.25cm}
    \caption{Effect of using reduced numbers of sampling steps in inference on the SE performance, in terms of PESQ. RestoreGrad demonstrates the strongest resistance to the reduction in reverse sampling steps for inference.} 
\label{fig: tolerance to reduced interence sampling steps appendix}
\end{figure}

\noindent\textbf{Performance with Using Different Numbers of Inference Steps:}
In Figure \ref{fig: tolerance to reduced interence sampling steps appendix}, we show how the trained diffusion models perform with respect to using different numbers of reverse steps for inference. Specifically, in each case of CDiffuSE, PriorGrad, and RestoreGrad, we trained the model for 96 epochs and then inferenced with $S\in\{3,4,5\}$ reverse steps to compare with the originally adopted $S=6$ steps in \citet{lu2022conditional}. We used $\beta^{\text{infer}}_t=[10^{-4}, 10^{-3}, 0.05, 0.2, 0.35]$ for $S=5$, $\beta^{\text{infer}}_t=[10^{-4}, 0.05, 0.2, 0.35]$ for $S=4$, and $\beta^{\text{infer}}_t=[0.05, 0.2, 0.35]$ for $S=3$. These choices were selected from the subsets of the original noise schedule for $S=6$, i.e., $\beta^{\text{infer}}_t=[10^{-4}, 10^{-3}, 0.01, 0.05, 0.2, 0.35]$, that resulted in best performance of the models. For the figure we can see that as $S$ becomes smaller, the baseline CDiffuSE degrades considerably, while PriorGrad shows certain resistance, and RestoreGrad manages to maintain the high performance. We present more comparison in Table \ref{table: tolerence to infer reduct extra} in terms of SI-SNR, CSIG, CBAK, and COVL metrics. The results further support that RestoreGrad is much more robust to the reduction in sampling steps, achieving the best quality scores in all the metrics over the baseline DDPM and PriorGrad across all sampling steps considered.

\begin{table}[!t]
\centering
\begin{small}
\setlength{\tabcolsep}{3pt} 
\caption{Performance comparison of RestoreGrad with the baseline DDPM (CDiffuSE) and PriorGrad for using various numbers of sampling steps $S$ during inference.}
\vspace{0.25cm}
\label{table: tolerence to infer reduct extra}
\resizebox{0.9\columnwidth}{!}{%
\begin{NiceTabular}{lcccc|cccc|cccc|cccc}
\toprule 
 \multirow{2}{*}{Methods} & \multicolumn{4}{c}{SI-SNR $\uparrow$} & \multicolumn{4}{c}{CSIG $\uparrow$} & \multicolumn{4}{c}{CBAK $\uparrow$} & \multicolumn{4}{c}{COVL $\uparrow$}  \\
 \cmidrule(lr){2-5}
 \cmidrule(lr){6-9}
 \cmidrule(lr){10-13}
 \cmidrule(lr){14-17}
 & \multirow{1}{*}{$S$=6} & \multirow{1}{*}{$S$=5}  & \multirow{1}{*}{$S$=4}  & \multirow{1}{*}{$S$=3}  & \multirow{1}{*}{$S$=6} & \multirow{1}{*}{$S$=5}  & \multirow{1}{*}{$S$=4}  & \multirow{1}{*}{$S$=3}  & \multirow{1}{*}{$S$=6} & \multirow{1}{*}{$S$=5}  & \multirow{1}{*}{$S$=4}  & \multirow{1}{*}{$S$=3}  & \multirow{1}{*}{$S$=6} & \multirow{1}{*}{$S$=5}  & \multirow{1}{*}{$S$=4}  & \multirow{1}{*}{$S$=3} \\
 \midrule
 CDiffuSE \citep{lu2022conditional} & 11.84 & 11.46 & 11.32 & 11.28 & 3.52 & 3.44 & 3.15 & 3.13 & 2.76 & 2.72 & 2.64 & 2.63 & 2.89 & 2.82 & 2.60 & 2.58 \\
 + PriorGrad \citep{lee2021priorgrad} & 14.21 & 13.98 & 13.93 & 13.93 & 3.67 & 3.61 & 3.56 & 3.54 & 2.93 & 2.90 & 2.88 & 2.88 & 3.02 & 2.97 & 2.93 & 2.92 \\
 + RestoreGrad (ours) & \textbf{14.74} & \textbf{14.66} & \textbf{14.64} & \textbf{14.65} & \textbf{3.80} & \textbf{3.77} & \textbf{3.75} & \textbf{3.75} & \textbf{3.00} & \textbf{2.99} & \textbf{2.99} & \textbf{2.99} & \textbf{3.14} & \textbf{3.12} & \textbf{3.11} & \textbf{3.11} \\
 \bottomrule
\end{NiceTabular}
}
\\
\vspace{0.1cm}
*Best values are indicated with bold text.
\end{small}
\vspace{0.1cm}
\end{table}

\vspace{0.2cm}

\begin{figure}[!th]
    \centering
    \includegraphics[width=0.9\linewidth]{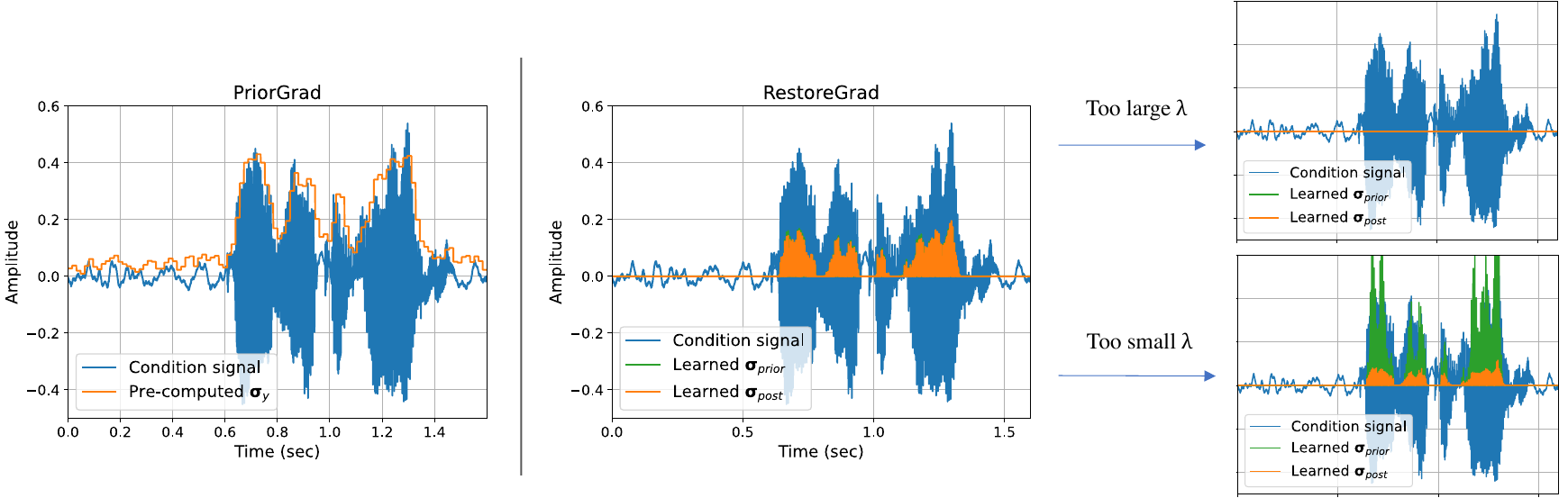}
    \caption{An example of learned latent distribution variances, $\boldsymbol{\Sigma}_{\text{prior}}=\mbox{diag}\{\boldsymbol{\sigma}^2_{\text{prior}}\}$ and $\boldsymbol{\Sigma}_{\text{post}}=\mbox{diag}\{\boldsymbol{\sigma}^2_{\text{post}}\}$ by RestoreGrad, and the effect of the KL weight $\lambda$ of the prior matching loss $\mathcal{L}_{\text{PM}}$ on the resulting latent distribution variances. The pre-computed variance of the handcrafted prior using PriorGrad is also presented for reference purposes.} 
\label{fig: learned noise prior}
\end{figure}

\noindent\textbf{Visualizing the Learned Prior.} It would be interesting to see how the latent noise prior that has been learned by RestoreGrad looks like and how it compares with that of the PriorGrad. In Figure \ref{fig: learned noise prior} we present an example of a randomly chosen noisy speech waveform and the corresponding latent noise $\boldsymbol{\Sigma}_{y}=\mbox{diag}\{\boldsymbol{\sigma}^2_{y}\}$ of PriorGrad and that of RestoreGrad (with $(\eta, \lambda)=(0.1, 0.5)$ for (\ref{eq: elbo of restoregrad})). It can be seen that the variances of the pre-defined (PriorGrad) and learned (RestoreGrad) latent noise distributions are actually quite different, though both show the trend of following the variation of the conditioner signal level. This trend indicates that both latent distributions aim to better approximate the true signal distribution in a more informative manner for improved efficiency, as against the standard Gaussian prior used in the original DDPM. Note that in the RestoreGrad training, we have chosen a proper KL weight $\lambda$ so that the Prior Net distribution matches the Posterior Net distribution reasonably well without harming the reconstruction performance of the DDPM model. On the other hand, using a too large $\lambda$ might lead to a collapsed latent space as the optimization could have put too much emphasis on matching the prior and posterior distribution, discarding the contribution of the reconstruction loss term. In contrast, using a too small $\lambda$ might result in large discrepancy between the learned prior and posterior distributions, as also illustrated in Figure \ref{fig: learned noise prior}. 
Empirically, we found a naive choice of 1 works reasonably well and also for similar values, e.g., 0.5, 10, etc., as similarly observed in the VAE-type model of \citet{kohl2018probabilistic}.

\vspace{0.2cm}

\noindent\textbf{Comparison to Existing Waveform-Domain Generative SE Models:} 
In Table \ref{table: se sota comp 2} we benchmark RestoreGrad with several generative SE approaches. Note that although RestoreGrad performs slightly inferior to DOSE, a recent SE model also based on DiffWave \citep{kong2020diffwave}, it was actually achieved with 4.6$\times$ fewer training epochs.

\begin{table}[!th]
\centering
\begin{small}
\setlength{\tabcolsep}{5pt} 
\caption{Comparison with existing time-domain, generative SE models.}
\vspace{0.1cm}
\label{table: se sota comp 2}
\resizebox{0.5\linewidth}{!}{%
\begin{NiceTabular}{lcccc}
\toprule 
 Methods & PESQ$\uparrow$ & CSIG$\uparrow$ & CBAK$\uparrow$ & COVL$\uparrow$ \\
\midrule
 Unprocessed & 1.97 & 3.35 & 2.44 & 2.63 \\
 \midrule
 SEGAN \citep{pascual2017segan} & 2.16 & 3.48 & 2.94 & 2.80 \\
 DSEGAN \citep{phan2020improving} & 2.39 & 3.46 & \underline{3.11} & 2.90 \\
 SE-Flow \citep{strauss2021flow} & 2.28 & 3.70 & 3.03 & 2.97 \\
 DOSE \citep{tai2024dose} & \textbf{2.56} & \textbf{3.83} & \textbf{3.27} & \textbf{3.19} \\
 \midrule
 CDiffuSE \citep{lu2022conditional} & 2.44 & 3.66 & 2.83 & 3.03 \\
 + RestoreGrad (ours) & \underline{2.51} & \underline{3.80} & 3.00 & \underline{3.14} \\
 
 \bottomrule
\end{NiceTabular}
}
\\
\vspace{0.1cm}
\scriptsize
*Best values in bold and second best values underlined. 
\end{small}
\end{table}

\vspace{0.2cm}

\noindent\textbf{Evaluation Using Automatic Speech Recognition (ASR):} Following \citet{benitadiffar} who perform evaluation of diffusion-based speech generation using ASR, we evaluate the SE model as a front-end denoiser for ASR under noisy environments. To this end, we pre-process the noisy VoiceBand+DEMAND test data samples through the well-trained SE model and feed the denoised audio separately to two pre-trained ASR engines taken from the NVIDIA NeMo toolkit\footnote{https://github.com/NVIDIA/NeMo}: \textit{Conformer-transducer-large} \citep{gulati2020conformer} and \textit{Citrinet-1024} \citep{majumdar2021citrinet}. We report the word error rate (WER) and character error rate (CER) for each ASR engine outcome, where the lower WER / CER indicate better SE performance. The results are presented in Table \ref{table: asr} with all the SE models trained after 96 epochs, inferred using 6 steps. It is interesting to see that CDiffuSE and PriorGrad actually lead to worse performance than the unprocessed speech case for Citrinet ASR. Our RestoreGrad is able to achieve the lowest WER and CER for both ASR models, demonstrating its efficacy for enhancing machine learning capabilities under noisy environments.

\begin{table}[!t]
\centering
\begin{small}
\setlength{\tabcolsep}{2.5pt} 
\caption{Following \citet{benitadiffar} who perform evaluation of diffusion-based speech generation using ASR, we evaluate SE models on two ASR engines (Conformer, Citrinet) for the VoiceBand+DEMAND test set. The results further confirm the superiority of RestoreGrad over the baseline DDPM (CDiffuSE) and PriorGrad.}
\vskip 0.1in
\label{table: asr}
\begin{NiceTabular}{lcc}
\toprule
 \multirow{3}{*}{SE model} & \multicolumn{2}{c}{ASR: WER $\downarrow$ (\%) / CER $\downarrow$ (\%)}  \\
\cmidrule{2-3}
     & Conformer \citep{gulati2020conformer} & Citrinet \citep{majumdar2021citrinet} \\
\midrule
    Unprocessed & 6.62 / 6.15 & 8.69 / 6.86 \\
\midrule
    CDiffuSE \citep{lu2022conditional} & 6.55 / 6.01 & 9.77 / 7.41 \\
    + PriorGrad \citep{lee2021priorgrad} & 6.13 / 5.70 & 9.15 / 7.00 \\
    + RestoreGrad (Ours) & \textbf{5.07} / \textbf{5.27} & \textbf{8.15} / \textbf{6.51} \\
\bottomrule
\end{NiceTabular}
\\
\vspace{0.1cm}
*Best values are indicated with bold text.
\end{small}
\vspace{0.5cm}
\end{table}

\vspace{0.2cm}

\begin{figure}[!t]
    \centering
    \includegraphics[width=0.8\linewidth]{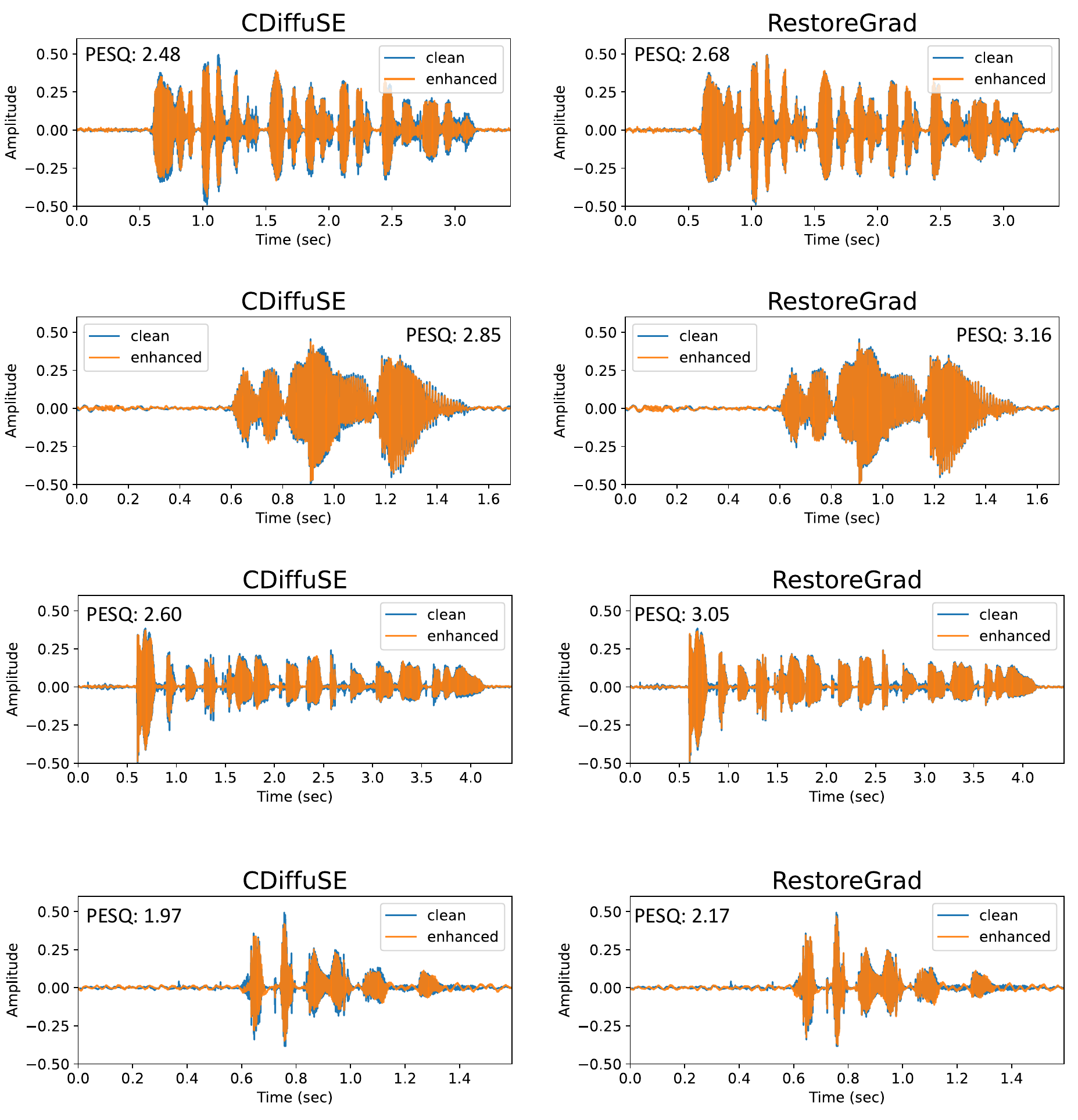}
    \vspace{-0.4cm}
    \caption{Enhanced speech examples of the baseline DDPM (CDiffuSE) and the proposed RestoreGrad for several noisy samples taken from the VoiceBank+DEMAND test set.} 
\label{fig: audio examples}
\end{figure}

\noindent\textbf{Enhanced Speech Examples:}
We present several audio examples in Figure \ref{fig: audio examples} to facilitate the comparison of the baseline DDPM and our RestoreGrad. It can be seen the RestoreGrad is able to recover a better speech signal closer to the target clean speech, which is also reflected by the higher PESQ scores obtained.

\vspace{0.2cm}

\subsection{Additional Results on IR}
\label{sec: additional results on IR appendix}

\vspace{0.2cm}

\noindent\textbf{Comparison to Existing IR Models on RainDrop Dataset:}
We compare our method with existing IR models including AttentiveGAN \citep{qian2018attentive}, DuRN \citep{liu2019dual}, RaindropAttn \citep{quan2019deep}, IDT \citep{xiao2022image} and RDDM \citep{liu2024residual} in Table \ref{table: ir sota comp raindrop} on the RainDrop dataset \citep{qian2018attentive}, where the models were all trained and tested on the same training and test samples. The results of the compared models were taken from \citet{ozdenizci2023restoring} and \citet{liu2024residual}, where the baseline RainDropDiff was trained for 37,042 epochs. \textit{Our RestoreGrad was only trained for 9,261 epochs (4$\times$ fewer than RainDropDiff), and has achieved higher PSNR and SSIM scores than RainDropDiff} (here we report $\text{mean}\pm \text{std}$ of RestoreGrad based on results of 10 independent samplings). Moreover, our performance is comparable to the recent state-of-the-art approach of RDDM, further suggesting the potential of our method to effectively improve baseline DDPM approaches (e.g., RainDropDiff).

\begin{table}[!th]
\centering
\begin{small}
\setlength{\tabcolsep}{5pt} 
\caption{Weather-specific (RainDrop dataset) model comparison.}
\vspace{0.1cm}
\label{table: ir sota comp raindrop}
\resizebox{0.6\linewidth}{!}{%
\begin{tabular}{lcc}
\toprule 
 \multirow{2}{*}{Methods} & \multicolumn{2}{c}{RainDrop}  \\
 \cmidrule(lr){2-3}
 & \multirow{1}{*}{PSNR $\uparrow$} & \multirow{1}{*}{SSIM $\uparrow$} \\
 \midrule
 AttentiveGAN \citep{qian2018attentive} & 31.59 & 0.9170 \\
 DuRN \citep{liu2019dual} &  31.24 & 0.9259 \\
 RaindropAttn \citep{quan2019deep} & 31.44 & 0.9263 \\
 IDT \citep{xiao2022image} & 31.87 & 0.9313 \\
 RDDM \citep{liu2024residual} &  \underline{32.51} & \textbf{0.9563} \\
 \midrule
 RainDropDiff \citep{ozdenizci2023restoring} & 32.29 & 0.9422 \\
 + RestoreGrad (ours) & \textbf{32.69}$\pm$0.03 & \underline{0.9441}$\pm$7e-5 \\
 \bottomrule
\end{tabular}
}
\\
\vspace{0.1cm}
\footnotesize
*Best values in bold and second best values underlined. 
\end{small}
\vspace{0.5cm}
\end{table}

\vspace{0.2cm}

\begin{figure}[!th]
    \centering
    \includegraphics[width=0.7\linewidth]{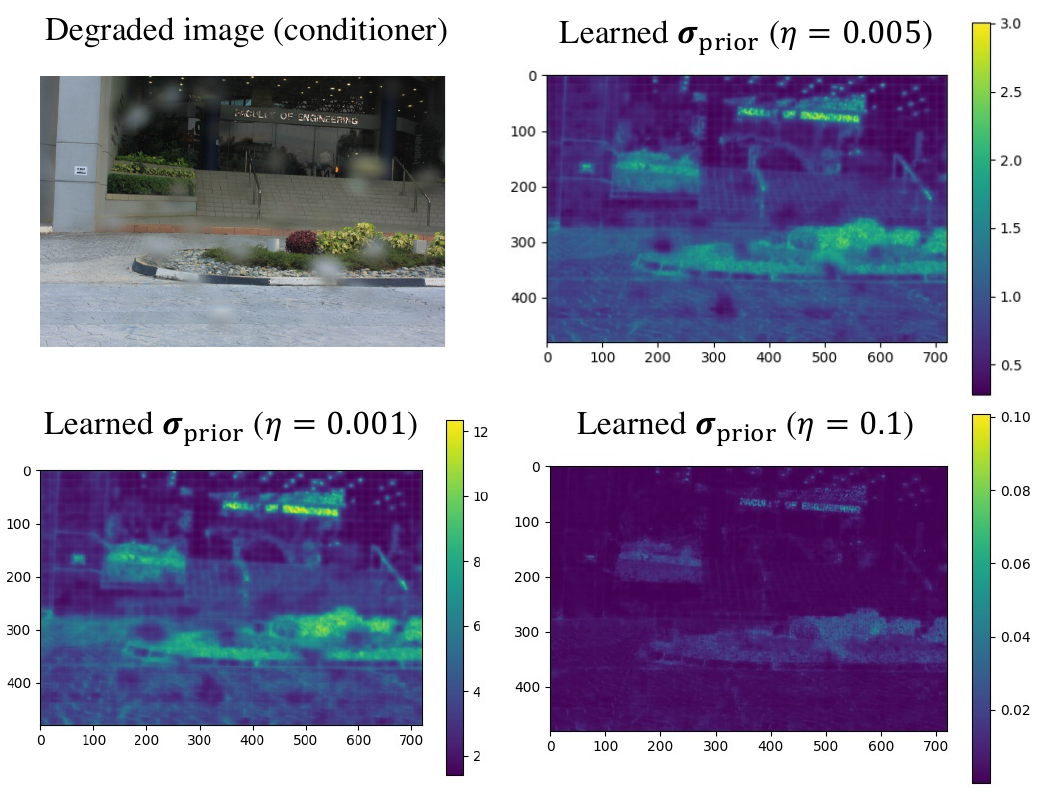}
    \vspace{-0.4cm}
    \caption{Visualization of learned prior distribution variances with various $\eta$ for a sample image taken from the RainDrop test set \citep{qian2018attentive}. Mind the magnitude color bar of each figure. We can see that a larger $\eta$ results in smaller variance of the prior distribution, while a smaller $\eta$ leads to larger variance.} 
\label{fig: learned prior vs eta image}
\vspace{1cm}
\end{figure}

\noindent\textbf{Visualizing the Learned Prior:}
We visualize the learned prior distribution variances for a chosen image input with various $\eta$ values in Figure \ref{fig: learned prior vs eta image} since we are interested in the effect of this newly introduced hyperparameter.
We plot the results for the first channel of the image. The original contaminated image (i.e., the conditioner $\mathbf{y}$ to the DDPM model) is also presented for reference purposes. As expected for the latent space regularization effect, a large $\eta$ results in smaller variances as enforcing stronger regularization, while a small $\eta$ leads to larger variances, as observed in the plots. Moreover, the learned prior appears to preserve the structure of the image, indicating that it tends to learn a prior distribution that approximates the data distribution.

\vspace{0.2cm}

\noindent\textbf{Restoration Performance vs. $\eta$ and $\lambda$:}
We also study the IR performance of the RestoreGrad models trained across various combinations of $\eta$ and $\lambda$ in Table \ref{table: ir eta lambda}, where the models were trained and tested on the RainDrop dataset. The results show that RestoreGrad works effectively for a wide range of $\eta$ and $\lambda$ values to outperform the baseline DDPM model, RainDropDiff from \citet{ozdenizci2023restoring}, which utilizes the standard Gaussian prior for the diffusion process.

\begin{table}[!th]
\centering
\begin{small}
\setlength{\tabcolsep}{5pt} 
\caption{RestoreGrad performance for various $\eta$ and $\lambda$, where the models were trained on the RaindDrop training set \citep{qian2018attentive} for 9,261 epochs and evaluated on the test set. The baseline RainDropDiff model scores reported in the original paper of \citet{ozdenizci2023restoring} (which was trained for 37,042 epochs, 4 times more than our RestoreGrad models) are also presented here for comparison purposes.}
\vspace{0.5cm}
\label{table: ir eta lambda}
\begin{NiceTabular}{cllcc}
    \toprule 
    Model & $\eta$ & $\lambda$ & PSNR $\uparrow$ & SSIM $\uparrow$ \\
    \midrule
    \multirow{5}{*}{RestoreGrad (ours)}
     & $0.05$ & \multirow{5}{*}{$0.1$} & \textbf{32.55} & \textbf{0.9440} \\
     & $0.01$ & & \textbf{32.73} & \textbf{0.9448} \\
     & $0.005$ & & \textbf{32.69} & \textbf{0.9441} \\
     & $0.001$ & & \textbf{32.63} & 0.9404 \\
     & $0.0005$ & & \textbf{32.50} & 0.9405 \\
     \midrule
     \multirow{4}{*}{RestoreGrad (ours)}
     & \multirow{4}{*}{0.005} 
       & 10 & \textbf{32.74} & \textbf{0.9442} \\
      & & 1 & \textbf{32.72} & \textbf{0.9441} \\
      & & 0.1 & \textbf{32.69} & \textbf{0.9441} \\
      & & 0.01 & \textbf{32.41} & 0.9417 \\
      \midrule
      RainDropDiff \citep{ozdenizci2023restoring} & - & - & 32.29 & 0.9422 \\
     \bottomrule
    \end{NiceTabular}
\\
\vspace{0.1cm}
*Values in bold text indicate better scores than the baseline ReainDropDiff model.
\end{small}
\vspace{0.5cm}
\end{table}

\vspace{0.2cm}

\begin{figure}[!thp]
    \centering
    \includegraphics[width=\textwidth]{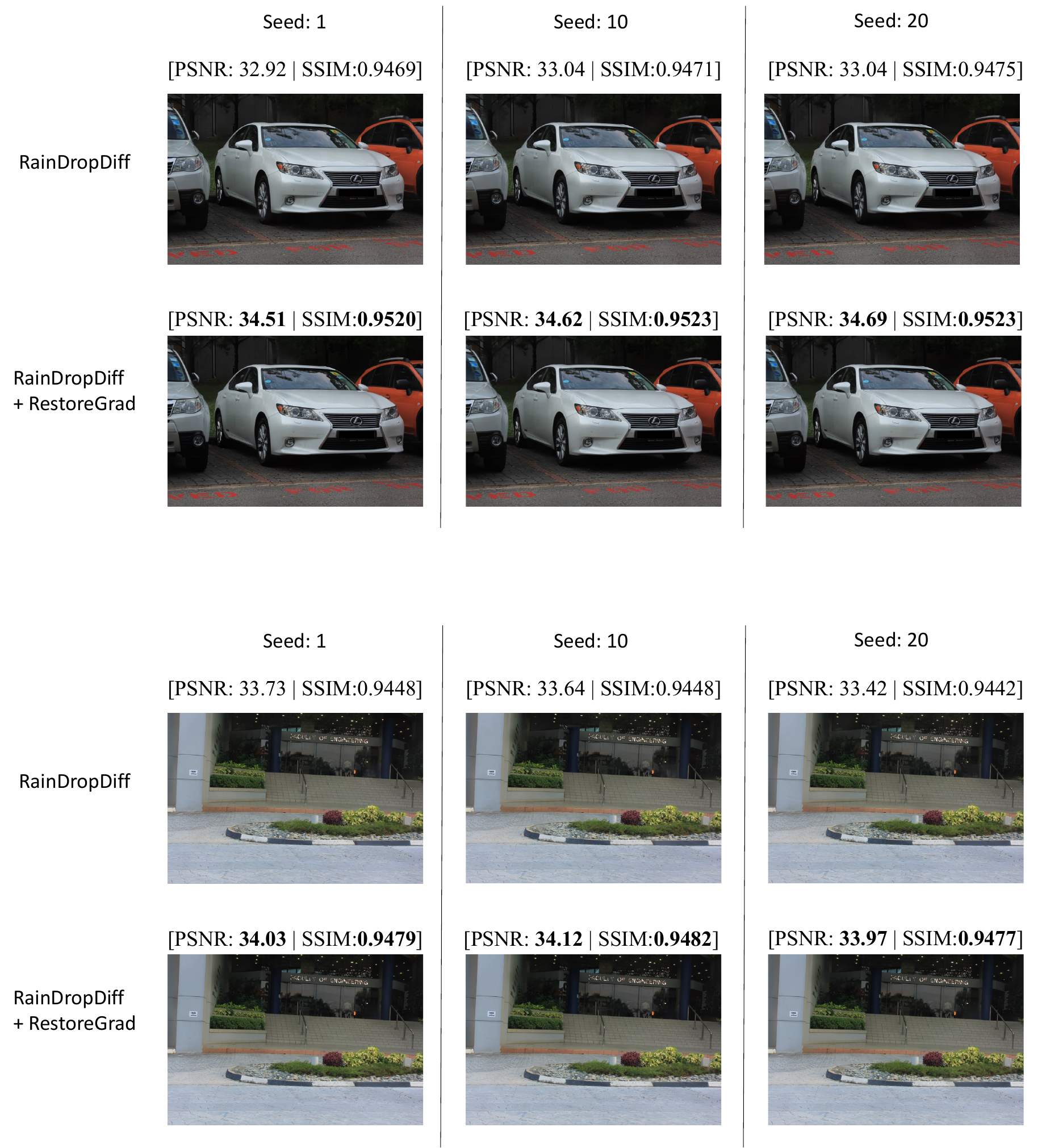}
    \vspace{-0.2cm}
    \caption{Example restored images from different prior noise samples (by using different random seed values) for the baseline DDPM (RainDropDiff) and our approach (RestoreGrad).}
    \label{fig: images_different_seeds}
\end{figure}

\textbf{Images Generated from Different Prior Noise Samples:}
In Figure \ref{fig: images_different_seeds}, we present example images generated for the same conditioner $\mathbf{y}$ using different random seed values (1, 10, and 20) when sampling the latent noise $\boldsymbol{\epsilon}$, for both the baseline DDPM (RainDropDiff) and our RestoreGrad. In the figure, although it is challenging to perceive the difference between the results of different seed values simply by inspecting the images visually, the quality (PSNR and SSIM scores) of the images produced by RestoreGrad is consistently better than the baseline DDPM among the different seed values used, further demonstrating the superiority of our method in restoring higher fidelity signals. 
 
\vspace{0.2cm}

\noindent\textbf{Signal Quality and Encoder Complexity Trade-Offs:} Similar to the computational complexity analysis provided in Table \ref{table: se quality complexity trade-off} for the SE task, we also present the results for the IR task on the RainDrop dataset in Table \ref{table: ir complexity}. From the table, we again observe that the performance of our approach slightly improves with the use of a larger encoder, while also with increasing complexity. However, the computational overhead of our approach, i.e., the additional complexity due to the encoder module(s), is again relatively smaller compared to the adopted DDPM module (i.e., RainDropDiff \citep{ozdenizci2023restoring}), in terms of processing latency ($<$1.3\% of DDPM) and memory usage ($<$30\% of DDPM). 

\begin{table}[!th]
\centering
\begin{small}
\setlength{\tabcolsep}{5pt} 
\caption{IR comparison of RestoreGrad models (on RainDrop dataset) using encoder modules of different sizes and the corresponding latency and GPU memory usage (measured on one NVIDIA Tesla V100 GPU) presented as the ratio of encoder to DDPM.}
\vspace{0.2cm}
\resizebox{0.5\columnwidth}{!}{%
\begin{tabular}{lcc|cc}
\toprule
  Encoder size & PSNR$\uparrow$ & SSIM$\uparrow$ & Proc. Time & Memory \\
\midrule
    Base (0.27M) & 32.65 & 0.9414  & 0.9\% & 16\% \\
    Large (1.9M) & 32.77 & 0.9444  & 1.3\% & 30\% \\
\bottomrule
\end{tabular} 
}
\label{table: ir complexity}
\end{small}
\vspace{0.5cm}
\end{table}

\vspace{0.2cm}

\begin{figure}[!thp]
    \centering
    \includegraphics[width=\textwidth]{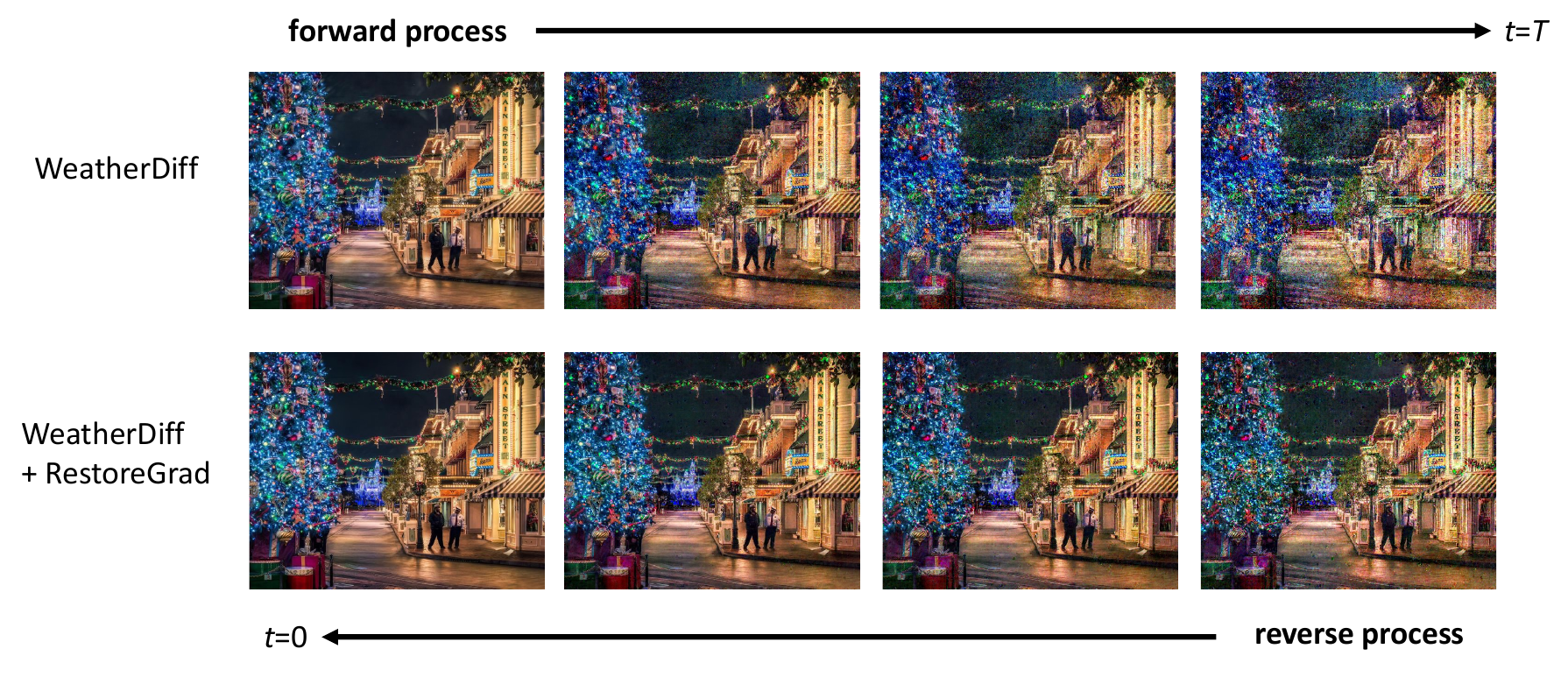}
    \vspace{-0.3cm}
    \caption{Visualization of diffusion processes for the baseline DDPM (WeatherDiff) and our method (RestoreGrad).}
    \label{fig: psnr_vs_infer_steps}
\end{figure}

\noindent\textbf{Visualization of Diffusion Processes:} We present a qualitative analysis of the diffusion processes by showing a sequence of images from $t=0$ to $t=T$ for the baseline conditional diffusion model (WeatherDiff) and with the proposed method (RestoreGrad) in Figure \ref{fig: psnr_vs_infer_steps}. It can be seen that, at a given timestep $t$, the generated image by using RestoreGrad is cleaner than that of WeatherDiff, indicating a better diffusion trajectory developed by using our approach.

\newpage

\noindent\textbf{Experiments on Image Deblurring:}
We apply RestoreGrad to the baseline conditional DDPM (cDDPM) which implements the same architecture as the patch-based DDPM of \citet{ozdenizci2023restoring} used for weather degradations for deblurring. We trained the baseline cDDPM and RestoreGrad models and validated their performance on the RealBlur dataset \citep{rim_2020_ECCV}, a large-scale dataset of real-world blurred images captured both in the camera raw and JPEG formats, leading to two sub-datasets: \textit{RealBlur-R} from the raw images and \textit{RealBlur-J} from the JPEG images. Each training set consists of 3,758 image pairs and each test set consists of 980 image pairs. In Table \ref{table: ir sota comp deblur}, we present results of the baseline cDDPM and RestoreGrad models trained after 853 epochs. We also include scores of two existing models, SRN-DeblurNet \citep{tao2018scale} and DeblurGAN-v2 \citep{kupyn2019deblurgan}, which performed similarly to the baseline cDDPM (taken from results by \citet{rim_2020_ECCV}), as references for comparison. We can see that, except for LPIPS and FID on RealBlur-J, RestoreGrad is able to achieve improved scores than the baseline cDDPM, and outperform the compared methods.

\begin{table}[!th]
\centering
\begin{small}{
\setlength{\tabcolsep}{5pt} 
\caption{Image deblurring of realistic blurred images.}
\vspace{0.1cm}
\label{table: ir sota comp deblur}
\resizebox{0.7\columnwidth}{!}{%
\begin{NiceTabular}{lcccccccc}
\toprule 
 \multirow{2}{*}{Methods} & \multicolumn{4}{c}{RealBlur-J} & \multicolumn{4}{c}{RealBlur-R}  \\
 \cmidrule(lr){2-5}
 \cmidrule(lr){6-9}
 & \multirow{1}{*}{PSNR$\uparrow$} & \multirow{1}{*}{SSIM$\uparrow$} 
 & \multirow{1}{*}{LPIPS$\downarrow$} & \multirow{1}{*}{FID$\downarrow$} 
 & \multirow{1}{*}{PSNR$\uparrow$} & \multirow{1}{*}{SSIM$\uparrow$} & \multirow{1}{*}{LPIPS$\downarrow$} & \multirow{1}{*}{FID$\downarrow$} \\
 \midrule
 SRN-DeblurNet & \underline{31.38} & \underline{0.9091} & - & - & \underline{38.65} & 0.9652 & - & - \\
 DeblurGAN-v2 & 29.69 & 0.8703 & - & - & 36.44 & 0.9347 & - & - \\
 \midrule
 Baseline cDDPM & 30.69 & 0.9043 & \textbf{0.220} & \textbf{15.17} & 37.71 & \underline{0.9777} & \underline{0.126} & \underline{14.46} \\
 + RestoreGrad (ours) & \textbf{31.51} & \textbf{0.9095} & \underline{0.224} & \underline{15.53} & \textbf{38.78} & \textbf{0.9796} & \textbf{0.122} & \textbf{13.61} \\
 \bottomrule
\end{NiceTabular}
}
\\
*Bold text for best and underlined text for second best values.
}
\end{small}
\vspace{0.5cm}
\end{table}

\vspace{0.2cm}

\noindent\textbf{Experiments on Image Super-Resolution:}
We further study the benefits of RestoreGrad over the baseline conditional DDPM (cDDPM) model on image super-resolution tasks with the DIV2K dataset \citep{Agustsson_2017_CVPR_Workshops,Timofte_2017_CVPR_Workshops}. We compare RestoreGrad with the baseline cDDPM model (the same architecture of the patch-based DDPM of WeatherDiff \citep{ozdenizci2023restoring}) for $\times 2$ and $\times 4$ downscale factor subsets (with bicubic downgrading operators). There are 800 images for training and 100 images for validation in each subset. For both subsets, we trained a baseline cDDPM and the RestoreGrad models for 2000 epochs on the training set and evaluated their performance on the corresponding validation set. The results are presented in Table \ref{table: ir sota comp sr}, where we can see that except for the LPIPS metric, RestoreGrad is more beneficial then the baseline cDDPM in terms of achieving better scores in the other three metrics.

\begin{table}[!th]
\centering
\begin{small}{
\setlength{\tabcolsep}{5pt} 
\caption{Comparison of baseline conditional DDPM (cDDPM) and the RestoreGrad on image super-resolution tasks.}
\vspace{0.1cm}
\label{table: ir sota comp sr}
\resizebox{0.7\columnwidth}{!}{%
\begin{NiceTabular}{lcccccccc}
\toprule 
 \multirow{2}{*}{Methods} & \multicolumn{4}{c}{DIV2K $\times$2} & \multicolumn{4}{c}{DIV2K $\times$4}  \\
 \cmidrule(lr){2-5}
 \cmidrule(lr){6-9}
 & \multirow{1}{*}{PSNR $\uparrow$} & \multirow{1}{*}{SSIM $\uparrow$} & \multirow{1}{*}{LPIPS $\downarrow$} & \multirow{1}{*}{FID $\downarrow$} & \multirow{1}{*}{PSNR $\uparrow$} & \multirow{1}{*}{SSIM $\uparrow$} & \multirow{1}{*}{LPIPS $\downarrow$} & \multirow{1}{*}{FID $\downarrow$} \\
 \midrule
 Baseline cDDPM & 27.40 & 0.9291 & \textbf{0.127} & 7.577 & 25.18 & 0.8064 & \textbf{0.269} & 7.849 \\
 + RestoreGrad (ours) & \textbf{27.56} & \textbf{0.9341} & 0.136 & \textbf{7.547} & \textbf{25.56} & \textbf{0.8228} & 0.290 & \textbf{7.839} \\
 \bottomrule
\end{NiceTabular}
}
}
\\
\vspace{0.1cm}
*Better values are indicated with bold text.
\end{small}
\vspace{0.5cm}
\end{table}

\newpage

\begin{figure}[!htp]
    \centering
    \includegraphics[width=\linewidth]{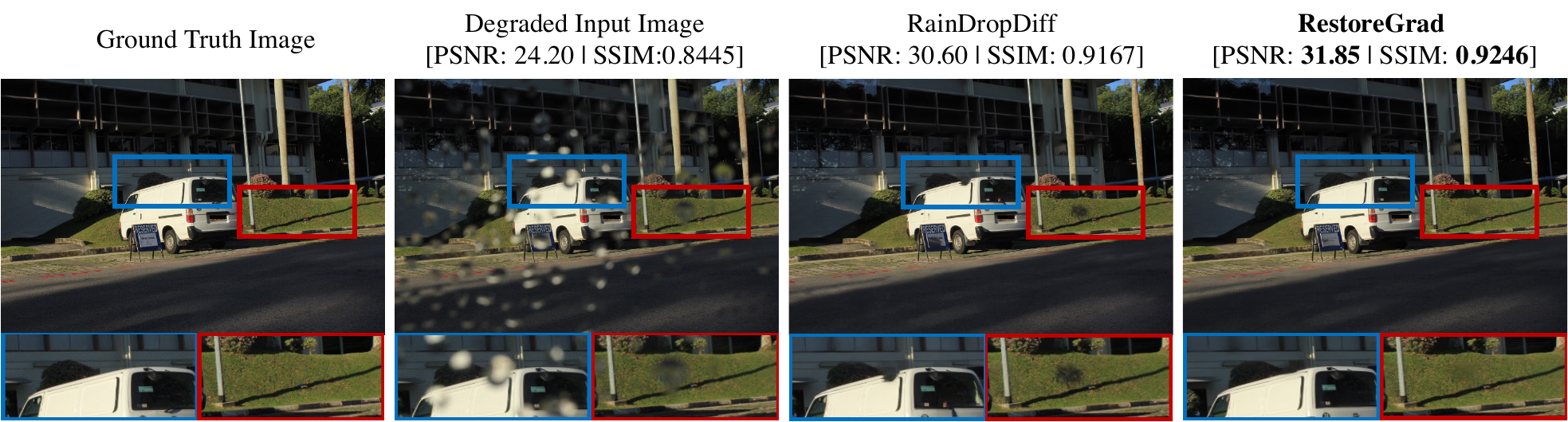}
    \vspace{-0.5cm}
    \caption{Restored images by RainDropDiff \citep{ozdenizci2023restoring} and RestoreGrad (ours) for a test sample from the RainDrop test set.} 
\label{fig: ir example randropdiff}
\end{figure}

\begin{figure}[!htp]
    \centering
    \includegraphics[width=\linewidth]{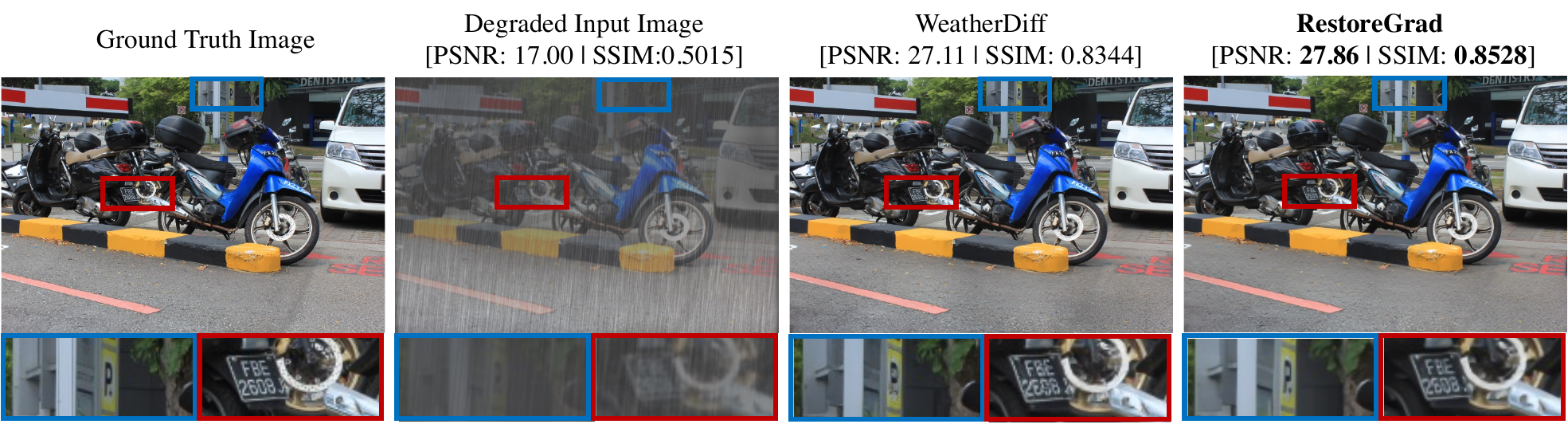}
    \vspace{-0.5cm}
    \caption{Restored images by WeatherDiff \citep{ozdenizci2023restoring} and RestoreGrad (ours) for a test sample from the Outdoor-Rain test set.} 
\label{fig: ir example rainfog}
\end{figure}

\begin{figure}[!htp]
    \centering
    \includegraphics[width=\linewidth]{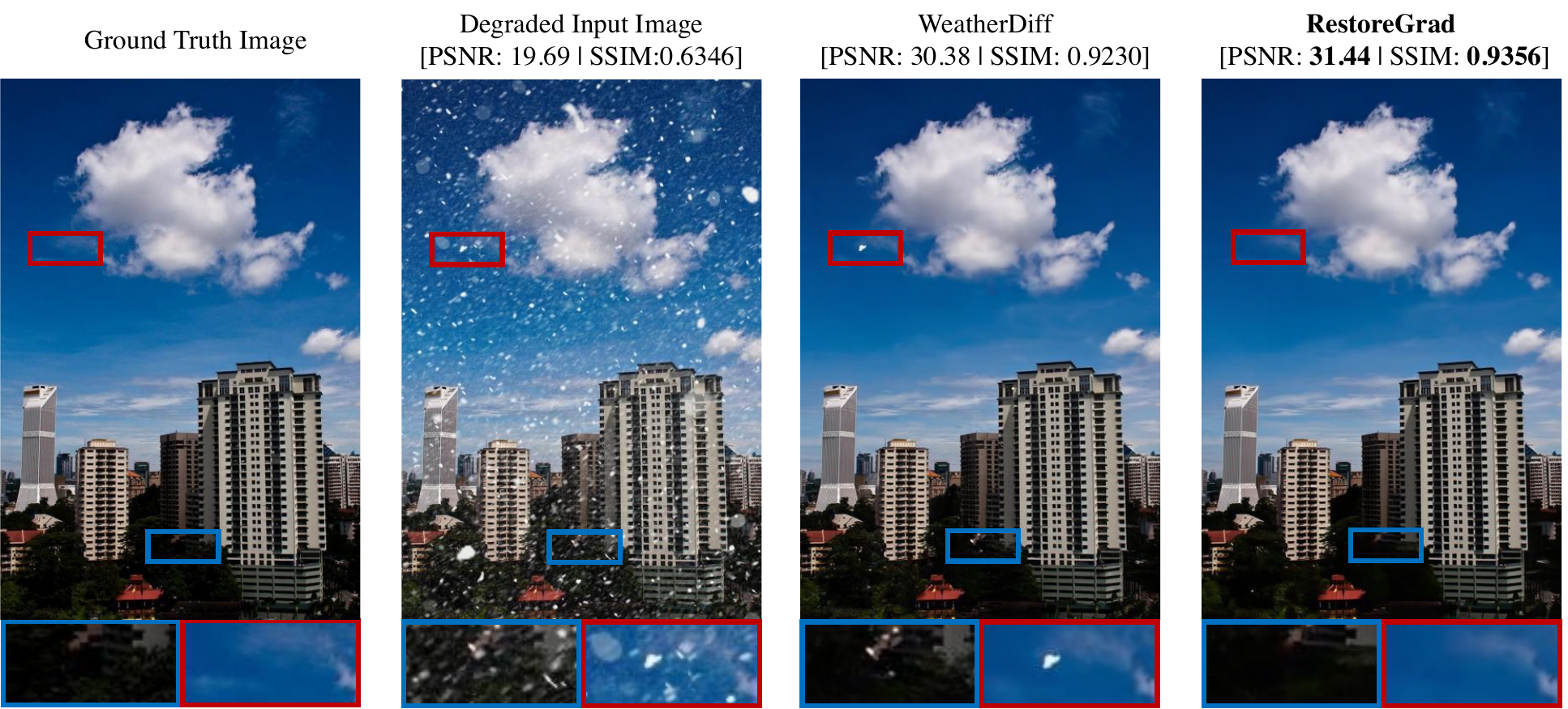}
    \vspace{-0.5cm}
    \caption{Restored images by WeatherDiff \citep{ozdenizci2023restoring} and RestoreGrad (ours) for a test sample from the Snow100K-L test set.} 
\label{fig: ir example snow 2}
\end{figure}

\begin{figure}[!htp]
    \centering
    \includegraphics[width=\linewidth]{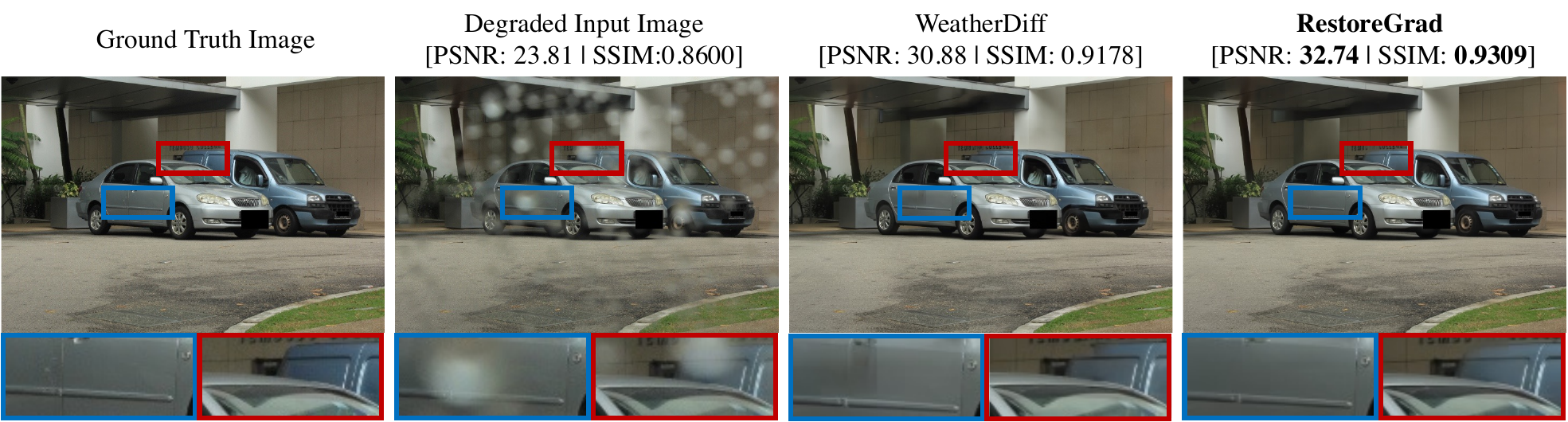}
    \vspace{-0.5cm}
    \caption{Restored images by WeatherDiff \citep{ozdenizci2023restoring} and RestoreGrad (ours) for a test sample from the RainDrop test set.} 
\label{fig: ir example raindrop}
\end{figure}

\begin{figure}[!htp]
    \centering
    \includegraphics[width=\linewidth]{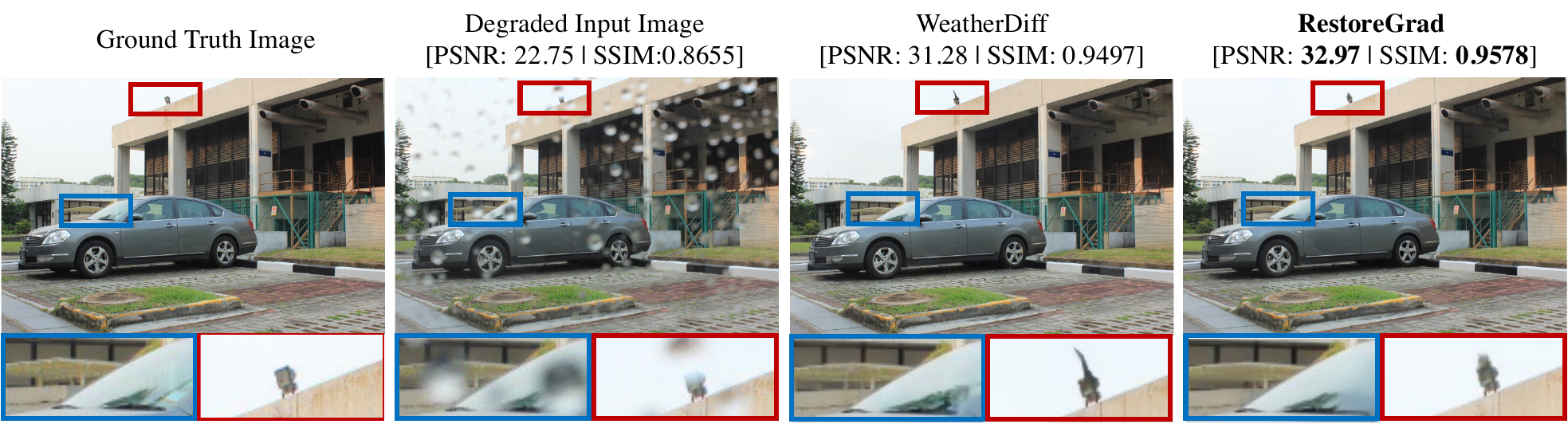}
    \vspace{-0.5cm}
    \caption{Restored images by WeatherDiff \citep{ozdenizci2023restoring} and RestoreGrad (ours) for a test sample from the RainDrop test set.} 
\label{fig: ir example raindrop 2}
\end{figure}

\begin{figure}[!htp]
    \centering
    \includegraphics[width=\linewidth]{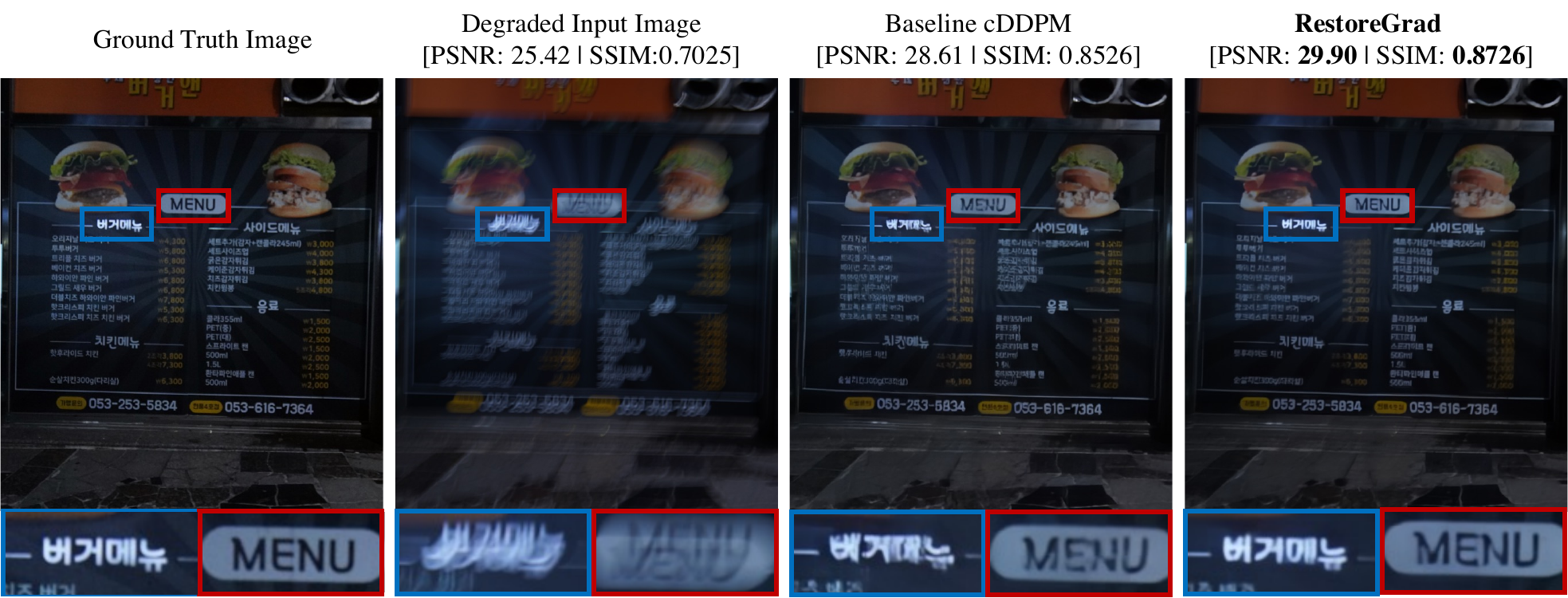}
    \vspace{-0.5cm}
    \caption{Image deblurring examples using a test image taken from the RealBlur test set.} 
\label{fig: ir example blur j}
\end{figure}

\begin{figure}[!htp]
    \centering
    \includegraphics[width=\linewidth]{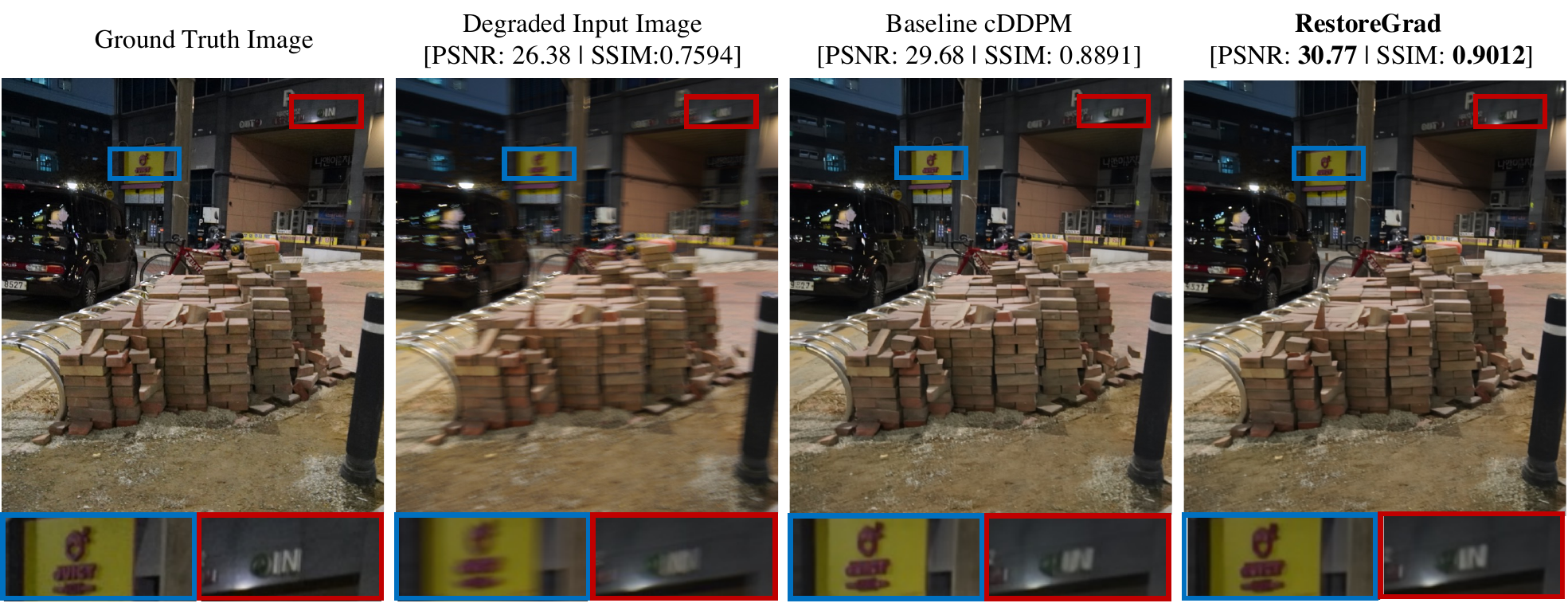}
    \vspace{-0.5cm}
    \caption{Image deblurring examples using a test image taken from the RealBlur test set.} 
\label{fig: ir example blur j 2}
\end{figure}

\noindent\textbf{More Image Restoration Examples:}
We provide more examples in Figures \ref{fig: ir example randropdiff}, \ref{fig: ir example rainfog}, \ref{fig: ir example snow 2},  \ref{fig: ir example raindrop}, \ref{fig: ir example raindrop 2}, for comparing our RestoreGrad with the baseline DDPM approaches (i.e., RainDropDiff, WeatherDiff) of \citet{ozdenizci2023restoring}. The restored image of RainDropDiff in Figure \ref{fig: ir example randropdiff} was obtained by using the model weights trained by ourselves. The restored images of WeatherDiff in Figures \ref{fig: ir example rainfog}, \ref{fig: ir example snow 2}, \ref{fig: ir example raindrop}, \ref{fig: ir example raindrop 2} were obtained by using the trained model weights provided by \citet{ozdenizci2023restoring} at \url{https://github.com/IGITUGraz/WeatherDiffusion}.  We also provide examples of image deblurring in Figures \ref{fig: ir example blur j} and \ref{fig: ir example blur j 2}.

\vspace{0.5cm}

\vspace{0.5cm}

\vspace{0.5cm}

\vspace{0.5cm}

\vspace{0.5cm}

\vspace{0.5cm}
\newpage

\vspace{0.5cm}

\end{document}